\documentclass[acmsmall]{acmart}

\usepackage{graphicx}
\usepackage{url}
\usepackage{subcaption}
\usepackage{code}
\usepackage{soul}

\newcommand{\mycomment}[1]{}
\setcopyright{rightsretained}
\begin{document}

\title{Using Relative Lines of Code to Guide Automated Test Generation for Python}

\author{Josie Holmes}

\affiliation{School of Informatics, Computing \& Cyber Systems, Northern Arizona University}

\email{josie.holmes@nau.edu}

\author{Iftekhar Ahmed}

\affiliation{Donald Bren School of Information and Computer Sciences, University of California, Irvine}
\email{iftekha@uci.edu}

\author{Caius Brindescu}
\affiliation{School of Electrical Engineering and Computer, Oregon State University}
\email{brindesc@oregonstate.edu}

\author{Rahul Gopinath}
\affiliation{Center for IT-Security, Privacy and Accountability (CISPA), University of Saarbr\"ucken}
\email{rahul@gopinath.org}

\author{He Zhang}
\affiliation{School of Electrical Engineering and Computer, Oregon State University}
\email{zhangh7@oregonstate.edu}

\author{Alex Groce}
\affiliation{School of Informatics, Computing \& Cyber Systems, Northern Arizona University}
\email{agroce@gmail.com}

\ccsdesc[500]{Software and its engineering~Software testing and debugging}

\keywords{automated test generation, static code metrics, testing heuristics}

\setcopyright{acmcopyright}
\acmJournal{TOSEM}
\acmYear{2020} 
\acmVolume{1} 
\acmNumber{1} 
\acmArticle{1} 
\acmMonth{1} 
\acmPrice{15.00}
\acmDOI{10.1145/3408896}



\begin{abstract}
Raw lines of code (LOC) is a metric that does not, at first glance, seem
extremely useful for automated test generation.  It is both highly
language-dependent and not extremely meaningful, semantically, within a
language: one coder can produce the same effect with many fewer lines than
another.  However, \emph{relative LOC}, between components of the same project, turns
out to be a highly useful metric for automated testing. In this paper, we make use
of a heuristic based on LOC counts for tested functions to dramatically
improve the effectiveness of automated test generation.  This approach
is particularly valuable in languages where collecting code coverage data
to guide testing has a very high overhead.  We apply the
heuristic to property-based
Python testing using the TSTL (Template Scripting Testing Language) tool.  In our experiments, the simple LOC heuristic can improve branch and
statement coverage by large margins (often more
than 20\%, up to 40\% or more), and improve fault detection by an
even larger margin (usually more than 75\%, and up to 400\% or more).  The LOC
heuristic is also easy to combine with other approaches, and is comparable to, and possibly more effective than, two well-established approaches for guiding random testing.
\end{abstract}

\maketitle


\section{Introduction}

Lines of code (LOC) is an extremely simple way to measure the size or
complexity of a software system or component.  It has clear
disadvantages. First, unless care is taken, the measure itself is
ambiguous, in that ``lines of code'' may mean number of carriage
returns or number of statements, may include comments, and so forth.
Second, lines of code are not comparable across languages: 10 lines of
C and 10 lines of Haskell are not the same, which is evident even in
the size of faults in these languages \cite{GopinathMutants}.
Finally, even within the same language, two different programmers may
express the same functionality using different amounts of code.  E.g.,
in a language like Python, the same list may be constructed using a
five LOC loop or a single LOC list comprehension.  In some cases, such
differences in LOC for the same functionality will signify a
difference in complexity, but in other cases the conceptual and
computational complexity will be identical, despite LOC differences
(e.g., chaining vs. sequential styles in DSLs, as
discussed by Fowler \cite{Fow10}).

Is measuring lines of code, then, pointless, except for very coarse
purposes such as establishing the approximate size of
software systems:  e.g., Mac OS X at 50 million LOC is much
larger than Google Chrome at 5 million LOC, which is much
larger than an AVL tree implementation at 300 LOC? We argue that, to
the contrary, counting LOC is the basis for a powerful heuristic for
improving automated test generation methods based on random testing.
In particular, we show that measuring \emph{relative} LOC between
components of a software system does provide useful information for
test generation.  By
relative LOC we mean that we are not so much concerned with the
absolute LOC size of a program element, but with whether one program
element is larger or smaller than another, and by how much.  Our claim
is that, while LOC is certainly imprecise as a measure of code
complexity or importance, the assumption that relatively larger functions are usually
more complex, more error-prone, and more critical for exploring system
state is actionable: we show that using LOC to bias random testing is
an effective heuristic approach for generating
tests for Python APIs.

\subsection{Small-Budget Automated Test Generation}

QuickCheck \cite{ClaessenH00} and other increasingly popular
\emph{property-based testing tools} \cite{PROPER,Hypothesis} offer
very rapid automatic testing of software, on the fly, based on random
testing \cite{Hamlet94,RandFormal}.  For the Software Under Test (SUT), a property-based testing
tool allows a user to specify some correctness properties (and usually
includes some default properties, such as that executions do not throw
uncaught exceptions), and generates random input values for which the
properties are checked.  Developers seem to expect such
tools to conduct their testing within at most a minute, in order to
provide rapid feedback on newly introduced faults during development,
when the fault is easiest to identify and fix.  The claim that a
minute is a typical expectation derives the
default one minute timeout for the very widely adopted Python Hypothesis \cite{Hypothesis}
testing tool (one of the most sophisticated QuickCheck variants, used
in more than 500 open source projects \cite{HypAdopt}),
and the fact that the original QuickCheck and many imitators such as
ScalaCheck \cite{ScalaCheckDoc}, PropEr \cite{PROPER}, and the Racket version of QuickCheck  use a default of only 100
random tests, which will typically require far \emph{less} than a minute to
perform.   
 Tools for generating Java
unit tests, such as EvoSuite \cite{FA11} and Randoop \cite{Pacheco}
also have default timeouts of one minute and 100 seconds per class to be
tested, respectively. In fact, to our knowledge,
\emph{all} automated testing tools in wide adoption use a default budget
close to 60 seconds or 100 tests, with the exception of fuzzers like {\tt afl-fuzz}
\cite{aflfuzz} intended to detect subtle security
vulnerabilities.  Testing with a limited budget is critical
for using property-based testing in a continuous integration setting,
where testing time per-task on a large project is limited \cite{TravisDoc} to ensure rapid
feedback \cite{HarmanSCAM}.

 Unfortunately, one minute is often not enough
time to effectively generate tests for an SUT using pure random testing.  It is unlikely that low-probability
faults will be exposed.  Moreover, for techniques relying on
genetic algorithms \cite{McMinn04search-basedsoftware} or other machine-learning techniques \cite{ISSRE,ISOLA12},
the overhead of learning, or lack of sufficient training data, may still
result in poor coverage or fault detection in a short testing run.
Even if developers sometimes perform hour-long or overnight automated testing
runs, it is still desirable to find faults or cover code as
quickly as possible; poor 60 second
performance is in a sense equivalent in property-based testing to having a very slow
compiler in code development.

\subsection{The High Cost of Code Coverage}

A further key issue in \emph{lightweight} automated test generation
\cite{ISSRE}, is that many programmers use languages that lack
sophisticated or efficient coverage instrumentation
\cite{yang2007survey,chilakamarri2004reducing}.  In
Python, computing coverage using the {\tt coverage.py}
\cite{Coveragepy} library (the only mature coverage tool for the
language) to guide testing often adds a large overhead, despite
its use of a low-level C tracing implementation.  Collecting coverage
in Python often results in performing \emph{far} less testing for the same
computational budget; in Section \ref{sec:codecov} we show that
turning off code coverage often results in performing \emph{at least
  10\% more, and up to 50
  times as many test actions (e.g. method calls)} in practice, with
median improvement in SUTs we studied of 2.03x (and mean improvement of 6.12x).  Is the advantage of coverage-directed testing
sufficient to overcome this cost?  Even if the answer is affirmative
for C or Java, with fast coverage tools, the answer may often be
``no'' for languages with higher overheads.
Python is not even the worst case: a newly popular language may lack
any effective coverage tool at all; for a long time Rust lacked any convenient way to measure coverage \cite{rustcov}.  Even ``good'' coverage tools may not have a low
enough overhead \cite{ohmann2016optimizing,Tikir:2002:EIC:566172.566186,Agrawal:1994:DSB:174675.175935} or conveniently provide
fast enough access to on-the-fly coverage for efficient testing.

Moreover, testing methods that use coverage information, or even more
expensive (and powerful) tools such as symbolic execution
\cite{GodefroidKS05,KLEE} face an inherent limitation.  As B\"{o}hme
and Soumya \cite{AutoEfficiency} argue, given even a perfect method
for partitioning system behavior by faults, if the method has a cost
(over that of random testing), it will be less
effective than random testing, for some test budgets.
Real-world techniques are not perfect in their defect targeting, and
often impose considerable costs---this is why performing symbolic
execution only on seed tests, generated by some other
method, is now a popular approach in both standard automated test generation
and security-based fuzzing
\cite{issta14,Marinescu:2012:MTS:2337223.2337308,Person:2011:DIS:1993498.1993558,TrailBitsSeeded}.
Small-budget automated test generation, therefore, stands in need
of more methods that  improve on pure random testing but require no
additional computational effort.
Ideally, such methods should be able, like random testing, to work
even without code coverage support.  How can
we discover such methods?  In a sense, we are searching for the
testing equivalent of a ``credit score''---while less accurate than
simply making a loan (that is, running a test) to see how well it performs, it
is also much less costly.  The bound on the cost of computing a credit
score, or measuring LOC, is constant and proportionally much smaller
than the cost of making a (large) loan or performing extensive
testing.  A credit score or rough LOC count is also likely more stable
over time than the details of each proposed loan or set of test
executions.  In essence, we want to equate examining program source
code in simply ways with performing a (fast, approximate) credit check.

The problem is most easily understood when simplified to its essence.
Imagine that you have two functions, {\tt f} and {\tt g}.
Furthermore, imagine that you can only test one of
these functions, once.  Which do you test?  Knowing nothing further, you
have no way to rationally choose.  What might you know about {\tt f}
and {\tt g} that would allow you, in the sense of expected-value, to
make a more intelligent decision?  You might, of course, wish to know
things such as how calling {\tt f} and {\tt g} would typically
contribute to improving code coverage for
the SUT, or which is more closely related to critical aspects of
the specification, or (most ideally) which one contains a bug.  These are
usually, unfortunately, very expensive things to discover, and we have already
stipulated that you have very little time---only time to run one test
of either {\tt f} or {\tt g}.  \emph{If} we can propose a very
inexpensive-to-compute heuristic for the ``{\tt f} or {\tt g}?'' question, we have a plausible way to bias small-budget automated test
generation.  Of course, we will seldom be faced with testing \emph{only}
{\tt f} or {\tt g}, but we will always face the question of which
functions to test \emph{more often}, in such a setting, and we will always
have to choose some final function to test when our testing budget is
about to run out.  Even techniques that are more complex than pure random testing,
such as those used in EvoSuite and Randoop, rely on the basic
building block of choosing an arbitrary method to call.

\subsection{Solution: Count Relative Lines of Code}

The central proposal of this paper is that, if you know only that {\tt
  f} has more LOC than {\tt g}, you should prefer testing {\tt f} to
testing {\tt g}.  LOC allows you to approximate some kind of
expectation (though not a lower bound--it might be easy to cover most
of {\tt g}'s code, and hard to cover more than a line or two of {\tt
  f}) of gain in code coverage, of course, but it should, more
importantly, approximate complexity and influence on
program state.  Not being a lower
bound is \emph{useful} here:  we do not want to bias against functions that
have hard-to-cover code; they are precisely the functions we
may we need to test \emph{most}.

Longer functions are generally more complex, and presumably have more
room for programmers to make mistakes.  This is assumed in, for
example, mutation testing \cite{mutant,Mut2000}, where the LOC size of a function is
strongly correlated with the number of mutants generated
for that function.  However, even when a longer function does not have
any faults, it is still, we claim, usually more important to test.
Longer functions, we expect, perform more computation. In stateful
systems, longer functions tend to modify system state more
than shorter ones.  Even if a modification is
correct, it may cause other, incorrect, functions to fail.  Even for functions with no side effects,
we hypothesize that longer pure functions, on average, either take more complex data
types as arguments, or return more complex data types as results than
shorter ones.  In fact, very short functions in many cases are getters
and setters.  These need to be tested, and
sometimes need to be called to detect faults in more complex code, but
are seldom, we suspect, themselves faulty.

Of course, other than the general correlation detected in various
studies between defects and LOC at the function, module, or file level
\cite{Zhang2009ICSM, Ostrand2005TSE,Fenton2000TSE, Olague2007TSE,
  Andersson2007TSE}, it is difficult to know to what precise extent
length matters.  However, \emph{if} our {\tt f}/{\tt g} answer is reasonable, it
follows that biasing the probabilities for calling functions/methods
in random testing based on the relative LOC in those functions/methods should
improve the effectiveness of random testing, for most SUTs.  Some
caution is required: if a function {\tt f} is itself short,
but always or almost always calls {\tt h}, which is long and complex,
then in practice perhaps {\tt f} is a ``long'' function.  Or, one may argue that since {\tt f} is longer,
it also likely takes more time to run than {\tt g}, making it
``correct'' to choose {\tt f}, but a different problem than
selecting a next action in random testing.  The alternative to {\tt
  f} may not be testing {\tt g}, but perhaps
testing {\tt g} three times, or testing {\tt g}, {\tt h}, and {\tt i},
all of which are much shorter than {\tt f}.  Whether our
proposed bias is actually useful in practice is an empirical question,
despite having a sound analytical basis, thanks to these confounding factors,
and can be resolved only by experimentation.

The experiments in this paper, based on a simple linear bias in favor
of test actions with relatively more LOC, demonstrate that our proposed solution to the ``{\tt f} or {\tt g}?''
question is a useful one in that it often improves both coverage and fault detection.  We also thus demonstrate the general method
of moving from a plausible answer to the ``{\tt f} or {\tt g}?'' question to
improving the effectiveness of automated test generation.

In fact, we show that for many 
Python SUTs we examined, the
LOC heuristic, despite not using expensive coverage information, is
better than a coverage-based approach, and has better mean coverage over all SUTs, \emph{even though we impose the
  (unnecessary) cost of collecting coverage on LOC-guided testing}; further, incorporating the LOC heuristic into the coverage-based approach improves results over using coverage alone, for a large majority of SUTs.
The basic technique is very simple: we first run (pure, unguided) random testing on the SUT once, for a
short period (two minutes, in our experiments) and, for each function
or set of function calls that constitute a single testing action, compute the mean total LOC in all SUT functions and
methods executed while taking that type of test action.  Note that this is not the
same as measuring coverage: we measure (only counting each function
once) how large each function executed is, in total LOC, \emph{even if the
  majority of the code is not executed.}
The ``sizes'' in LOC are
then used to bias probabilities for selecting test actions so that
actions with higher LOC counts are chosen more often in all future
testing.  Unlike dynamic code coverage-based methods, these sizes do
not need to be recomputed for each test run; we show that the
technique is robust to even very outdated LOC estimates.

\subsection{Contributions}

We propose a novel and powerful heuristic for use in (small-budget)
automated generation of property-based Python unit tests, based on counting lines of code in
functions under test.  We evaluate the heuristic (and its combination
with other testing methods) across a set of 14 Python libraries, including widely-used real-world systems.
Overall, the LOC heuristic improves testing effectiveness for most
subjects.  The LOC heuristic also combines well
with other test generation heuristics to increase their
effectiveness.  It is often more effective than unbiased random
testing by a large margin (20\%-40\% or more improvement in
branch/statement coverage, 30\%-400\% or more gain in fault detection
rates).  The LOC heuristic, or a combination of the LOC heuristic and
another approach, is
the best method for testing more SUTs than any non-LOC approaches,
and is worse than random testing for fewer SUTs than the other two (established and widely used) test generation methods we tried.  Even if the overhead for coverage were negligible, a user
would be best off using the LOC heuristic or the LOC heuristic plus a coverage-guided
genetic algorithm, for most SUTs.  Our results also present a strong
argument for exploring the use of simple,
almost static (thus available without learning cost \emph{at testing
time}) metrics of source code to guide small-budget automated test generation,
especially in languages such as Python, where coverage instrumentation
is either very costly or
unavailable.

\section{LOC-Based Heuristics}
 
We present our basic approach in the context of the TSTL \cite{tstlsttt,NFM15} tool for property-based unit testing of Python programs for several reasons.  First, Python is a language with expensive (and coarse-grained:  there is no support for path coverage or coverage counts) code coverage tools.  Second, TSTL is the only tool, to our knowledge, that is focused on generating unit tests (sequences of value choices, method/function calls, and assertions) yet is essentially a  property-based testing tool \cite{ClaessenH00}, where users are expected to provide guidance as to what aspects of an interface are to be tested and frequently define custom generators or implement complex properties, in exchange for fast random testing to quickly detect semantic faults in code. A property-based testing tool is seldom seen as a test suite generator (unlike Randoop or EvoSuite), even though most property-based tools can also be used to produce test suites.  QuickCheck, Hypothesis, PROPER, or TSTL is usually executed to generate new tests after every code change.  It is in this setting, where generating new, effective tests within a small test budget is a frequently performed task, that the need for better heuristics is largest.

Before proceeding to the detailed Python implementation, we define the general class of LOC-based heuristics:  {\bf a LOC based heuristic biases the probability of method or function call choices in random testing proportionally to the measured lines of code in the method(s) or function(s) called.}  In this paper, we present one instantiation of this general idea, tuned to Python test generation.

\subsection{General LOC Heuristic Definition}

The general algorithm for LOC heuristics is most easily described by understanding how it changes the selection of a \emph{test action}.  Assume that test actions $\{a_1 \ldots a_n\}$ are normally chosen with uniform probability, i.e., $P(a) = \frac{1}{n}$.  Using LOC, instead, the probabilities are determined based on both an action set $a_1 \ldots a_n$ and a LOC-mapping, $m : a \in \{a_1 \ldots a_n\}  \Rightarrow k$, where $m(a)$ is a measure (possibly approximate) of the \emph{lines of code} \emph{potentially} (rather than actually) covered by executing $a$.  If $a$ is a simple function $f$ that calls no other functions, then $m(a)$ should be the number of lines in the implementation of the function $f$.  Given $m$, when using LOC, $P(a) = h(m(a))$ where $h$ should be a \emph{monotonically increasing function}, except in the special case that $m(a) = 0$, as discussed below.  It is the monotonically increasing nature of $h$ that produces the desired bias.  In this paper, we consider only the case where $h$ is a simple linear mapping.

\subsection{Python Implementation}

\begin{figure}[t]
{\scriptsize
\begin{code}
@import avl
\vspace{0.02in}
pool: <int> 4
pool: <avl> 3
\vspace{0.02in}
property: <avl>.check\_balanced()
\vspace{0.02in}
<int> := <[1..20]>
<avl> := avl.AVLTree()
\vspace{0.02in}
<avl>.insert(<int>)
<avl>.delete(<int>)
\end{code}
}
\caption{TSTL harness for AVL trees}
\label{fig:example}
\end{figure}

In TSTL, tests are composed of a series of \emph{actions}.  An action is a fragment of Python code, usually either an assignment to a \emph{pool} value \cite{AndrewsTR} (variables assigned during the test to store either input values for testing or objects under test), or a function or method call, or an assertion.  Actions basically correspond to what one might expect to see in a single line of a unit test.  Constructing a test in TSTL is essentially a matter of choosing the lines that will appear in a constructed unit test.  Actions are grouped into \emph{action classes}, defined in one line of a TSTL test harness \cite{WODACommon} file.  Figure \ref{fig:example} shows part of a simple TSTL harness, for testing an AVL tree (with no properties beyond that the tree is balanced and does not throw any exceptions).  The line of TSTL code {\tt <int> := <[1..20]>} defines an \emph{action class} that includes many actions:   {\tt int0 = 2}, {\tt int1 = 3}, and {\tt int3 = 10}, and so forth.  The AVL harness defines 4 action classes (one for each line after the {\tt property}).  Of these, the first calls no SUT code (hence $m(a)$ would be 0), while the other three call the {\tt AVLTree} constructor or an {\tt AVLTree} object method, and $m(a)$ would be based on the code in those methods.  The same (top-level) method is called for each action in an action class, in most, but not all cases; we examined our SUTs for cases where this was not the case, and found that the top-level method called could vary in about 20\% of all actions.
Our heuristic is simple, and operates in two phases:  first, a measure of LOC for each action class (a construction of $m(a)$) is needed, and then the LOC measures must be transformed into probabilities for action classes, to bias random testing in favor of actions with higher LOC values (a function $h$ is defined).  

\subsubsection{Estimating LOC in Test Actions}

\begin{figure}
{\scriptsize
\begin{code}
def traceLOC(frame, event, arg):
    \# inputs are provided by the Python runtime
    \# - frame is the current stack frame
    \# - event is one of 'call', 'line', 'return', or 'exception'
    \# - arg is specific to the event type
    
    \# We care only about function 'call's; on a call, iff the function was not
    \# previously seen, we 1) mark it as seen in this action and 2) add the
    \# LOC count for the function to the global thisLOCs count, which is
    \# reset before each action in the TSTL testing.  Otherwise, we just return.
    \# In any case, we return this function, since Python's tracing requires
    \# the tracer to return the tracer to use in the future.
    
    global thisLOCs, seenCode
    
    if event != "call": return traceLOC
    co = frame.f\_code
    fn = co.co\_filename
    if (co.co\_name, co.co\_filename) in seenCode: 
        return traceLOC
    if fn == sut.\_\_file\_\_.replace(".pyc",".py"):
        return traceLOC
    seenCode[(co.co\_name, fn)] = True 
    thisLOCs += len(inspect.getsourcelines(co)[0])
    return traceLOC
\end{code}}
\caption{Portion of code for LOC measurement tracing in Python}
\label{fig:collect}
\end{figure}

To bias probabilities by LOC, we need to collect a mapping from action classes to the LOC in SUT code called by the actions in the class.  In theory, this could be based on static analysis of the call graph; however, in Python determining an accurate call graph can be very difficult, due to the extremely dynamic nature of the language.  Moreover, we are generally interested only in functions that have a non-negligible chance of being called during short-budget testing; calling some functions may not even be possible given the test harness and input ranges used.  Our efforts to collect reliable information statically, based on matching the textual names in actions to a static list of function sizes generated using Python's inspection tools, produced a large improvement in testing for one SUT, but it was generally not very helpful, and sometimes greatly reduced the effectiveness of testing compared to pure random testing; we report on this in more detail below.  The primary cause was simple inaccuracy,  e.g., if an action class varies which method or function it calls depending on the type of an object in a pool (a fairly common pattern in TSTL), the tool simply counted LOC for the wrong code.  The dynamic nature of Python, which is exploited heavily in TSTL harnesses, simply defeats a purely-static approach.

Inspecting the incorrect results also helped us see that simply counting LOC in a top-level function is inappropriate for TSTL/Python.  TSTL harnesses seldom include all methods of a class; instead the testing is focused on the high-level APIs actually used by users, not other functions, and these are often very small wrappers that dispatch to a more complex method.  In TSTL, most coverage can only be obtained \emph{indirectly} \cite{GayCoverage}.

We therefore used Python's system tracing and introspection modules to collect a one-time estimate\footnote{In Section \ref{sec:rq7} we show that the one-time aspect is likely not important; results appear to be robust even to large code changes, and thus certainly to mere sample variance.}  of ``total LOC'' for each action class for each SUT by detecting function calls during actions and then measuring LOC reported by Python's {\tt getsourcelines} for every such function, using Python's {\tt settrace} feature, as shown in Figure \ref{fig:collect}; comments in the code describe the algorithm and some implementation details.  It is important to understand that this does \emph{not} measure code coverage---it simply collects the total LOC (as counted by Python's notion of code lines, which includes blank spaces and comments\footnote{Including comments and blank lines is not a problem for our basic hypothesis:  we also expect that code with more comments (or even more blank lines) is, all things being equal, more interesting to test.}) for any Python function or method \emph{entered} during execution of a test action, even if almost none of the code for that function/method is executed.   The value recorded for an action is the \emph{sum} of function/method LOCs, with each one only counting once (e.g., if {\tt f} has 30 LOC and is called 40 times by an action, it only adds 30 to the LOC count for that action).  The tracing function is installed with {\tt sys.settrace} before each action is executed.  In order to ensure that all action classes are measured, the sampling tool always selects any enabled actions whose class has not yet been sampled at least once.  After each action class is sampled once, sampling is random until a fixed time limit is reached.   The value recorded for each action class's LOC count is the \emph{mean} of all samples: $m(a) = \frac{\sum_{s \in \mathit{samples}}LOC(s)}{|\mathit{samples}|}$, where $\mathit{samples}$ is the set of all values computed using {\tt traceLOC} for that action class.  Taking the mean is required because, again, due to the highly dynamic nature of TSTL test harnesses, the same action class may not even be calling the same top-level method in every case.  Since we cannot distinguish such cases statically, we want to know on average ``how big'' each action class is in terms of LOC.  All action classes could be sampled effectively, multiple times, with 120 seconds of sampling for all of our experimental subjects.  Any action classes that cannot be sampled within 120 seconds, using a strong bias in favor of unsampled classes, are highly unlikely to ever be covered during small-budget testing, in any case.  After one such sampling run, the probabilities can be used in any number of future testing runs, as we show below.

\subsubsection{Biasing Probabilities for Action Classes}

Given this one-time mapping from action classes to LOC counts, we need to produce a biased probability distribution for action classes to be selected in future random testing.  Additionally, there must be some way to handle action classes that do not cover any SUT LOC.  These action classes cannot be excluded from testing:  most harnesses need to generate simple input data, such as integers or booleans, where generating data does not cover any code under test.  Our solution is simple:  we evenly distribute 20\% of the probability distribution among \emph{all} action classes that do not call any SUT code (where the LOC value is 0).  The remaining 80\% is distributed to action classes with a non-zero LOC count in proportion to their share of the total LOC count for all action classes.  This means that our heuristic does not care about absolute LOC at all, only relative LOC:  it does not matter how long a function is, only how much longer (or shorter) it is than other functions to be tested.  The core idea of our heuristic is this linear bias in favor of test choices proportional to their share of total LOC count. Formally, we define:

$$M_0 = |a \in \{a_1 \ldots a_n\} : m(a) = 0|$$

$$M_1 = \sum_{a \in \{a_1 \ldots a_n\} : m(a) > 0}m(a)$$

That is, $M_0$ is the number of actions (or, here, action classes) with 0 measured LOC, and $M_1$ is the total sum of all LOC measures for actions/action classes whose LOC estimate is non-zero (of course, since $m(a) = 0$ in these cases, we could also include them in the total).

We can then define $h$, for the special case of zero-LOC action classes and for other action classes, thus:

  $$h(0) = \frac{0.2}{M_0}$$

  $$h(c > 0) = \frac{0.8 c}{M_1}$$

For example, consider a TSTL harness with only three action classes:  {\tt <int> := <[1..20]>}, {\tt f(<int>)}, and {\tt g(<int>)}.  The first action class does not call any SUT code: it simply assigns a value to be used in later testing.  Assume that {\tt f} calls no functions, but has 30 LOC, and {\tt g} has 6 LOC itself and always calls {\tt h}, which has 14 LOC, twice.  Using the (mean) LOC counts of 0, 30, and 20, respectively, we get the following probabilities for the action classes, according to the LOC heuristic:

\vspace{0.1in}

{
\begin{tabular}{lccc}
Action class & Mean LOC & Formula & P(class) \\
\hline
\vspace{0.01in}
{\tt <int> := <[1..20]>} & 0 & $\frac{0.20}{1}$ & 0.20\\
\vspace{0.03in}
{\tt f(<int>)} & 30 & $\frac{30}{50} \times 0.80$ & 0.48 \\
{\tt g(<int>)} & 20 &$\frac{6+14}{50} \times 0.80$ & 0.32 \\
\end{tabular}
}

If we add another action class that does not call any code, and a direct call to {\tt h} (with 14 LOC), the probabilities change to:

{
\begin{tabular}{lccc}
Action class & Mean LOC & Formula & P(class) \\
\hline
\vspace{0.01in}
{\tt <int> := <[1..20]>} & 0 & $\frac{0.20}{2}$ & 0.100\\
{\tt <ch> := <['r','w']>} & 0 & $\frac{0.20}{2}$ & 0.100\\
\vspace{0.03in}
{\tt f(<int>)} & 30 & $\frac{30}{64} \times 0.80$ & 0.375 \\
{\tt g(<int>)} & 20 &$\frac{6+14}{64} \times 0.80$ & 0.250 \\
{\tt h(<ch>)} & 14 &$\frac{14}{64} \times 0.80$ & 0.175 \\
\end{tabular}
}

One objection to this sampling approach is that it pays a non-negligible measurement cost, unlike purely static measurement.  This is true, but in another sense there is a fundamental difference between \emph{essentially constant-time} approaches (measuring LOC in source or fixed-time LOC sampling) and, e.g., coverage instrumentation that imposes a cost that will always be (at best) linear in the number of test actions executed. 
Even so, why not simply run for 120 seconds and measure code coverage instead of LOC, and use that measure?  There are two answers.  First, the general LOC idea remains static; in a less dynamic language than Python, it should even be possible to statically measure the LOC count for a method and the methods it calls, though this would lose the probability of calling non top-level methods.  Second, and more importantly, using coverage is worse, for our purposes, than using LOC:  any single short run is likely to only cover a small part of the code for any complex function with many hard-to-take branches.  LOC is a much better way to estimate \emph{maximum possible coverage}, since most runs will not cover the interesting (hard to cover) part of a function.  Of course, biasing exploration by coverage is a very useful test-generation method; however, coverage is so dependent on actual test sequence and values, unlike LOC, that it is only effective in an approach, such as the Genetic Algorithm (GA) we compare with below, using runtime context and online instrumentation.  To clarify the point, consider using our sampling approach to determine a ``size'' for a function {\tt f} that takes a list {\tt s} as an argument.  Even if the function is very complex and lengthy, in a single short test run, most calls to it may be made with an empty list as an argument.  That is, if the function looks like:

{
\begin{code}
 if len(s) == 0:
     return 0
 ...
 40 lines of complex destructuring and tabulation of the list    
\end{code}
}

\noindent then the mean coverage for the action calling {\tt f} will be very low, but the LOC count will reflect the fact that when called with a non-empty list the function will perform complex computation, \emph{even} if the sampling never calls the function with a non-empty list.  Traditional coverage-driven test generation using, e.g., a GA, relies on the context of a test with a non-empty list to identify the action calling {\tt f} as interesting; in fact, such methods usually don't identify a single action as interesting itself, but only a \emph{test} as interesting.  The price to be paid, however, is that coverage must be collected for every test at runtime.  Our approach only instruments test execution during a one-time sampling phase, and thereafter uses that data to bias test generation.  However, as we show, such a  contextless LOC-based overapproximation of size is often, even ignoring this price, a better way to bias testing than a GA, for small test budgets.

\section{Experimental Evaluation}

\subsection{Research Questions}

Our primary research questions concern the utility of the LOC heuristic.

\begin{itemize}
\item {\bf RQ1:}  How does test generation for 60 second budgets using
  the LOC heuristic
  compare to random testing, in terms of fault detection and code coverage?
\item {\bf RQ2:}  How does test generation for 60 second budgets using
  the LOC heuristic
  compare to coverage-guided testing using a Genetic Algorithm (GA) approach in the style of EvoSuite \cite{FA11},
  or to swarm testing \cite{ISSTA12}, for fault detection and code coverage?
\item {\bf RQ3:} How does combining orthogonal generation approaches
  (e.g., a Genetic Algorithm (GA), but using the LOC heuristic) for 60
  second budgets compare to using only one test generation heuristic,
  for fault detection and code coverage?
  Is biasing a more sophisticated heuristic by also applying the LOC heuristic useful?
\end{itemize}

Our hypothesis is that using LOC to guide testing will be useful,
outperforming, in terms of increased code coverage and/or fault
detection for a fixed, small, testing budget, random testing alone for
over 60\% of SUTs, and outperforming (by the same measure) established
heuristics/methods for at least 50\% of SUTs studied.   Further,
combining LOC with compatible heuristics will frequently provide
additional benefits, improving code coverage and/or fault detection
for a given budget.

We also provide limited, exploratory, experimental results to
supplement these primary results, covering a set of related issues; in particular:

\begin{itemize}
\item  What is the typical cost of measuring code coverage
  in Python
  using the best known tool, the {\tt coverage.py} library?
\item How does test generation for 60 second budgets using
  the LOC heuristic
  compare to using {\tt afl-fuzz} \cite{aflfuzz} via the {\tt
    python-afl} module?
\item Can the LOC heuristic improve the effectiveness (as
  measured by code coverage) of
  feedback-directed random testing \cite{Pacheco} in Java?
\item What is the impact of using outdated dynamic estimates of
  lines of code on the performance of the LOC heuristic?
\item Do the coverage advantages provided by the LOC heuristic
  over random testing persist over time, or are they limited to small
  budget testing?
\end{itemize}

\subsection{Experimental Setup and Methodology}

All experiments were performed on a Macbook Pro 2015 15'' model with
16 GB of RAM and 2.8GHz Quad-core Intel Core i7, running OS X 10.10
and Python 2.7; experiments only used one core.

\subsubsection{Evaluation Measures}

We report results for \emph{both coverage
and fault detection} for our core question, small (default) budget test generation
effectiveness.  The reasons for the use of two core measures are
simple. Fault detection essentially needs no justification: it is the
end-goal of software testing, in that a test effort that fails to
detect a fault in its scope has failed at its primary task.  However,
it is important to also measure code coverage for a number of
reasons.  First, small budget
testing aims to detect just-introduced problems, and in a setting
where simply covering all code is difficult, the best way to determine
which defects will be detected will often be code coverage: the kinds
of bugs detected in this way may not ever make it into
committed/released software and thus not be represented by defects in
externally visible code.   Given this context, developers
performing small budget testing are plausibly interested in simply covering
as much of the code under test as possible as quickly as possible.
For property-based testing, furthermore, developers often have sufficiently
strong oracles (e.g., differential \cite{Differential,ICSEDiff} ones)
that mere coverage is more often sufficient for fault detection \cite{progTestOracle}, especially for just-introduced faults,
which are often easy to trigger, we suspect.  Previous work on
weaknesses of automatically generated tests showed that failure to cover
the faulty code is often a critical problem:  36.7\% of non-detected faults
were simply never covered by any test \cite{AutoTestFaults}.

A second reason it is dangerous to rely on only fault detection for
evaluation is that defect sets are unfortunately typically quite small (e.g., Defects4J
\cite{just2014defects4j} covers just 
6 projects and about 400 bugs), and, more importantly for our
purposes, are heavily
biased towards bugs that lasted long enough to appear in bug
databases, and usually only contain Java or C programs.  As Dwyer et al. \cite{PathSensitive} showed, using a small
set of defects can produce unreliable evaluations of testing methods,
since so much depends on the exact
faults used.  Where there are
significant differences in branch or statement coverage, correlation with
fault detection for (what we use in every case) \emph{fixed-size} test
suites is known to be strong.  Even work questioning the value of 
  coverage \cite{NotCorrelated} tends to confirm the relationship for
  suites of the same fixed size, with Kendall $\tau$ usually 0.46 or
  better,  often 0.70 or better \cite{notcorweb}; i.e. higher coverage is highly likely to
  indicate higher bug/mutant detection \cite{ISSTA13,SuiteEval}. We
therefore demonstrate effectiveness using 
\emph{both} coverage and limited fault-based evaluations.

BugSwarm
\cite{DBLP:conf/icse/DmeiriTWBLDVR19} does provide a larger number of
faults, and includes (unlike other data sets) Python examples.
Unfortunately, when we examined the BugSwarm defects, the subjects and
bug types were seldom easily translatable to a property-driven test
harness, unless we were to undertake to essentially use our knowledge
of the bug to carve out a portion of the Python API to test, and
properties to check.  This does not accurately
reflect typical use of property-based testing, and would inevitably
introduce bias.  Using general-purpose harnesses for testing the
libraries, developed without bug knowledge (and with developer input
in some cases), is more realistic.  The TSTL harnesses used in this
paper were all, first and foremost, designed to reflect typical use of
property-based testing, rather than to detect a specific bug, or for
use as experimental subjects.  These are all \emph{realistic test
  harnesses a developer might produce}, in our opinion.

That is, while
(mostly) written by authors of
this paper, these harnesses were written A) to focus on important
parts of API, to find bugs, not automatically generated with no
understanding of the likely usage of the methods and B) with manually
constructed, but simple, oracles, similar in style and power to those
found throughout both the literature of property-based testing and
real-world usage.  The {\tt pyfakefs} harness has benefitted from
considerably commentary and examination by the {\tt pyfakefs}
developers, who made contributions to the TSTL code in order to
support better TSTL testing of {\tt pyfakefs}.  Feedback from SymPy
developers contributed in a lesser way to tuning that harness.  We
also examined a large number of real-world Hypothesis test harnesses
to better understand real-world developer uses of property-based
testing in Python.  Furthermore, one of the authors originally used
TSTL as a developer/tester only, not a researcher, in the course of
pursuing an MS in Geographic Information Systems, focusing on testing
the widely used ArcPy library for GIS, and either contributed to
harnesses or vetted them as similar to her own efforts in a purely
QA/development role.

Following the advice of Arcuri and Briand \cite{arcuri2014hitchhiker},
we do not assume that statistical distributions involved in random
testing are normal, and thus use the Mann-Whitney test for determining
significance of non-paired
comparisons (e.g., within-SUT runs), and the Wilcoxon test for
determining significance of paired comparisons (per-SUT overall
results, where the matching by SUT is important). We believe the 100
runs used for all experiments are (more than)
sufficient sample size to effectively compare the impacts of various
test generation methods, for each SUTs.  The large sample size of runs
per SUT,
relative to significant but not unbounded variance in results we observed, allows us to
detect, with very high probability, any differences between generation methods, and their effect size
and direction---in short, the expected distributions of coverage and
fault detection results for each SUT/method pair---unless the differences are very small.

We measured code coverage, both in our core experiments and in
measures of the overhead of code coverage, using version 4.5.2 of the widely-used,
essentially standard, {\tt coverage.py} Python module\footnote{GitHub
  reports that over 80000 Python projects use {\tt coverage.py}.}.  To our
knowledge, {\tt coverage.py} is practically the only Python coverage
library used in practice, and is at least as low-overhead and
efficient as any other Python coverage tool.  Version 4.5.2 is not the
most recent {\tt coverage.py} release, but the changes in the two most
recent 4.5 releases only concern packaging metadata and multiprocessing
support in Python 3.8, neither of which affects overhead or is
relevant to this paper's concerns
(\url{https://coverage.readthedocs.io/en/coverage-5.0/changes.html}).
The 5.0 release similarly does not appreciably change measured
overheads.  All versions we used are based on a fast native C tracing implementation.

\subsubsection{Experimental Subjects}

\begin{table}[t]
\centering
\caption{Python SUTs.  Size is lines of code as
  measured using cloc.  Stars / Uses reports GitHub repo stars and the GitHub
  ``Used by'' statistic, a rough measure of popularity.  GitHub does
  not always report the ``Used by'' statistic.}
\label{tab:subjects}
{\scriptsize
\begin{tabular}{l|lll}
SUT & Size & Source & Stars / Uses\\
\hline
{arrow} & 2707 & \url{https://github.com/crsmithdev/arrow} &
                                                                   5900
                                                                   / 10400\\
{AVL} & 225 & TSTL example \cite{avltree} & N/A \\
{heap} & 56 & Hypothesis example \cite{hypheaps} & N/A\\
{pyfakefs} & 2788 & \url{https://github.com/jmcgeheeiv/pyfakefs}
  & 281 / 398\\
{sortedcontainers} & 2017 
          &\url{http://www.grantjenks.com/docs/sortedcontainers/} &
                                                                    1600
                                                                    /
                                                                    Not
  reported\\
{SymPy} & 227959 & \url{http://www.sympy.org/en/index.html} &
                                                                     6500
                                                                     /
  14600\\
\hline
{bidict} & 569 &  \url{http://pythonhosted.org/bidict/home.html} & 535
  / Not reported\\
{biopython} & 81386 &
                      \url{https://github.com/biopython/biopython.github.io/}
                    & 2000 / 3500\\
{C Parser} & 5033 & \url{https://github.com/albertz/PyCParser} & 1900
                                                                 / Not
  reported\\
{python-rsa} & 1597 & \url{https://github.com/sybrenstuvel/python-rsa}
  & 219 / Not reported\\
{redis-py} & 2722 & \url{https://github.com/andymccurdy/redis-py} &
                                                                    8200
                                                                    / 62900\\
{simplejson} & 2811 &
                      \url{https://simplejson.readthedocs.io/en/latest/}
                    & 1200 / 35400\\
{TensorFlow} & 193374 & \url{https://github.com/tensorflow/tensorflow}
                    & 140000 / 60100\\
{z3} & 10501 & \url{https://github.com/Z3Prover/z3} & 5000 / 20\\
\end{tabular}
}
\end{table}

\begin{table}[t]
\centering
\caption{Fault Information}
\label{tab:bugs}
{\scriptsize
\begin{tabular}{l|c|p{8cm}}
SUT & \# Faults & Description of Faults or Link to GitHub Issues\\
\hline
  {arrow} & 3 & \url{https://github.com/crsmithdev/arrow/issues/492}
                plus two other {\tt ValueErrors} for unusual inputs, fixed since discovered \\
  \hline
{AVL} & 1 & Change of rotation direction produced improperly balanced
            trees.\\
  \hline
{heap} & 1 & Implementation bears little resemblance to a real heap,
             yet can pass many simple tests.\\
  \hline
  {pyfakefs} & 6 &
                   \url{https://github.com/jmcgeheeiv/pyfakefs/issues/256}
  \\
& &                   \url{https://github.com/jmcgeheeiv/pyfakefs/issues/272} \\  
& &                   \url{https://github.com/jmcgeheeiv/pyfakefs/issues/284} \\
& &  \url{https://github.com/jmcgeheeiv/pyfakefs/issues/299} \\
& &                 \url{https://github.com/jmcgeheeiv/pyfakefs/issues/370}\\  
& &                  \url{https://github.com/jmcgeheeiv/pyfakefs/issues/381}\\
  \hline
{sortedcontainers} & 2 &
                         \url{https://github.com/grantjenks/python-sortedcontainers/issues/55}
  \\
    & & \url{https://github.com/grantjenks/python-sortedcontainers/issues/61} \\
  \hline
  {SymPy} & 26 & \url{https://github.com/sympy/sympy/issues/11151}\\
    & & \url{https://github.com/sympy/sympy/issues/11157} \\
    & & \url{https://github.com/sympy/sympy/issues/11159} \\
  & & 23 additonal problems identified by unique Exception structure,
      not reported due to being fixed before above, reported, issues
      were resolved.
\end{tabular}
}
\end{table}

\begin{table}
\centering
\caption{Omitted Subjects}
\label{tab:omit}
{\scriptsize
  \begin{tabular}{l|p{8cm}}
    \hline
XML & Tests consistently hit exponential case or bug causing loop;
      TSTL does not support action timeouts so this makes experiments
      impossible to perform. \\
  \hline
arcpy & Each test action requires more than 60 seconds on average to
        perform; also most code under test is compiled C++ without
        coverage instrumentation, and only available on Windows. \\
    \hline
bintrees & harness detects bug that consistently causes timeout; also,
           development stopped and replaced by {\tt
           sortedcontainers}. \\
    \hline
    danluuexample & Toy example, with only one function to call. \\
    \hline
datarray\_inference & Coverage saturates in 60 seconds, and bug is
                      only detectable using nondeterminism checks. \\
    \hline
    dateutil & Bug consistently causes timeout.\\
    \hline
eval & Actual test of SUT is via subprocess execution, so coverage not
       possible, also only one test function.\\
    \hline
    gmpy2 & Code under test is almost entirely C code, not Python code.\\
    \hline
kwic & Toy example from the classic Parnas problem, used in a software engineering class.  Coverage saturates easily in 60s with random testing, and there are no bugs. \\
    \hline
    maze & Toy example with only one function to call. \\
    \hline
    microjson & Only call to SUT is in a property. \\
    \hline
numpy & Average test action requires more than 60 seconds to run, and
        timeouts exceeding test budget are extremely frequent, along with
        crashes due to memory consumption. \\
    \hline
nutshell & Toy example for TSTL Readme, with no actual code to
           test. \\
    \hline
oldAVL & This is simply a less-readable version of the AVL harness
         included in our SUTs. \\
    \hline
    perfsort & Only calls one SUT function. \\
    \hline
    pysplay & Bug consistently causes timeout. \\
    \hline
    pystan & Only calls one SUT function (essentially a compiler
             test). \\
    \hline
    solc & Only calls one SUT function (essentially a compiler
           test). \\
    \hline
    stringh & Code under test is C code, not Python. \\
    \hline
    tictactoe & Toy example, code of interest only in property, and
                saturates coverage/detection of ``fault'' in 60
                seconds. \\
    \hline
    trivial & Toy/contrived examples to test corner cases of test
              reduction.  Saturates coverage/detection of ``fault'' in 60
              seconds. \\
    \hline
    turtle & Toy example, no ``testing'' involved but an effort to
             produce interesting random art using test generation. \\
    \hline
    water & Toy example, no actual function calls at all. \\
    \hline             
\end{tabular}
}
\end{table}

We applied the LOC heuristic, pure random testing, and two established
test generation heuristics (a coverage-driven Genetic Algorithm (GA)
and swarm testing) discussed below, to testing a set of Python
libraries (Table \ref{tab:subjects}) with test harnesses provided in
the TSTL GitHub repository \cite{tstl}.  Two of our SUTs ({\tt AVL}
and {\tt heap}) are toy
programs with hard-to-detect faults, used in TSTL or Hypothesis
documentation and benchmarking; the remaining programs are
popular Python libraries, with many GitHub stars indicating
popularity.  We could have omitted the ``toy'' examples as
unrealistic, but included them because, while not of real-world code,
both are extremely similar to property-based harnesses for real-world
containers, which are frequently the target of property-based testing
efforts.  Moreover, the relatively small APIs in both cases made
understanding the differences between test generation method
performance easier (and thus made it easier to construct hypotheses
about causes of method effectiveness for
more complex SUTs).

Table \ref{tab:bugs} provides information on the faults in SUTs for
which we investigated fault detection.  Most faults in this table were
detected by at least one of our core experimental runs.  The exception
is for {\tt SymPy}, where the most faults detected by any single run
was 2, and there were only 14
detected faults for our core experiments.  We know of 12
additional detectable faults not found by any of the core
experimental runs, as well; the set of 26 noted in the table is for all saved TSTL output
for that version of {\tt SymPy}, e.g., including our hour
long experimental runs.  A single recursion-depth
error accounted for the largest fraction of detected faults, about half of all
detections.  Many faults were detected only three times out of 1250
experimental runs.

To avoid bias, we attempted to apply our approach to \emph{all of the example
test harnesses included with the TSTL distribution at the time we performed our experiments}, omitting \emph{only}
those where the experiments would be meaningless or give our approach
an unfair advantage.  Reasons for omission were limited to:

\begin{enumerate}
\item Python is almost completely an interface to
underlying C code or an executable (e.g., {\tt gmpy2}, {\tt eval}),
and so LOC of Python functions contains
no useful information.
\item The SUT consistently enters an infinite
loop (which makes it impossible to perform our 60 second budget
experiments, since TSTL locks up and does not produce coverage
summaries).
\item The minimum reasonable budget for testing is much longer
than 60 seconds (e.g., for ESRI ArcPy, simple GIS operations often
take more time than the entire test budget).
\item The harness is
clearly a toy, without any real testing functionality (e.g., an
implementation of a puzzle, or a tool for producing random turtle
art), relevant coverage targets, or any (even fake) bugs.
\item All methods consistently completely saturate coverage within 60
  seconds, and detect all faults.
\item Either no actions call SUT
code (possible when all SUT interactions are via the property), or
only one action calls SUT code (which would usually give an unfair advantage
to our approach, which would automatically give that call 80\% of the
test budget).
\end{enumerate}
Table \ref{tab:omit} explains the reason for omission for
every harness in the TSTL examples directory at the time we performed
our experiments for which we did not report results; in some cases
several of the above reasons apply.  Obviously, harnesses introduced
after we performed our experiments were not included (and most would
be rejected for other reasons as well).

TSTL has been used to report real faults (later corrected by the developers) for
{\tt pyfakefs}, {\tt SymPy}, and {\tt sortedcontainers}.  The {\tt
  pyfakefs} effort is ongoing, with more than 80 detected and
corrected defects to date, and one error discovered not in {\tt pyfakefs}
but in Apple's OS X implementation of POSIX rename.  Note that LOC for
each SUT is usually much greater than coverage below.  This is in part
due to our focus on 60 second testing, in part due to the coverage
tool not considering function/class definition code (e.g., {\tt def}
or {\tt class} lines) as covered, and in part due to a more complex cause:  most of the test harnesses only focus on easily specified,
high-criticality interface functions, and omit functions whose output
cannot effectively be checked for correctness, that are very
infrequently used in practice, or that are easily completely verified by simple
unit tests.  These harnesses, for the portion of each SUT's API tested,
usually provide considerable oracle strength beyond that offered by
Randoop or EvoSuite style test generation.  The {\tt AVL}, {\tt pyfakefs}, {\tt
  bidict}, and {\tt sortedcontainers} harnesses provide complete
differential testing with respect to a reference implementation in the
standard Python library or the operating system, and the {\tt heap},
{\tt SymPy},
{\tt python-rsa}, and {\tt simplejson} harnesses provide round-trip or
other semantic correctness properties.

A full implementation of the LOC heuristic approach evaluated in this
paper has been available
in the release version of TSTL since 2017, using the {\tt
  --generateLOC} and {\tt --biasLOC} options.  In general, to reproduce
our results, no special replication package is needed; using the
current release of TSTL plus appropriate versions of tested Python modules is all that is required; experiments were performed on
mac OS X, but should work in any Unix-like environment. The GitHub repository
\url{https://github.com/agroce/LOCtests} contains our raw TSTL output
files used for all analysis in this paper, and the exact configuration
of TSTL used can be extracted by examining the first line of those files, plus version information to help with installation of appropriate versions of SUT libraries. The {\tt scripts} directory in this repository contains our
analysis scripts, though we warn the user that these are tuned to a
local TSTL install and Python environment, and were not developed to
be re-usable.  The state of the code means that using your own analysis
scripts may be more useful; they do show how to parse TSTL output,
however. Note that the swarm
dependency computation recently changed (\url{https://blog.regehr.org/archives/1700}), so the {\tt
  --noSwarmForceParent} option must be used to match the older swarm
results.  

\subsection{Results Comparing Test Generation Methods Supported by
  TSTL (RQs 1, 2, and 3)}
\label{sec:results}

\begin{table}[b]
\centering
\caption{Gain or loss in coverage/faults detected vs. random testing}
{
\begin{tabular}{@{}l|r|r|r|r}
\hline
\hline
SUT & branch & stmt & faults & =branch\\
\hline
\hline
\multicolumn{5}{c}{with detectable faults}\\
\hline
{arrow} & -6.47\% & -5.20\%  & {\bf +75.9\%} & 9.5\\
{\bf AVL} & {\bf +2.95\%}    & {\bf +3.30\%} & {\bf +37.5\%} &{\bf 98.2}\\
{\bf  heap} & \emph{0.00\%} & \emph{0.00\%} & {\bf +190.0}\%& \emph{60.0}\\
{\bf  pyfakefs} &  \emph{+0.09\%}  & {\bf +0.13\%} &  {\bf +403.7\%}& \emph{95.2}\\
{\bf sortedcontainers} &   {\bf+33.03\%}  & {\bf +33.34\%} &  {\bf +INF*\%}&{\bf 186.6}\\
{\bf  SymPy} & {\bf  +25.28\%} &  {\bf+24.87\%}    & \emph{-15.4\%} &{\bf 216.5}\\
\hline
\multicolumn{5}{l}{* using older version of {sortedcontainers} with two faults;}\\
\multicolumn{5}{l}{random testing never detected these faults; hence \% gain INF;}\\
\multicolumn{5}{l}{LOC produced 0.1 mean faults / 60s}\\
\hline
\hline
\multicolumn{5}{c}{without detectable faults}\\
\hline
{bidict} & -6.18\% & -7.21\% & N/A&8.8\\
{biopython} & -13.52\% & -13.66\% & N/A&9.9\\
{\bf  C Parser} & {\bf +41.17\%} &  {\bf +40.35\%}   & N/A&{\bf 1108.9}\\
{\bf  python-rsa} &   \emph{+0.40\%} &  {\bf +0.41\%} & N/A& {\bf 66.1} \\
{\bf  redis-py} & {\bf +16.95\%} &  {\bf +15.52\%} & N/A&{\bf 237.0}\\
{simplejson} & -15.45\% & -13.81\% & N/A & 19.4\\
{\bf sortedcontainers} & {\bf   +35.43\%} &  {\bf +35.27\%} & N/A&{\bf 268.7 }\\
{\bf TensorFlow} & {\bf   +7.71\%} &  {\bf +7.26\%} & N/A&{\bf 121.5}\\
{\bf   z3} & {\bf +11.20\%} & {\bf+8.48\%}&N/A&{\bf 979.9}\\
\hline
\end{tabular}
}
\label{pygainloss}
\end{table}

{
\begin{figure*}
\begin{subfigure}{0.692\columnwidth}
\includegraphics[width=\columnwidth]{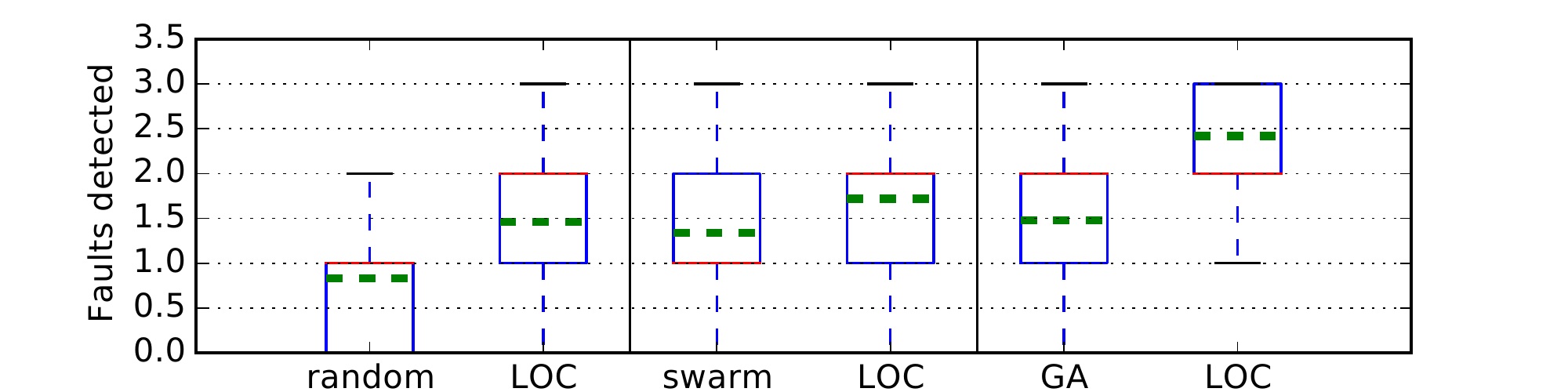}
\caption{arrow (3 faults)}
\end{subfigure}
\begin{subfigure}{0.692\columnwidth}
\includegraphics[width=\columnwidth]{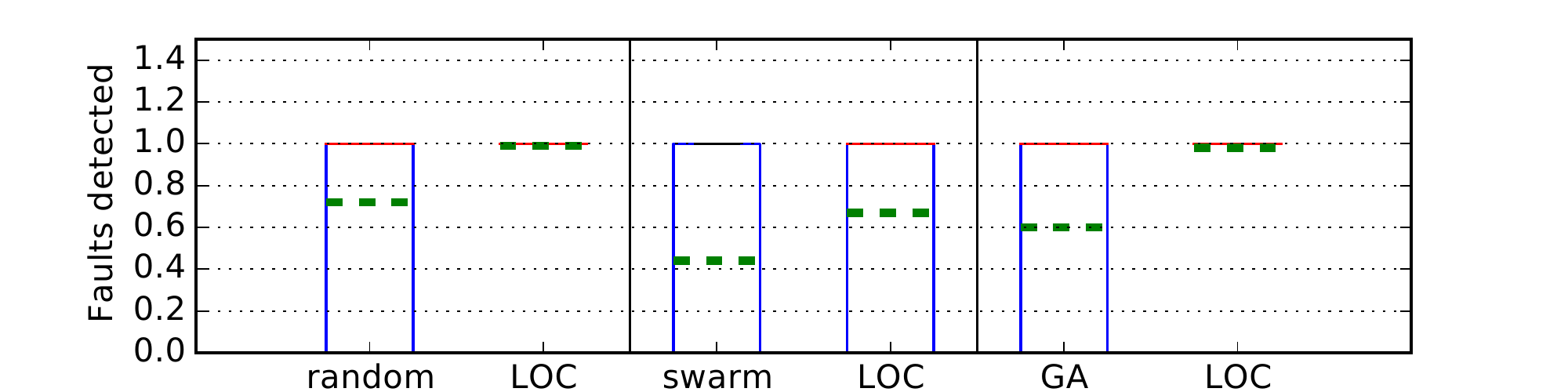}
\caption{AVL (1 fault)}
\end{subfigure}
\begin{subfigure}{0.692\columnwidth}
\includegraphics[width=\columnwidth]{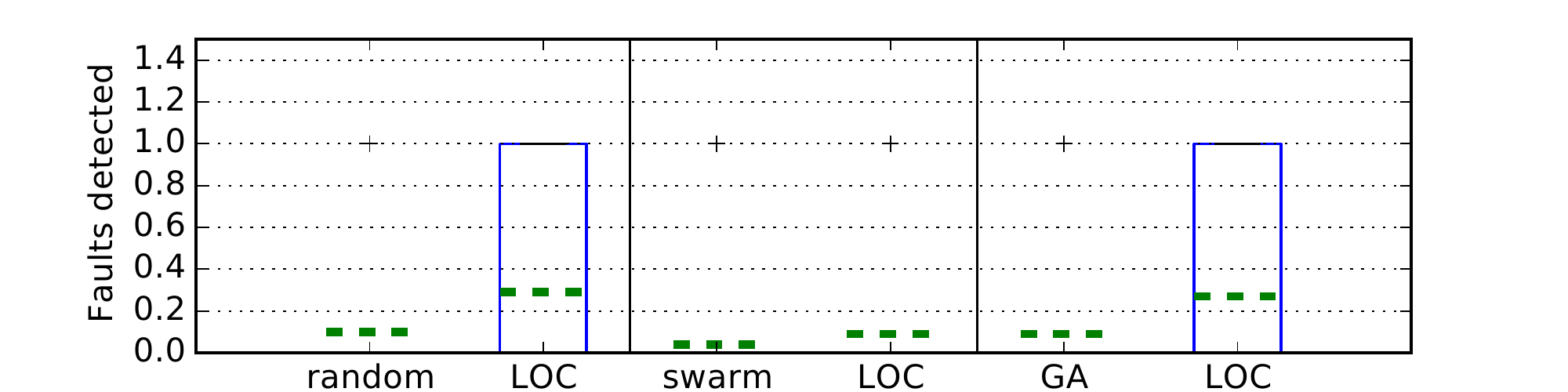}
\caption{heap (1 fault)}
\end{subfigure}
\begin{subfigure}{0.692\columnwidth}
\includegraphics[width=\columnwidth]{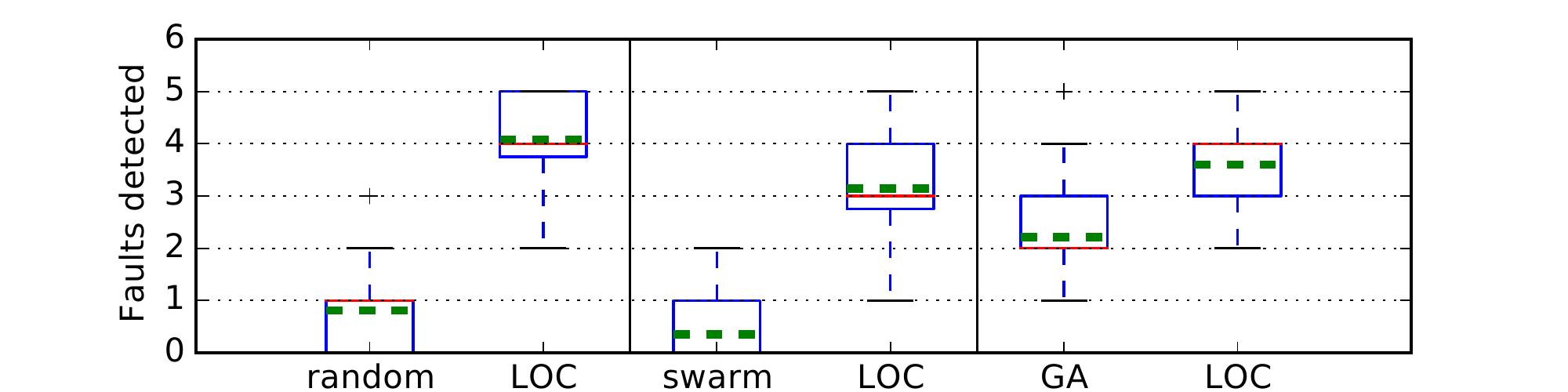}
\caption{pyfakefs (6 faults)}
\label{faults:pyfakefs}
\end{subfigure}
\begin{subfigure}{0.692\columnwidth}
\includegraphics[width=\columnwidth]{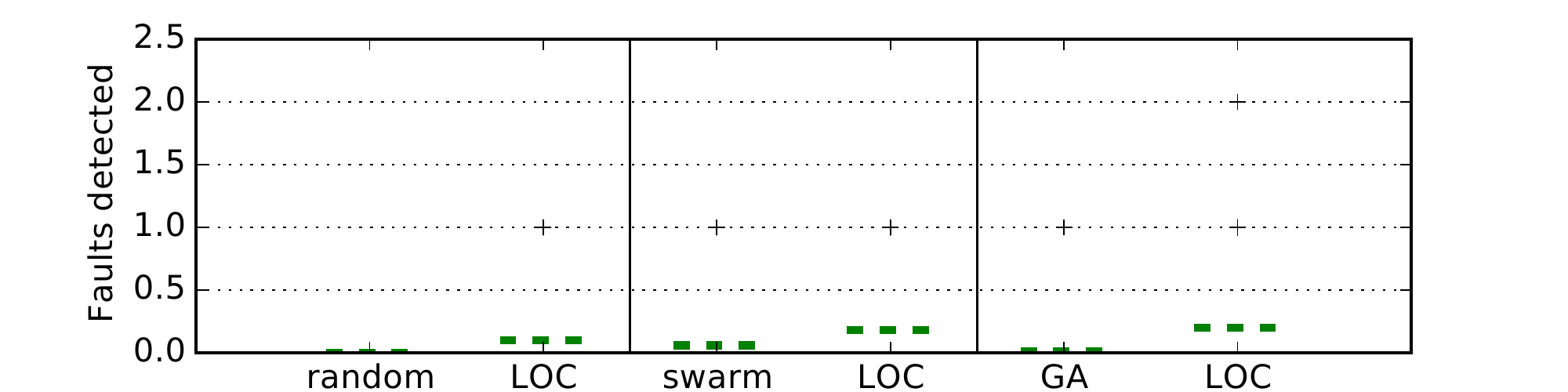}
\caption{sortedcontainers (2 faults)}
\end{subfigure}
\begin{subfigure}{0.692\columnwidth}
\includegraphics[width=\columnwidth]{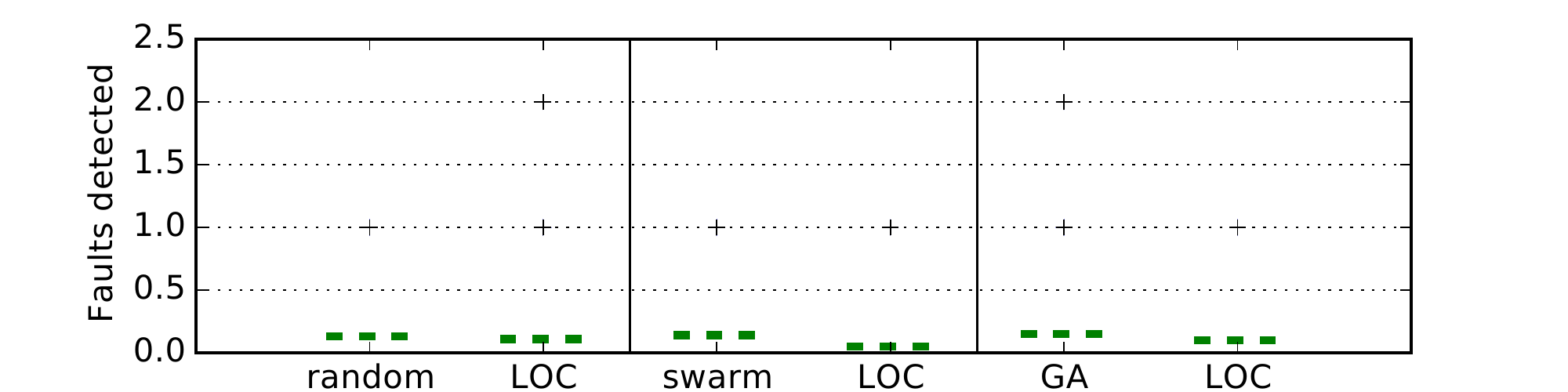}
\caption{sympy (14 faults)}
\end{subfigure}
\caption{Fault detection results.  Caption indicates total \# of
  distinct faults detected over all runs.}
\label{fig:failures}
\end{figure*}
}

{
\begin{figure*}
\begin{subfigure}{0.3\columnwidth}
\includegraphics[width=\columnwidth]{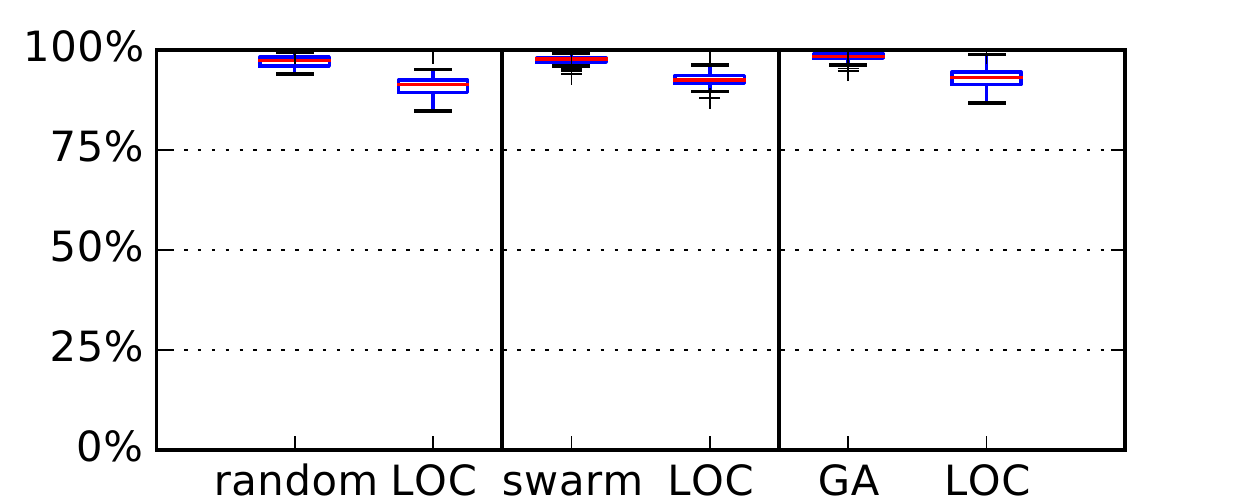}
\caption{arrow}
\end{subfigure}
\begin{subfigure}{0.3\columnwidth}
\includegraphics[width=\columnwidth]{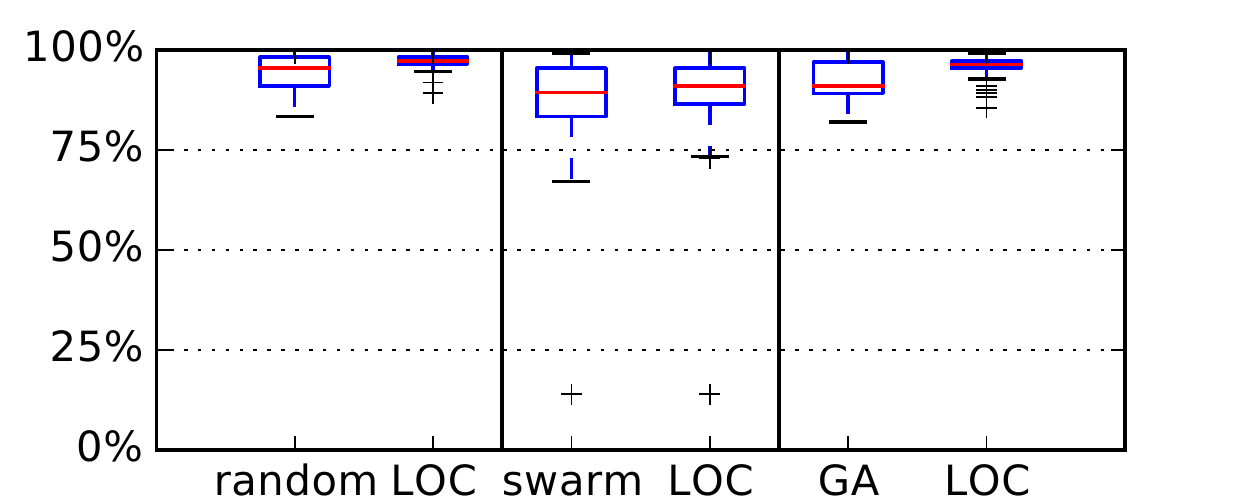}
\caption{AVL}
\end{subfigure}
\begin{subfigure}{0.3\columnwidth}
  \includegraphics[width=\columnwidth]{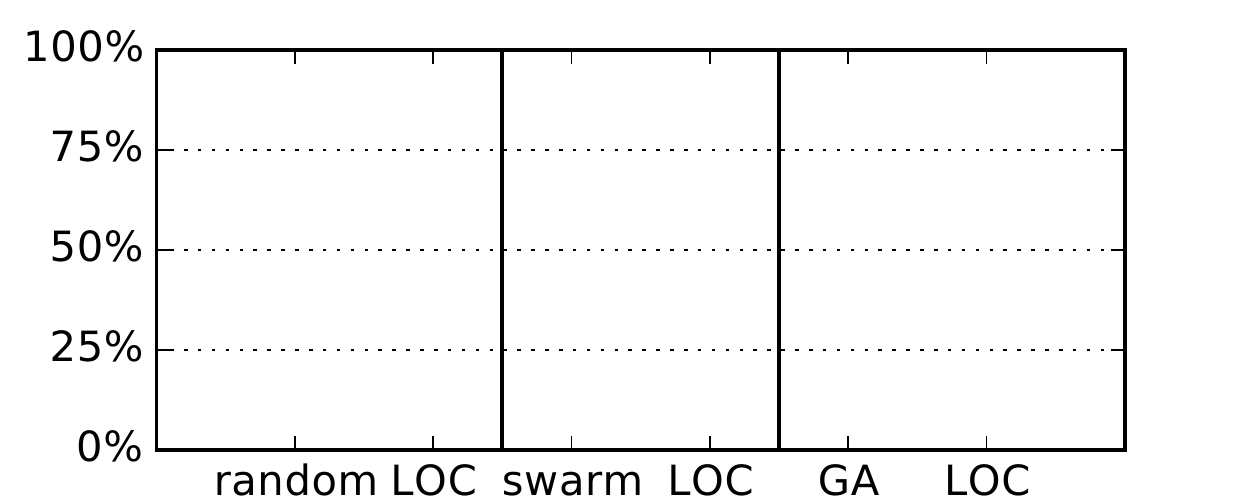}
  \caption{heap}
  \label{heapcov}
\end{subfigure}
\begin{subfigure}{0.3\columnwidth}
\includegraphics[width=\columnwidth]{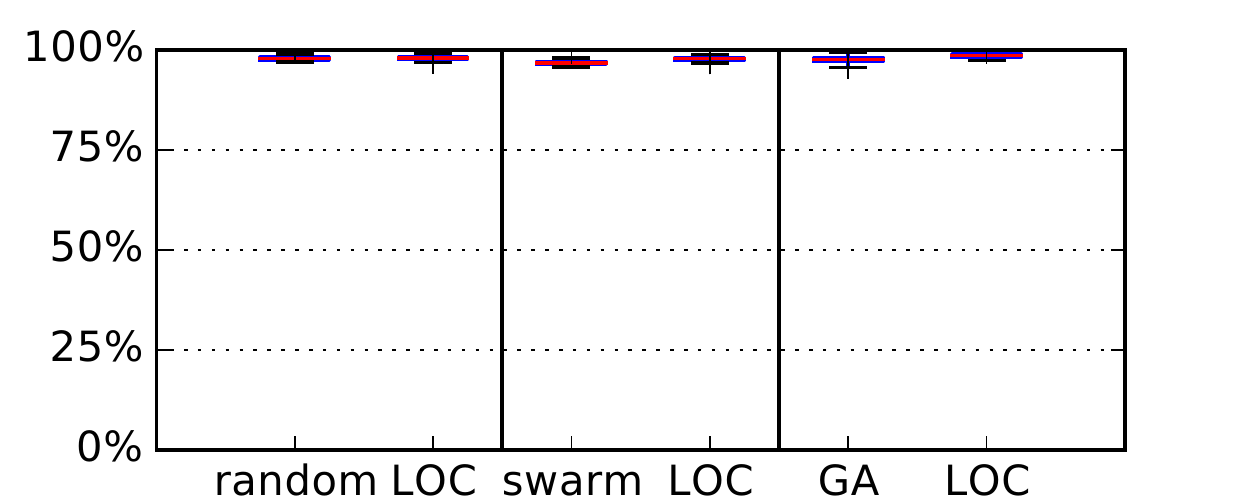}
\caption{pyfakefs}
\end{subfigure}
\begin{subfigure}{0.3\columnwidth}
\includegraphics[width=\columnwidth]{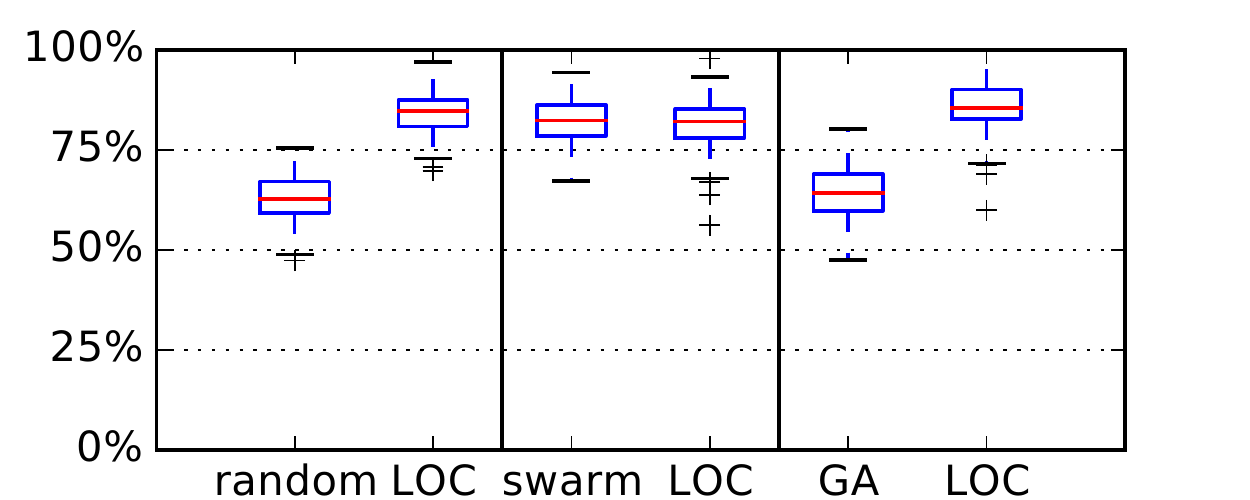}
\caption{sortedcontainers (old)}
\end{subfigure}
\begin{subfigure}{0.3\columnwidth}
\includegraphics[width=\columnwidth]{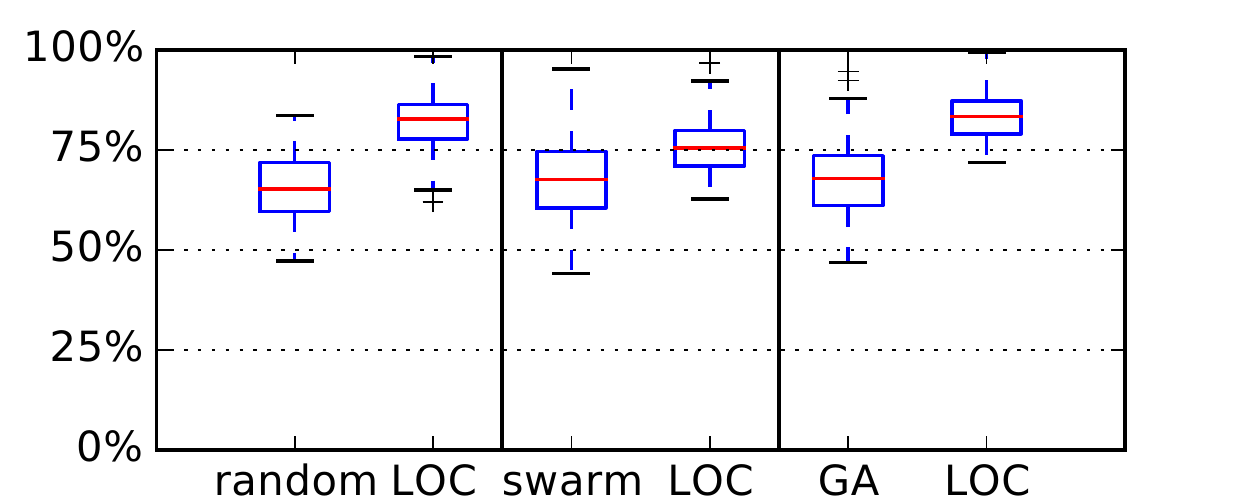}
\caption{sympy}
\end{subfigure}
\begin{subfigure}{0.3\columnwidth}
\includegraphics[width=\columnwidth]{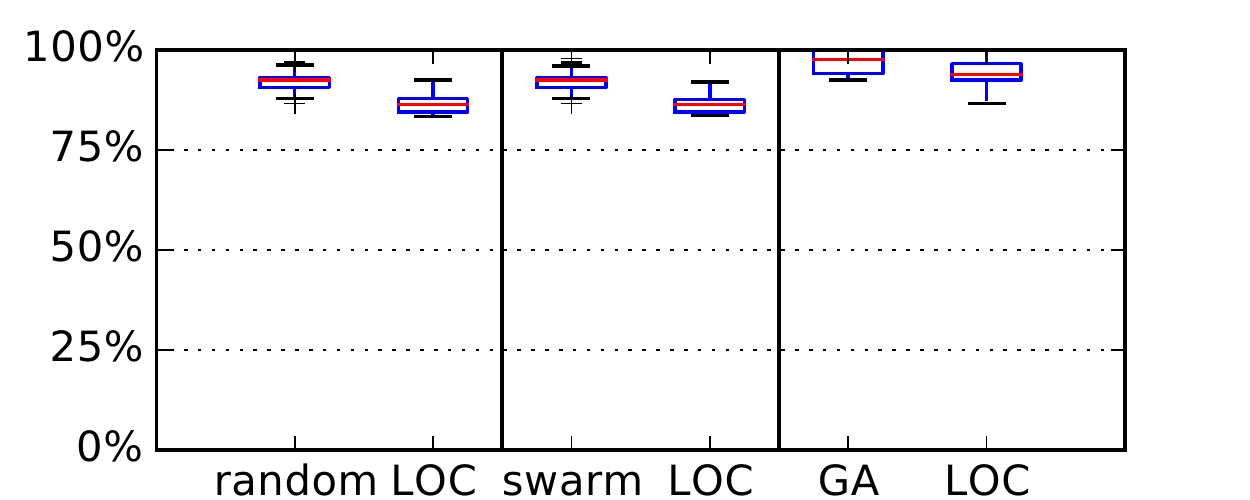}
\caption{bidict}
\end{subfigure}
\begin{subfigure}{0.3\columnwidth}
\includegraphics[width=\columnwidth]{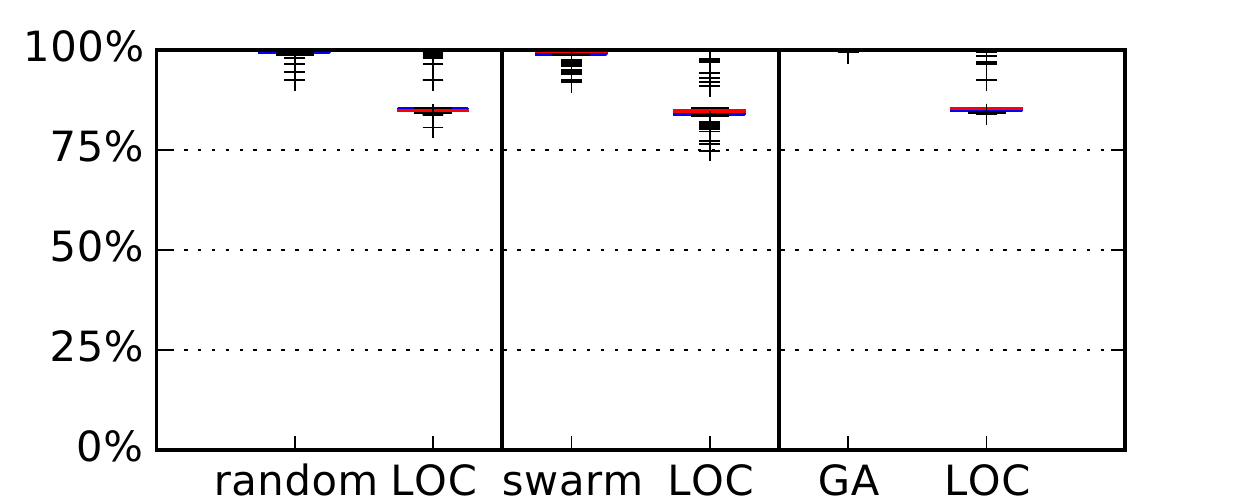}
\caption{biopython}
\end{subfigure}
\begin{subfigure}{0.3\columnwidth}
\includegraphics[width=\columnwidth]{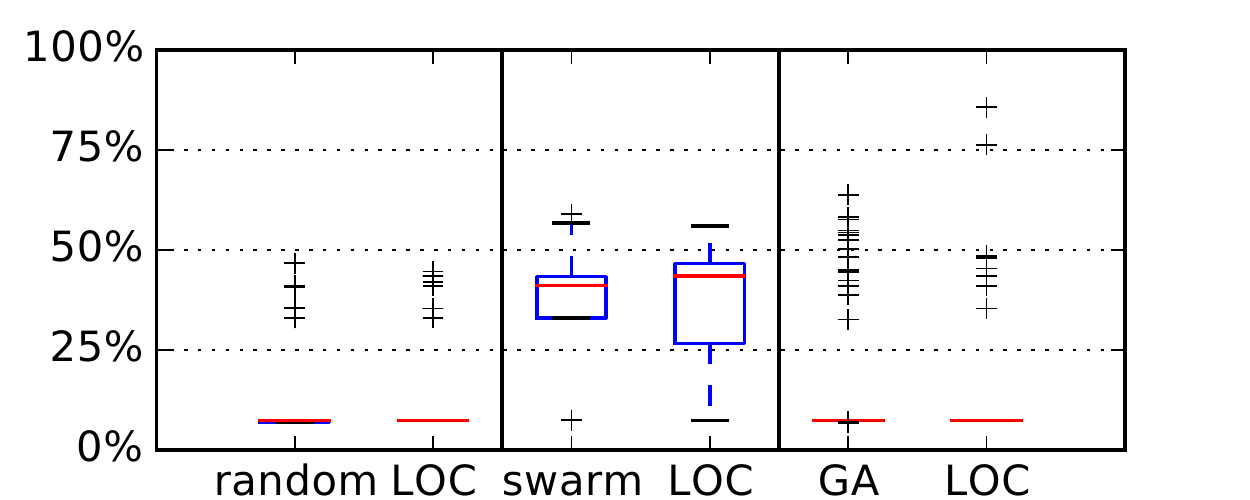}
\caption{C Parser}
\end{subfigure}
\begin{subfigure}{0.3\columnwidth}
\includegraphics[width=\columnwidth]{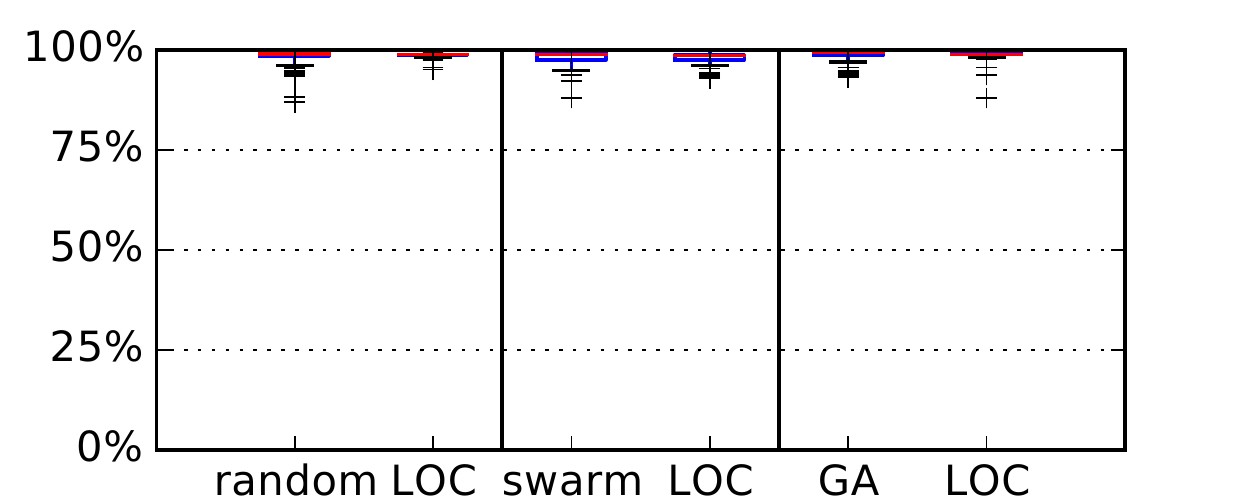}
\caption{python-rsa}
\end{subfigure}
\begin{subfigure}{0.3\columnwidth}
\includegraphics[width=\columnwidth]{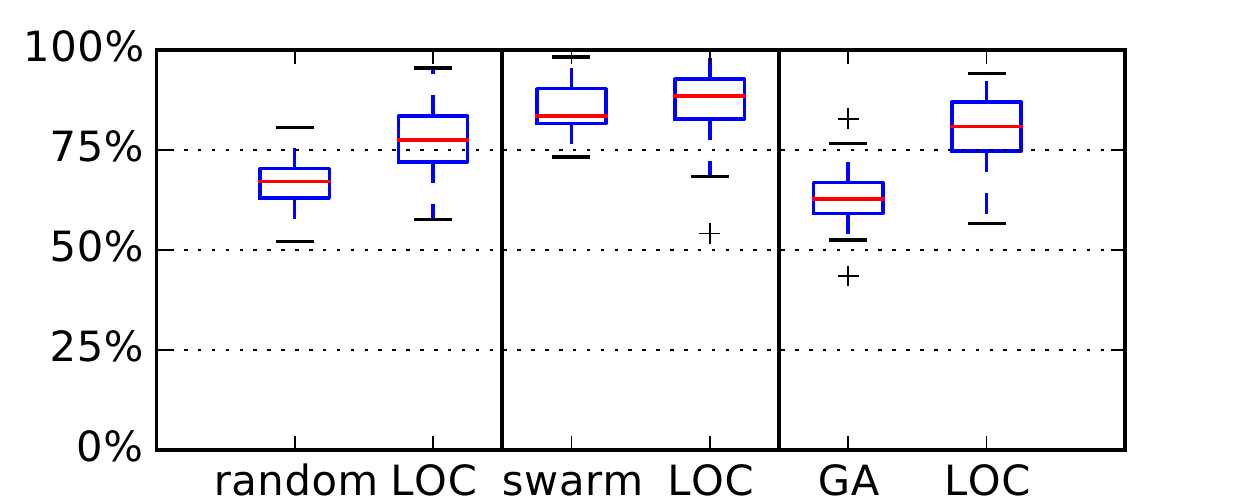}
\caption{redis-py}
\end{subfigure}
\begin{subfigure}{0.3\columnwidth}
\includegraphics[width=\columnwidth]{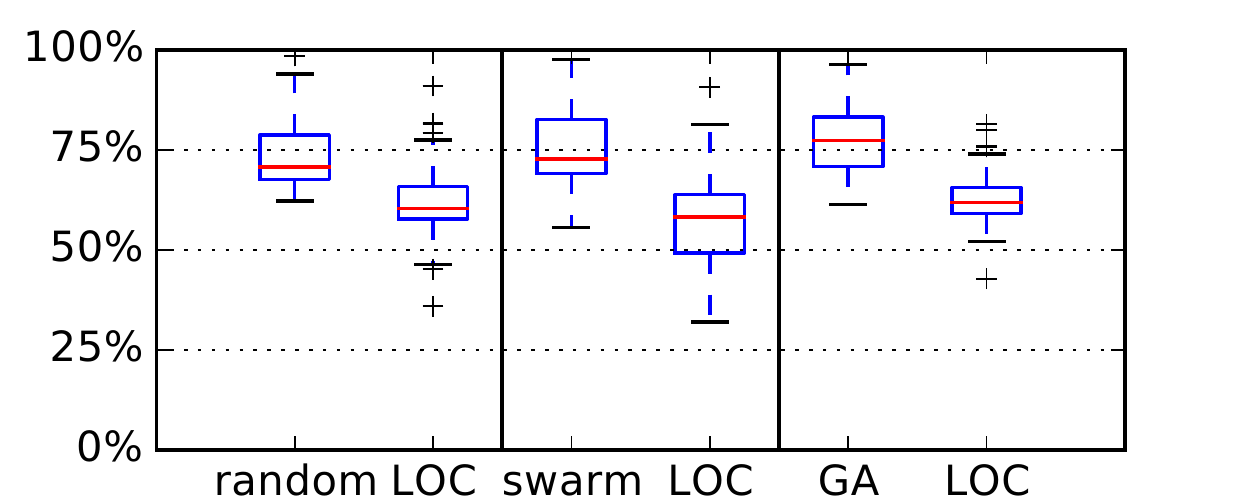}
\caption{simplejson}
\end{subfigure}
\begin{subfigure}{0.3\columnwidth}
\includegraphics[width=\columnwidth]{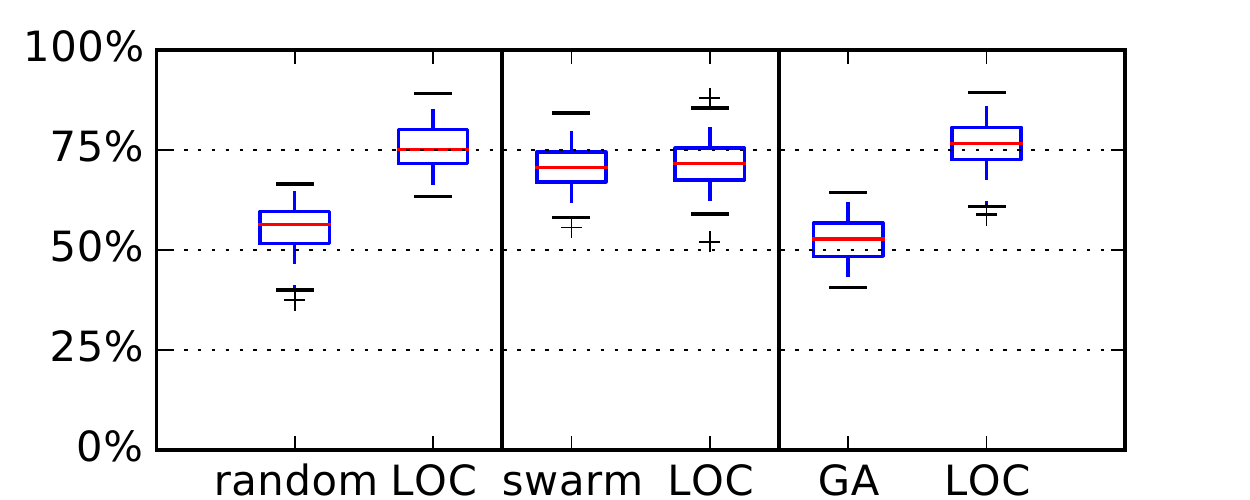}
\caption{sortedcontainers}
\end{subfigure}
\begin{subfigure}{0.3\columnwidth}
\includegraphics[width=\columnwidth]{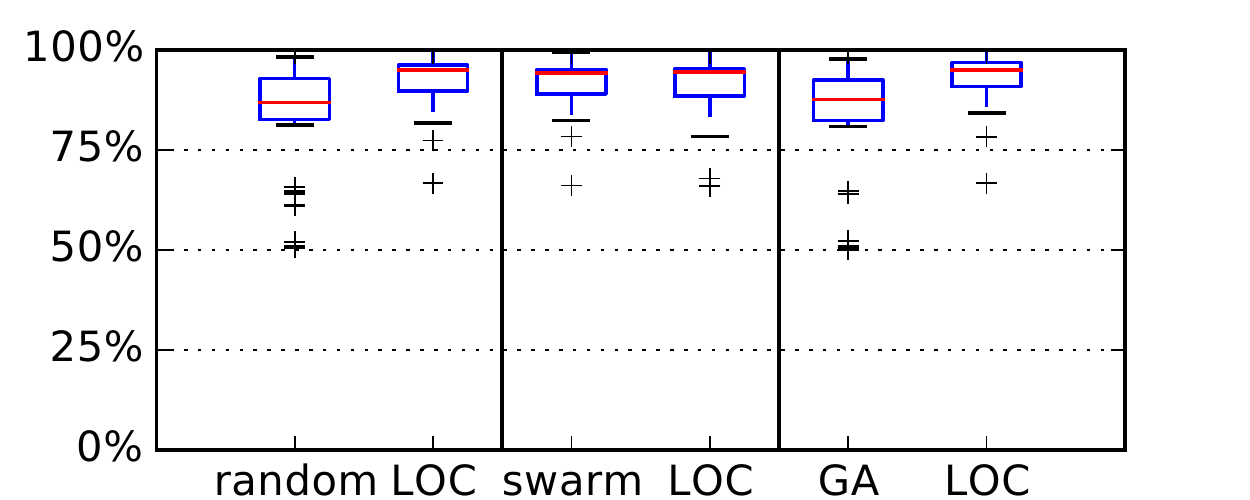}
\caption{TensorFlow}
\end{subfigure}
\begin{subfigure}{0.3\columnwidth}
\includegraphics[width=\columnwidth]{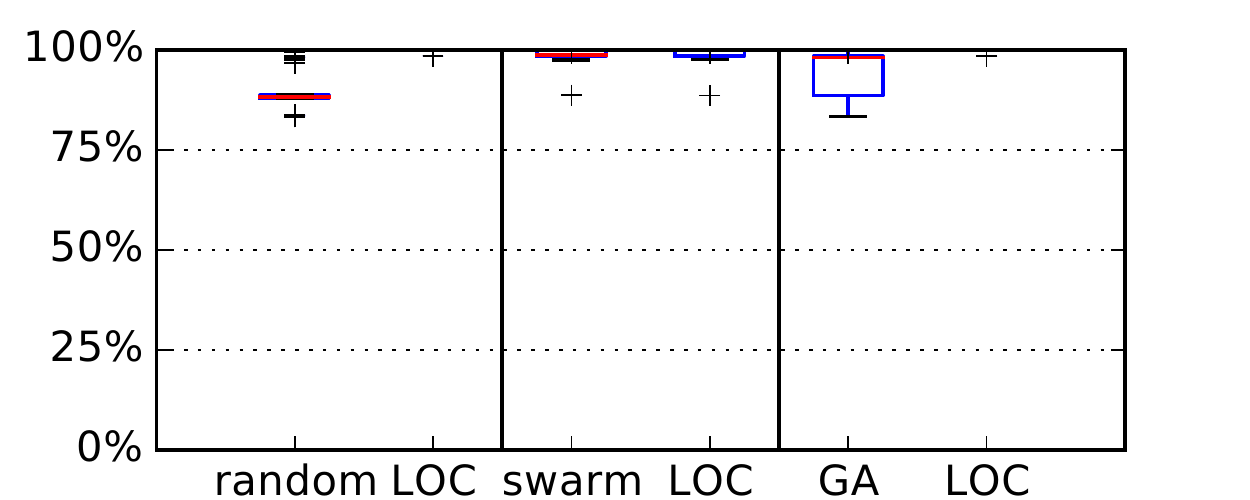}
\caption{z3}
\end{subfigure}
\begin{subfigure}{0.3\columnwidth}
\includegraphics[width=\columnwidth]{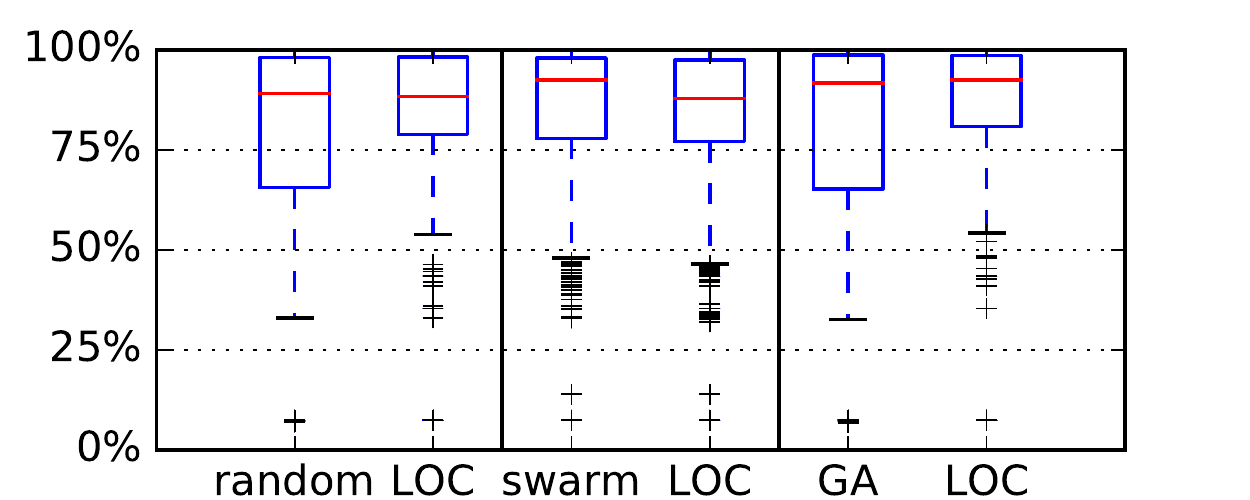}
\caption{All SUTs combined}
\label{allsuts}
\end{subfigure}
\caption{Branch coverage results for individual Python SUTs, plus summary.}
\label{fig:covallpy}
\end{figure*}
}

We ran each test generation method 100 times for 60 seconds on each
SUT, and used those runs as a basis for evaluation. Table \ref{pygainloss} gives an overall picture of the difference
between unguided random testing and LOC heuristic-guided testing,
divided between SUTs with detectable faults and those without.  The second
and third columns show mean changes in coverage. For those SUTs with faults
detectable by the test harness, the {\bf faults} column shows changes in number
of faults detected.  
The final column, {\bf =branch} shows the time required to obtain the mean 60
second branch coverage achieved using LOC, using pure random testing.
Very roughly speaking, 60 seconds of testing with the LOC heuristic is
as effective (at least in terms of branch coverage) as this amount of
pure random testing.  This is computed by taking the LOC coverage,
then repeatedly running random testing until it reaches the same
coverage, and taking the mean of the runtimes over 30 runs.  This value is more informative than the simple
percent improvement in coverage, since different branches are not
equally hard to cover: for many SUTs 60-80\% or more of all branches
covered by any test are covered in the first few seconds of every
test, but covering \emph{all} branches covered by any test may take more than
an hour, even with the most effective methods.  For example, while
{\tt z3} has one of the smaller percent improvements in coverage, it
takes \emph{more than 16 minutes on average} for random testing to cover the missing 11\% of
branches. For all columns, values in italics indicate differences that were not statistically
significant (by Mann-Whitney U test \cite{arcuri2014hitchhiker});
values in bold indicate a statistically significant improvement over
random testing, and SUTs in bold indicate that all significant changes
for that SUT were improvements.  There are two versions of
{\tt sortedcontainers}; the latest, and an older version before TSTL
testing, with two very hard-to-detect faults (first detected by
hours of random testing).

For 11 of the 15 SUTs and 9 of the 13 non-toy SUTs, \emph{all}
statistically significant differences from random testing were
improvements, and often very large improvements ({\bf RQ1}).  In every case where
there was a significant change in fault detection rate, there was an
improvement, and the improvements were all larger than the single (not
significant) negative change.  Note that because of the time taken to
process failing tests, when LOC improves fault
detection, it pays a price in coverage
proportional to the gain.

We also tried using the purely static method for estimating LOC
discussed above.  While this still often improved on random testing, it
was statistically significantly much worse (by over 1000 branches/statements, and a large corresponding decrease in fault detection) than random testing for {\tt
  pyfakefs}, and statistically significantly worse (but in a less dramatic fashion) than the dynamic sampling approach for both kinds of coverage for \emph{all other SUTs except {\tt
  redis-py}}.  For {\tt redis-py}, the static method was actually
better than the dynamic approach, perhaps because there is almost no
dynamic element to the types in the harness and the interesting code
is primarily in top-level wrappers; inspection suggests that it also
may simply be that an inaccurate estimate here has a beneficial
effect, due to the dependencies between methods.  Such a result for static LOC is, we see, unusual, at
least among our SUTs, and the cost of dynamic sampling seems
acceptable, given the benefits for all but one SUT.

Figures \ref{fig:failures} and \ref{fig:covallpy} show more detailed results for applying
the LOC heuristic, with each graphic structured to show pairs of
methods, with the LOC-biased version in the second (right) position of each
pair, to visualize results for {\bf RQ1-RQ3}\footnote{The odd graph
  for Figure \ref{heapcov} means all runs always hit 100\% branch
  coverage.}.  We omit statement coverage results, as the graphs are
essentially not distinguishable from branch coverage
results\footnote{The graph for {\tt heap} is difficult to read because all methods
  always obtain 100\% coverage.}.  The
leftmost pair (random and LOC) corresponds to the data in Table \ref{pygainloss}.
Coverage
results in Figure \ref{fig:covallpy} are normalized to \% of maximum coverage
obtained in any run, while faults are shown as actual number of faults
detected.  For fault detection, we added a dashed green line showing
mean, since the small range of values (0-3) for most subjects produces
similar median values even when results are quite different.
In addition to pure random testing, these graphs show results for combining LOC
with two
 other heuristics provided by TSTL.  Because the LOC heuristic simply biases the actions
chosen by the core random selection mechanism, it can be combined with
many other testing methods, so long as they do not require setting
action (class) probabilities.  The middle and rightmost comparisons
are for, without and with LOC bias:

\begin{itemize}
\item {\bf Swarm} \cite{ISSTA12}, before each test, randomly disables
a set of action classes (with 50\% probability) for each test.  The TSTL swarm testing
implementation also uses dependencies between action classes to
improve the performance of swarm testing over the previously published method.

\item {\bf GA} is a typical coverage-driven
  genetic algorithm \cite{FA11,McMinn04search-basedsoftware,GuidedGA}, mixing 20\% random generation with mutation, crossover, and extension of high-fitness (as
  measured by code coverage, without branch distances) tests.  It
  allows us to directly compare with
  search-based testing/mutational fuzzing driven by dynamic coverage measures.
  The TSTL GA
implementation is tuned to small-budget testing:
20\% of the time, or if there are no high-fitness tests in the
population, it generates a new random test instead of choosing a test
from the pool, and can extend a test rather than mutating it. This mixes initial population selection with
population refinement and allows the
system to escape local minima.
\end{itemize}

All three methods (LOC, swarm, and GA) are useful; it is never best to
stick to pure random testing.
Swarm often made testing worse{\footnote{This is because swarm
  testing increases test diversity, which may make
  individual tests less effective, and not pay off in only 60
  seconds.}, but when it was effective, it was highly
effective, for three of the most difficult to test SUTs, the C parser,
{\tt redis-py}, and {\tt z3} (swarm's origins in compiler testing show
in that it helps with constructing structured, program-like, inputs).
However, in all three cases, also using LOC makes swarm perform even
better, so that swarm is never ``the best'' method.

For fault detection, all methods detected all faults \emph{in at least one
run} for {\tt arrow}, {\tt AVL}, {\tt heap}, and {\tt pyfakefs}.  For
{\tt pyfakefs}, note that while every method found every fault at
least once, only
LOC-based methods were reliably able to detect most faults, and only
LOC alone consistently found most of the faults, most of the time
(see Figure \ref{faults:pyfakefs}).  Using LOC alone is statistically
significantly better than all other methods (all differences with LOC
and with random are significant, with $p<1.0\times10^{-5}$), detecting a mean of 4.07 faults per
run. The next best method, the GA with LOC, only detected a mean of 3.6
faults/run.  No method not using LOC detected more than 2.3 of the
faults per run, on average.  Random testing and swarm without LOC
detected less than one fault, on average.

Using LOC also made detection
rates and
mean number of faults found higher for {\tt arrow}, {\tt AVL}, and
{\tt heap}, but in a less dramatic fashion (see Figure \ref{fig:failures}).  For all subjects with
detectable faults, except {\tt SymPy}, the method detecting the most
faults was either the LOC heuristic or a combination of the LOC
heuristic with the GA, and the difference in fault detection
rates for {\tt SymPy} was not statistically significant between
methods ({\bf RQ2-3}).

For
{\tt sortedcontainers} and {\tt sympy} total fault detection results were
more interesting.  Pure random testing never discovered either fault
in {\tt sortedcontainers}.
LOC alone discovered one fault only, 30 times
out of all runs; GA discovered only the other fault, and only 3 times;
swarm testing found both faults, but detected a fault at all about
half as often as LOC (6 detections for the fault the GA found, and 12
for the fault LOC found).  Combining methods, swarm with the LOC found
the fault LOC detected 66 times, and using the GA with LOC also found
that same fault 66 times, and additionally detected the fault detected
by GA 3 times.  The story for {\tt SymPy} was even more complex.  Pure
random testing found the most diverse fault set, but still only 7 of
the 14 total detected faults.  The LOC heuristic only found 3 different faults,
but two of these were ones not found using pure random testing; one
was not detected by any other approach.  The
GA found 4 different faults, two of which were not found using
pure random testing, and one of which only it detected.  Swarm testing found 4 different faults, again
including one not found by random testing, but none unique among
methods.  Combining the GA with LOC made it possible to detect 5
faults, none of which were unique to that approach, and combining
swarm with LOC found 3 different faults, of which one was unique.
These results suggest that for finding faults, diversity of approach
may be critical, a generalization of the reasoning behind swarm
testing and swarm verification \cite{swarmIEEE}, but these overall
detection results are not
statistically validated in any sense: lumping all runs together essentially
produces one large run.

Absolute differences in coverage varied for {\bf RQ1}, but were
often large.   The mean mean gain in branches covered, for LOC vs. pure
random testing, was 219.1 branches, with a median mean gain of 146.1
branches; mean loss when LOC was ineffective was only 27.9 branches
(mean) or 29 branches (median).  For 8 of 11 branch coverage
improvements the gain was more than $80$ branches.  

\begin{table}
\centering
\caption{Best Methods}
\label{tab:best}
\begin{tabular}{l|l|l|l|l|l|l}
  & random & LOC & GA & GA+LOC & swarm & swarm+LOC \\
  \hline
  branch coverage & 0 & 1 & 5 & 7 & 0 & 2 \\
  statement coverage & 0 & 2 & 4 & 7 & 0 & 2 \\  
  fault detection & 0 & 2 & 1 & 3 & 0 & 0 \\
\end{tabular}
\end{table}

Perhaps the simplest way to compare methods for all of {\bf RQ1-RQ3} is to ask ``For
how many SUTs was a particular method the best approach for 
coverage?'' and ``For
how many SUTs was a particular method the best approach for fault
detection?''  ignoring statistical significance.  
Table \ref{tab:best} shows the results.  This analysis also reflects probabilites
that, based on trial runs, a user would select each method for use in
testing, an important point we consider further in Section \ref{sec:disc}.  The combination of a GA with the LOC heuristic is
clearly the best method, but we should always recall that this imposes the
costly overhead of collecting coverage; below we attempt to estimate
how the methods would compare if LOC were given the advantage of not
having to compute coverage.  Moreover, when LOC improved on 
random testing, it almost always improved GA and swarm to use LOC, as
well ({\bf RQ3}).  The exceptions were that for branch coverage, adding LOC did not
improve on GA for the C parser, did not improve on swarm for the buggy version of {\tt
  sortedcontainers}, and did not improve either GA or swarm for {\tt python-rsa}.
For {\tt python-rsa}, however, all values were so similar that this disagreement
was not statistically significant.  For fault detection, improvement over random was always accompanied
by improvement when added to GA and swarm.  It is clear that
the LOC heuristic is a low-cost way to improve the performance of a GA
in most cases, and that, ignoring the cost of code coverage, combining
methods is often very useful.

We summarize the answers to {\bf RQ1-RQ3} in terms of individual SUTs more succinctly in Section \ref{sec:rq13summary} below, as a prelude to discussing the overall meaning and possible causes for our results.

\subsection{Analysis Combining All Subjects (RQ1-RQ3)}
Using normalized coverage data allows analysis of all subjects together, both
those where LOC helped and those where it was harmful (Figure
\ref{allsuts}, which includes subjects with and without faults).
Considering the impact of LOC on each SUT, and how often it was
helpful is generally a more important way to understand the results,
but the combined analysis provides some additional insights into the
effect sizes for the various heuristics, and makes the comparison with
pure random testing ({\bf RQ1}) even clearer.  The means for
LOC were 83.0\% of maximum branch coverage and 83.2\% of maximum statement coverage.  The means for pure random testing were
79.3\% branch coverage and 79.6\% statement coverage ({\bf RQ1}).  All differences were significant by Wilcoxon
test at $p<1.0\times 10^{-18}$.  LOC was also better for branch and
statement coverage than GA (80.2\% branch coverage, 80.4\%
statement coverage, $p < 1.0 \times 10^{-23}$) ({\bf RQ2}).  This is particularly
notable: despite also paying the (unnecessary) overhead of coverage,
using random testing with the LOC bias outperformed a GA using code
coverage results to drive testing.  Combining the GA and LOC ({\bf RQ3}) produced
mean branch coverage of 84.2\% and mean statement coverage of 84.5\%.  It is not clear that
combining LOC and GA would even improve on LOC, if LOC did not pay the
(high) overhead of code coverage; that is, the advantage of adding the
GA may be overwhelmed by coverage costs for many SUTs.  Swarm without LOC had the best
coverage means (85.2\% for both kinds), despite never being the method
for best branch or statement coverage for any SUT ({\bf RQ2}).  Swarm with LOC performed slightly better than LOC alone in terms of coverage (83.2\% branch
coverage, 83.4\% statement coverage) ({\bf RQ3}).  Again, we emphasize that for the very subjects where swarm
testing was neccessary for producing good results, swarm with LOC was \emph{always} better
than swarm alone ({\bf RQ3}).  Swarm's higher mean is entirely due to the
\emph{very} poor performance of LOC (or swarm with LOC) on a few
subjects, and for these subjects swarm was also always less effective than
GA ({\bf RQ2}).

We similarly normalized fault detection by counting failed tests
(hence probability of detecting any faults at all), and using the
maximum number of failed tests as 100\% ({\bf RQ1-RQ3}).  Swarm testing had the worst
fault detection results by a large margin, with a mean of only 5.8\%
of maximum failures, faring badly compared to LOC (20.1\%), GA
(11.1\%), LOC+GA (24.8\%) and even pure random testing (6.3\%).
Presumably, this is partly because our faulty SUTs and
``compiler-like'' SUTs did not overlap.  Swarm with LOC improved this,
but only to 10.6\%.  LOC and LOC+GA, with 11.1\% and 21.6\%, respectively, were the \emph{only} methods with \emph{median} values better than 0.0\%, i.e., with
median detection of \emph{any} faults ({\bf RQ1-RQ3}).

There was not a compelling correlation between SUT size and either
branch coverage ($R^2=0.01$) or fault detection ($R^2=0.22$)
effectiveness for LOC compared to random testing.  The directions of
correlations are also opposite (positive for coverage, negative for
faults).  Using maximum statement coverage to measure ``effective size
of tested surface'' instead of actual SUT LOC, produced similar
results ($R^2=0.11$, $R^2=0.17$).

\subsection{The Cost of Coverage}
\label{sec:codecov}

A key assumption of this paper (and factor in choosing Python as the
target language) is that, despite Python's popularity, and years of
work on testing tools for Python, such as unit testing libraries, the
overhead of collecting coverage information in Python is large. 

Table \ref{tab:coverage} shows a simple measure of the cost of
coverage:  for each SUT we ran 60 seconds of testing with and without
coverage instrumentation, provided by the state-of-the-art {\tt
  coverage.py} tool, for the same random seeds, 10 times, and
recorded the total number of test actions taken in each case.  The table shows
the average ratio between total actions without coverage and total
actions with coverage.  The cost ranges from negligible ({\tt arrow}) to
exorbitant ({\tt biopython}).  Improving Python coverage costs is
non-trivial, as discussions of the issues (and the fact that these
overheads persist despite years of development on {\tt coverage.py}) suggest
\cite{covdiff1,covdiff2}.  

The second column of results in Table \ref{tab:coverage} is for
execution using the Python JIT (Just-In-Time compiler) {\tt pypy}.  Overhead for coverage is similar, except for {\tt biopython}, where the cost is still significant at well over 2x more test operations when running without coverage instrumentation; it is merely reasonable compared to the exorbitant 50x fewer test operations with coverage, when not using the JIT.   The mean is considerably reduced, due to {\tt
  biopython}, but median cost is similar.  We attempted to investigate
why the overhead for {\tt biopython} is so extreme when running
without a JIT, but were unable to determine what the source of the
problem is, based on profiler information.  Dropping it as an outlier,
considering only median overheads, it seems safe to say that with or
without a JIT, in Python, measuring code coverage will usually
effectively reduce the test budget considerably.  We also tried
re-running our experiments with the just released (Dec 14, 2019) stable
non-alpha  {\tt coverage.py} 5.0, though none of the changes seemed
likely to be relevant. As expected, results were within 10\% of those
for version 4.5.2, with no consistent improvement (or degradation).
It is difficult to say what the cost of coverage would be, using the
most sophisticated methods available in the literature; some are not
appropriate, in that coverage-driven methods need to check \emph{every} test
executed for at least lower-frequency targets, and are likely to
tolerate even infrequent ``misses'' poorly.  Furthermore, given the
popularity and development effort involved in {\tt coverage.py},
including effort spent reducing overhead, we do not expect to see any less costly
approaches available in Python for the forseeable future.  The dynamic
nature of the language, even under a JIT (as shown above) may preclude
reaching the low overheads sometimes seen for C and Java code.

It is important to note that GA is
the \emph{only} method in our experiments that actually requires measuring code
coverage during testing.  However, in order to report coverage results
as an evaluation, all methods were run with code coverage collection turned
on.  Again, we emphasize that in practice, a user interested in
actual testing 
would run without code coverage, when not using GA,
obtaining higher test throughput.  Determining the exact impact of coverage overhead is
difficult; we can record tests generated during execution
  without coverage instrumentation, and run the tests later to determine what coverage
  they would have obtained; however, the cost of recording all tests
  is itself very high, since some SUTs can
  generate thousands of tests in a minute, and storing thoses tests is expensive.  In this paper, we simply compare results as if
  all testing methods were
required to collect coverage data, but this over-reports the
effectiveness of GA compared to other methods, including pure random
testing.  While we cannot effectively compare the LOC
heuristic to GA for code coverage, without measuring code coverage, we
can examine fault detection.  LOC already detected significantly more
faults than any other
method for some SUTs ({\tt AVL}, {\tt heap}, and {\tt pyfakefs}), and
allowing for more test actions by disabling coverage only increased
the gap.  When we re-ran without code
coverage, LOC also significantly ($p = 0.04$) outperformed all GA-based methods for
{\tt SymPy}, detecting 0.24 mean faults per 60 second run, compared to
0.1 for the LOC+GA combination, or 0.15 for the GA alone (the best
performing method when coverage was collected).  For {\tt arrow} and
{\tt sortedcontainers}, on the other hand, the cost of coverage was
too low to enable LOC alone to outperform the previously
best-performing LOC+GA combination. The primary determination for
whether it is worth collecting code coverage in our experimental
results seems to be the cost of
coverage, rather than the effectiveness of GA.  That is, when coverage
is very cheap to collect, it is worth paying that cost to add GA to
LOC, and when coverage is expensive, at least in our results, it seems
that ``paying for GA'' may not be a good idea, even if GA is, ignoring
that cost, a useful
addition to LOC.

\begin{table}[b]
\centering
\caption{Gain in test operations when executing without coverage instrumentation}
\label{tab:coverage}
{\scriptsize
\begin{tabular}{l|r|r}
SUT & $\frac{\textrm{Actions without coverage}}{\textrm{Actions with
      coverage}}$ & $\frac{\textrm{Actions without     
                    coverage}}{\textrm{Actions with coverage}}$\\
& & (tests executed with {\tt pypy}) \\
\hline
\hline
{arrow} & 1.00 & 1.01\\
{AVL} & 6.41 & 4.66\\
{heap} & 4.98 & 5.07\\
{pyfakefs} & 2.03 & 2.11\\
{sortedcontainers} & 1.09 & 1.17\\
{SymPy} & 2.30 & 1.88\\
\hline
{bidict} & 1.48 & 1.24\\
{biopython} & 50.07 & 2.16\\
{C Parser} & 1.04 & 1.02\\
{python-rsa} & 1.14 & 1.22\\
{redis-py} & 1.02 & 1.06\\
{simplejson} & 1.47 & 1.09\\
{TensorFlow} & 9.54 & 9.62\\
{z3} & 2.14 & 2.11\\
\hline
\hline
Mean & 6.12 & 2.53\\
Median & 2.03 & 1.56 \\
\end{tabular}
}
\end{table}

\subsection{Comparison with {\tt python-afl}}

American Fuzzy Lop (AFL), commonly known as {\tt afl-fuzz} \cite{aflfuzz}, is an extremely popular
coverage-driven fuzzer (essentially using a GA over path coverage).  {\tt python-afl}
(\url{https://github.com/jwilk/python-afl}) makes it possible to fuzz
Python programs using AFL.  Because TSTL supports generating tests
using {\tt python-afl} in place of TSTL itself, we were able to
perform an additional comparison with the AFL algorithm (and {\tt
  python-afl}'s instrumentation, designed to be low-overhead, with a C
implementation) for test generation.  TSTL in this setting is
\emph{only} used for test execution and property checking, not for
test generation.  We know from past experience that {\tt python-afl}
can find faults that TSTL cannot (e.g., \url{https://github.com/jmcgeheeiv/pyfakefs/issues/378}).

Because AFL requires a corpus of
initial inputs on which to base fuzzing (it is a mutational fuzzer),
we gave AFL 60 seconds to fuzz after using pure TSTL random testing
for 20 seconds to produce an initial corpus.  This in theory gives
{\tt python-afl} a substantial advantage over our TSTL-only tests.

Pure random testing with TSTL essentially always dramatically
outperformed {\tt python-afl} in terms of branch coverage, even though
{\tt python-afl} had the advantage of incorporating results from 20
seconds of TSTL random generation.  For example, mean branch
coverage for the simple AVL example decreased by 11\%, while it decreased by almost 15\% for {\tt sortedcontainers}, and nearly 80\% for {\tt sympy}.  The path
coverage-based GA of AFL did produce improvements over pure random testing
for some of the toy examples, e.g. improving fault detection rates by about
1\% over random testing for the AVL example, and from 10\% to 29\% for
{\tt hypothesis\_heaps}.  However, this was still much worse than the
LOC heuristic fault detection rates of 99\% and 71\%, respectively.
Unsurprisingly, given the huge loss in code coverage, {\tt python-afl}
was unable to find the faults in {\tt
  sortedcontainers} and {\tt sympy}.

Rather than elaborate on these results, we simply note that it is
unfair to compare against {\tt python-afl} under our experimental
settings and for the use case considered in this paper.  Modern
mutation-based fuzzers are primarily intended to be used in runs of at
least 24 hours \cite{Hicks18}.  They are, despite very sophisticated algorithms,
extensive tuning, and high-performance instrumentation, not useful for
quick turnaround property-based testing.  Random testing is, at least in the
Python setting, for this problem, a better baseline.  As would be
expected, given that LOC generally performs much better than random,
it also outperforms {\tt python-afl}, including in some cases where it
performs worse than pure random generation.

\subsection{Comparison with Feedback-Directed Random Testing}

We would also like to compare to the feedback-directed random testing
algorithm of Pacheco et al. \cite{Pacheco}.  Unfortunately, perhaps
due to reset or object equality overheads, it and related methods
\cite{FeedControl} are known to perform poorly in Python
\cite{Kazuki}.  We therefore performed a preliminary experiment using
the Randoop Java implementation.

\begin{figure}[t]
\includegraphics[width=\columnwidth]{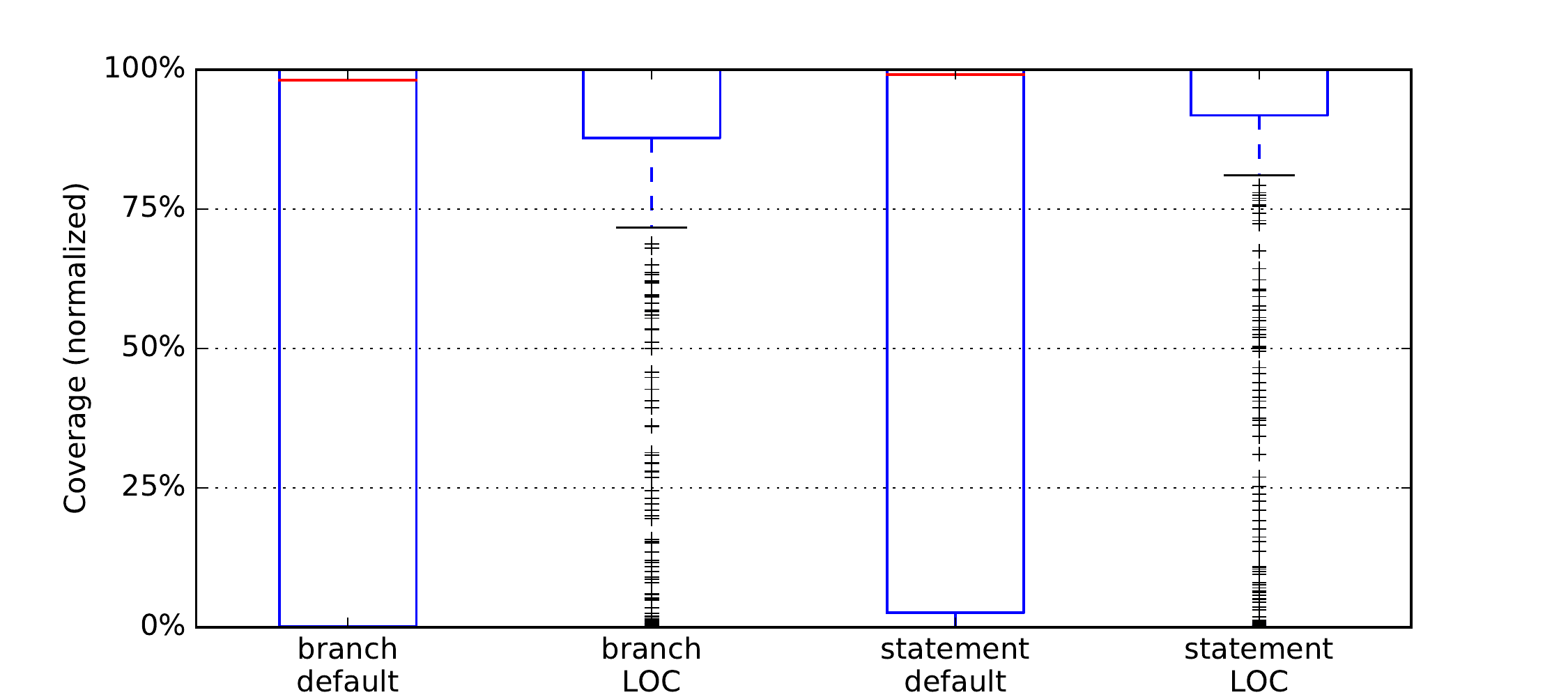}
\caption{Randoop coverage, default and with LOC heuristic}
\label{javacov}
\end{figure}

The implementation was simple:  we measured LOC (unlike with Python, only for non-comment, non-blank lines) for each method in the Class Under Test (CUT) using the Understand  \cite{_understand_2017} tool, and then used Randoop's ability to take as input a list of methods to test to bias probabilities accordingly.  That is, in place of simply listing all methods of the CUT, we duplicated each method in the list a number of times equal to its LOC, thus biasing the probability in favor of larger methods in exactly the same way as with the Python testing.   Unlike with Python, there was no dynamic sampling, consideration of additional methods called by a top-level method, or need to specify probabilities for ``actions'' not calling CUT code.  Simply measuring top-level method LOC is reasonable in the Randoop context, since Randoop is not performing property-based testing with complex actions, but testing at the method level.

For our experiments, we used a set of 1177 Java classes taken from 27
projects hosted on GitHub and used in previous work on measuring
testedness \cite{ahmed_testedness}\footnote{Thus projects known to
  compile, pass all tests, and be analyzable by Understand.}.  We
randomly selected projects until we had $\geq 1000$ classes, then
applied Randoop to all of each project's classes, first using the
standard Randoop settings, and then again with no changes except use
of LOC to bias method choice.  Note that here measurement of LOC is a
purely static, nearly zero cost, activity, and there is no coverage
measurement during test generation for any approach.  Coverage was
measured by instrumenting and executing generated unit tests.

The results are, at a high level, similar to those for Python: the LOC
heuristic is sometimes harmful, but more often produces an improvement
in test effectiveness.  Using the LOC heuristic increased mean branch
coverage in unit tests for these classes produced by Randoop from 1640.7 branches to  1,924.1 branches (a 17.3\% improvement), and mean statement coverage from 16010.8 statements to 18241.4 statements (a 13.9\% improvement).  The changes in median coverage were from 72.0 to 479.0 branches and from 234.0 to 6963.0 statements.  
These results were significant by Wilcoxon test \cite{arcuri2014hitchhiker}, with $p < 1.5 \times 10^{-6}$.   Figure \ref{javacov} shows coverage, over all classes, normalized.  Normalization means that we consider the maximum coverage for either the default Randoop or LOC heuristic suite to be ``100\% coverage.''  This allows us to show results for very different class sizes using a consistent scale.  The graph makes it clear that, while there were many classes where LOC was not useful, overall the effect was striking, with coverage much more tightly clustered close to the maximum observed.  Median coverage (both kinds) for LOC was 100\%; default coverage fell to 98.1\% (branch) and 99.1\% (statement).   Mean branch coverage improved from 65.7\% to 78.1\% using the LOC heuristic, and mean statement coverage from 67.5\% to 81.3\%.  Normalized results are significant with $p < 1.5 \times 10^{-11}$.

At the project level, coverage change was significant for only six projects (in part because most projects do not have very many classes).  For five of these projects, branch (+4.9\%, +10.2\%, +12.5\%, +33.0\%, +112.6\%) and statement coverage (+10.3\%, +11.6\%, +10.1\%, +33.8\%, +87.6\%) both improved significantly, and absolute gains were larger than a mean of 1,000 branches/statements (in one case more than 10K statements) for all but one project.  The other project had a significant decrease of 23.6\% in branch coverage only, about 1,000 branches.  The largest gain from using LOC at project level was 200\% improvement in mean branch and statement coverage.  More than 47 individual classes (across 11 different projects) had gains of 2000\% or more, however.

\subsection{Using Outdated LOC Estimates}
\label{sec:rq7}

In order to estimate the impact of the quality of the LOC estimates,
where a large impact would force programmers to frequently re-analyze their code,
we ran the exact same experiment as for {\bf RQ1-RQ3},
except using probabilities
sampled from older versions of the system, for all of the SUTs where
(1) the LOC heuristic was more effective than random testing and (2)
there existed significantly older versions of the code compatible with
the test harness.  In each case, we
used as old a version as was compatible with the API of the
latest version of the system with respect to the test harness.

For {\tt SymPy} we were able to revert all the way from the 1.0 release
(2016-3-8) to the 0.7.6 release (2014-11-20); difference of 3559
commits with total diff
size, measured in lines, of 214125).  For {\tt python-rsa}
we based probabilities on version 3.1.1 (2012-06-18; 131 commits/diff
size 6338), nearly 4 years
older than the current version 3.4.2 (2016-03-29).  With {\tt redis-py} we
reverted to version 2.10.0 (2014-06-01; 90 commits/diff  size 1380) in place of the current
version, 2.10.5 (2015-11-2).  Finally, for {\tt sortedcontainers}, we only
reverted to version 1.5.2 (2016-05-28; 21 commits/diff size 1090) in place of version 1.5.7
(2016-12-22); earlier versions removed a few interesting functions to
test, and we wanted to see if a somewhat closer-to-latest version
changed results in an obvious way.  In all cases, the results were
either very similar and statistically indistinguishable ($p > 0.05$),
or, for {\tt SymPy}, superior to, results
using recent, more accurate, counts.

{
\begin{figure*}[t]
\centering 
\begin{subfigure}{1.0\columnwidth}
\centering
\includegraphics[width=0.8\columnwidth]{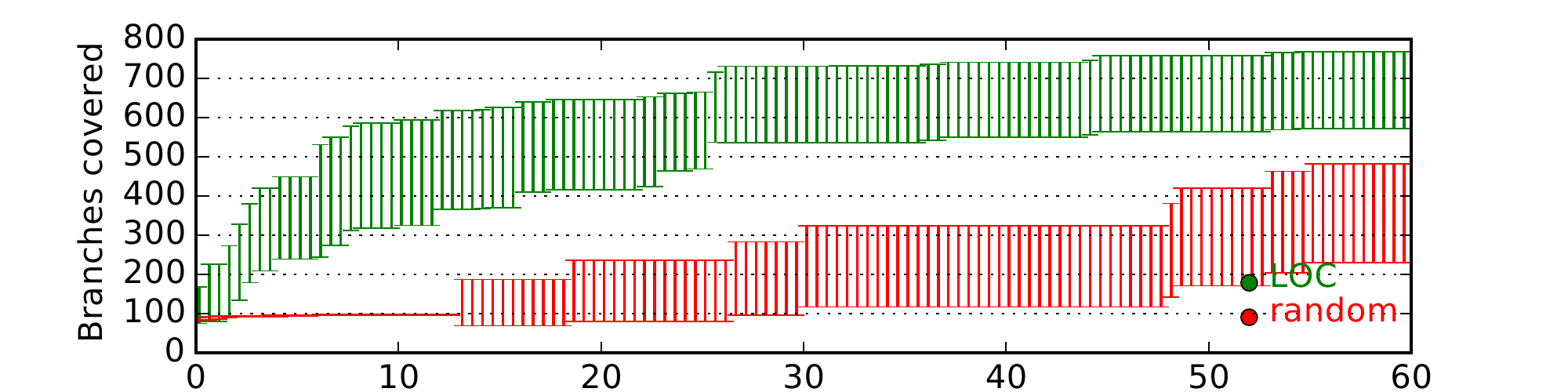}
\caption{C Parser}
\label{fig:graphC}
\end{subfigure}
\begin{subfigure}{1.0\columnwidth}
\centering
\includegraphics[width=0.8\columnwidth]{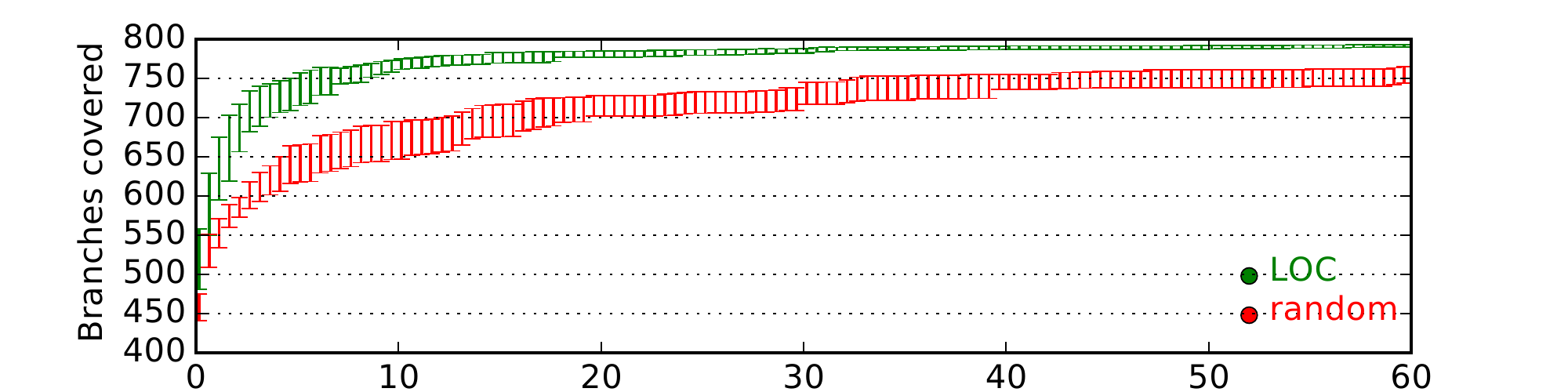}
\caption{redis-py}
\label{fig:graphredis}
\end{subfigure}
\begin{subfigure}{1.0\columnwidth}
\centering
\includegraphics[width=0.8\columnwidth]{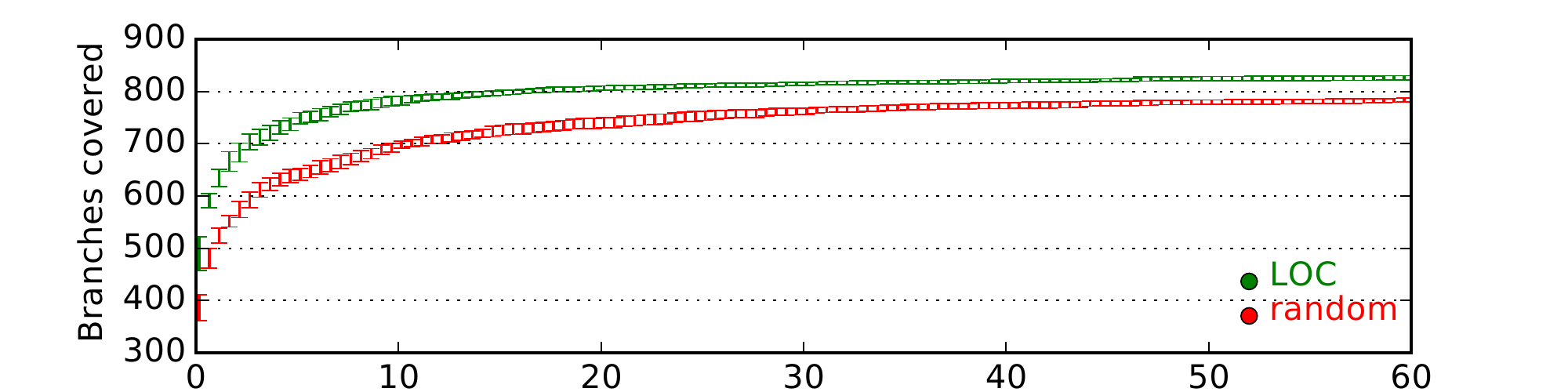}
\caption{sortedcontainers}
\label{fig:graphsorted}
\end{subfigure}
\begin{subfigure}{1.0\columnwidth}
\centering
\includegraphics[width=0.8\columnwidth]{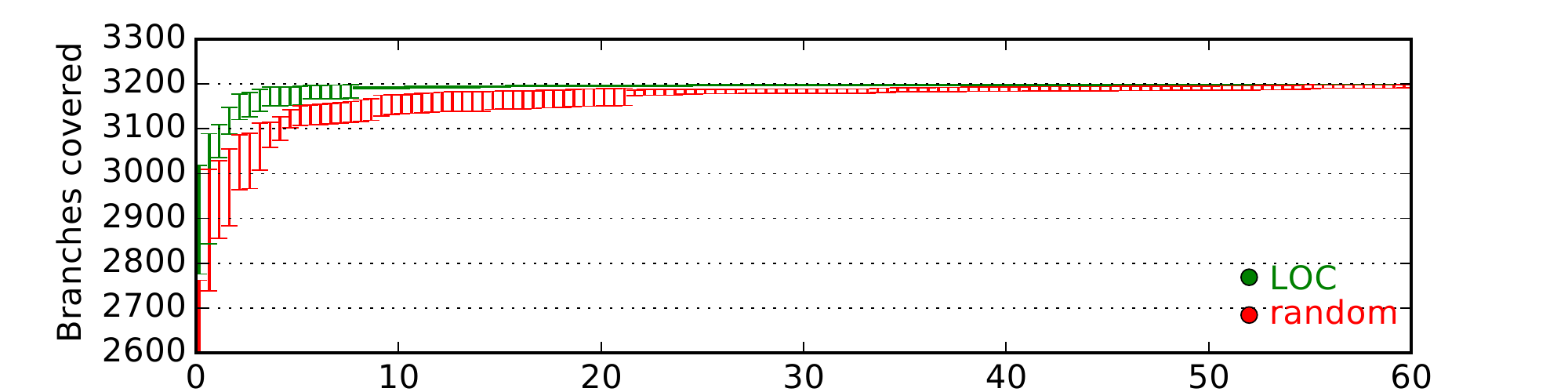}
\caption{TensorFlow}
\label{fig:graphtf}
\end{subfigure}
\begin{subfigure}{1.0\columnwidth}
\centering
\includegraphics[width=0.8\columnwidth]{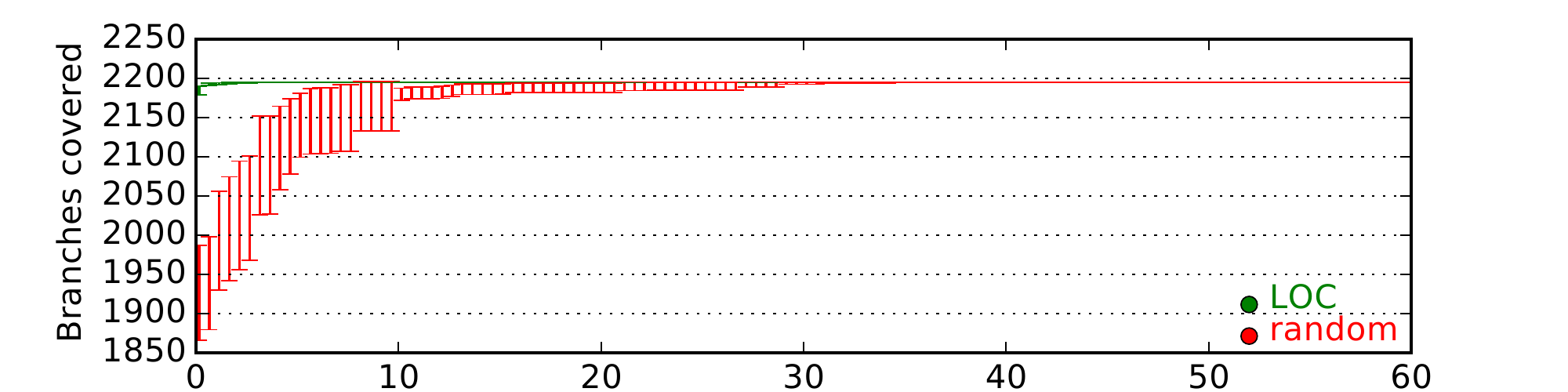}
\caption{z3}
\label{fig:graphz3}
\end{subfigure}
\caption{1 hour testing branch coverage results.}
\end{figure*}
}

\subsection{Using LOC with Larger Test Budgets}

In order to check whether the LOC heuristic remains viable for longer
test budgets, we also ran 1 hour of testing on the SUTs where
LOC was effective for small budgets, coverage is not close to 100\% at
60 seconds (and does not saturate reliably within two minutes, at least), and there are
no faults.  Our expectation was that while LOC is generally useful for
improving coverage of both code and state-space, its tendency to focus
on high-heuristic-value methods might eventually cause those portions
of the code to become saturated, and some low-LOC methods or functions
(especially ones that have few LOC but can, under unusual
circumstances, call high-LOC methods not included in the dynamic
estimate) to be under-covered. In all cases LOC continued to
improve on pure random testing for much larger test budgets.  Figures
\ref{fig:graphC}-\ref{fig:graphz3} show the results of 15 runs for
both LOC and pure random testing, with 95\% confidence
intervals\footnote{In order to make the saturation points more
  visible, we cut off the bottom of the very first confidence interval
  for {\tt TensorFlow}:  pure random can sometimes cover as few as
  2200 branches.}.  In
most cases, even after an hour, LOC was better than
pure random testing, often by a large, significant, margin.  With {\tt TensorFlow}
(Figure \ref{fig:graphtf}) and {\tt z3} (Figure \ref{fig:graphz3}), there \emph{was} saturation, in that
coverage reliably reached the maximum obtained within a few minutes
(10 for {\tt TensorFlow} and 2 for {\tt z3}); however, random testing required more than 30 minutes to
reliably hit the same set of branches.  Results for statement coverage
were (except for absolute numbers) essentially identical to those for
branch coverage. For {\tt SymPy}, not shown in the graphs, most 1 hour runs encountered an infinite 
loop fault, which may produce skewed results over completed runs; 
however, LOC covered 16,569 branches vs. random testing's 11,038 
branches, for the one successful run for each we eventually collected.

We confirmed that after 20 minutes (and usually
after less than 60 seconds), all action classes had been chosen many
times, so the differences here are not plausibly attributable to the
low-hanging-fruit nature of larger functions.  Instead, these gains
must be due to hard-to-cover branches being more common in longer
functions and the greater impact of longer functions on
system state.

\section{Summary and Discussion}

\subsection{Core Research Questions}
\label{sec:rq13summary}

For our core research questions (RQ1-3), we generally found that the
LOC heuristic was effective.  {\bf RQ1} could be clearly answered by saying
that, in general, the LOC heuristic performs better than random
testing without use of the heuristic to bias probabilities.  For five of
the six SUTs with faults, using the LOC heuristic significantly
improved fault detection over pure random testing.  The LOC heuristic
had a negative impact on fault detection for the remaining SUT, but
the change was \emph{not} (unlike the improvements) statistically
significant.  The effect sizes in improvements were large; ignoring
the one case where only using LOC allowed detection of any faults, the
mean improvement was 176.8\%.  Using the LOC heuristic significantly improved branch
coverage over pure random testing for 8 of the 15 SUTs, and
significantly decreased it for 4 of the SUTs.  The mean significant improvement
effect size (+21.7\%) was more than twice than the mean significant
decrease (-10.4\%).  Using the LOC heuristic also significantly improved
statement coverage for 10 of the 15 SUTs, and significantly decreased
it for 4 of the SUTs.  Mean significant improvement was 16.9\%,
compared to mean decrease of 10.0\%.  Our hypothesis was that LOC
would outperform random testing by either a code coverage or fault
localization meeasurefor 60\% of SUTs; LOC significantly improved on
random testing by some measure for 80\% of SUTs.

The results for comparing to more sophisticated strategies ({\bf RQ2})
were also good.  LOC was significantly better for fault detection than
the coverage-driven mutational GA for four SUTs and worse for only one
SUT, with mean effect sizes of +375.6\% and -90\%, respectively.  LOC
was significantly better than GA for branch coverage for 8 of the SUTs
and significantly worse for 6 of the SUTs, with mean effect sizes of
+17.4\% and -11.0\%, respectively.  For statement coverage, LOC was
significantly better for 9 SUTs and worse for 5 SUTs, with mean
significant effect sizes of +15.1\% and -12.9\%, respectively.  LOC
was significantly better than swarm for fault detection for three SUTs
and worse for no SUTs, with mean effect size of +134.9\%.  LOC was
significantly better than swarm for branch coverage for 7 of the SUTs
and significantly worse for 6 of the SUTs, with mean effect sizes of
+6.4\% and -34.3\%, respectively.  For statement coverage, LOC was
significantly better for 7 SUTs and worse for 6 SUTs, with mean
significant effect sizes of +6.7\% and -34.4\%, respectively.

For GA, the comparison is clearly favorable for the LOC heuristic; it
was better by some measure for more than 80\% of SUTs. For swarm
testing, LOC was better more often for all measures (improving on some
measure for 66\% of SUTs), and swarm was not
useful for finding the bugs in our faulty SUTs, but, as noted
above, when swarm is effective for coverage, it can be highly
effective, with a greater positive impact than LOC had.

As to why LOC performed better or worse than GA or swarm,
the explanation can be divided into two parts.   For the SUTs where
LOC is actually harmful, that is, worse than pure random testing, it
unsurprisingly is also worse than more effective testing methods than
random testing.  We discuss possible causes for LOC performing worse
than random testing below.  This explains most of the cases where LOC
performs worse than GA and swarm, simply:  it performs worse when it
is generally a bad idea to bias based on LOC, even compared to pure
random testing.  That is, the best way to predict if LOC will be more
or less useful than the other two methods is to see if it is helpful
compared to random testing. When LOC is worse than than
random, it is likely to lose to methods that tend to do better than
random.  This point is not quite as tautological as it might seem:
the other two cases for worse branch coverage are explained by GA or
swarm \emph{providing a large benefit LOC cannot match}, although LOC is
useful.  Swarm testing is, as noted, very powerful for compiler-like
SUTs, and a search-based mutational approach is sometimes the only way
to hit a hard-to-reach part of an SUT's code, unsurprisingly.  This is
why these methods are established; they provide substantial,
hard-to-duplicate benefits in some cases.  LOC could have performed
worse than GA and swarm in all cases, even if it was almost always a
useful method, because they are \emph{even more powerful} methods in general.
This was not what we observed; instead, when LOC was helpful, it was
often \emph{more helpful} than the other methods, but LOC also has cases
where it is not helpful, even compared to a ``bad'' method such as
pure random testing.

Finally, combining methods ({\bf RQ3}) was often useful.  The GA using the LOC
heuristic was the most effective method overall, for both branch
coverage and fault detection, and swarm with LOC was the best method
for branch coverage twice, while swarm alone was never the best method
for any SUT.  In particular, combining the GA and the LOC heuristic
was significantly better than LOC alone for 2 SUTs and worse for one
for fault detection, and better than LOC alone for 6 SUTs and worse
for 1 for both branch and statement coverage.  It was significantly better than the
GA alone for 3 SUTs and worse for 2 SUTs for fault detection, and significantly
better for 11 SUTs and worse for 3 SUTs for both branch and statement
coverage.   Combining with swarm testing was less effective; it was significantly
better than the LOC heuristic alone for 2 SUTs and worse for 3 for
fault detection, and significantly better for 3 SUTs and worse for 10 for both
branch and statement coverage.  Similarly, it was significantly better
than swarm alone for 1 SUT and worse for 3 SUTs for fault detection,
significantly better for 3 SUTs and worse for 7 SUTs for branch
coverage, and significantly better for only 2 SUTs and worse for 8
SUTs for statement coverage.  Thus, while swarm and LOC certainly \emph{can} cooperate (recall that LOC improved swarm performance for the three SUTS that were most improved by swarm, the C parser, {\tt redis-py}, and {\tt z3}), they often do seem to work poorly together.  We speculate that in some cases, LOC, by focusing testing on high-LOC functions, frustrates the increased test diversity provided by swarm testing; that is, if a configuration includes one high-LOC function, that function may consistently get the lion's share of testing, reducing the impact of swarm (tests look more alike, despite different configurations).  Swarm on the other hand frustrates LOC's goal, by often removing all high-LOC functions from the set of available actions.  However, given that LOC improved swarm performance for just those SUTs where swarm was most useful, this negative effect may only matter when swarm testing itself is not highly effective.

\subsection{Supplemental and Exploratory Results}

While these results are more exploratory than for our primary
questions, we also can draw some additional conclusions.  First, we believe that using the
current-state-of-the-art tool ({\tt coverage.py}) even in its latest
version, and with a JIT, there is a substantial overhead for computing
dynamic coverage in Python.  Not instrumenting for coverage allows a
testing tool to perform more than half again as much testing (that is,
to execute more test actions, by a factor of 1.5), in
the median case, even using a JIT.   This is not a small advantage.
Second, {\tt python-afl} shows that using an off-the-shelf fuzzer,
even a sophisticated one, does not compete with even a pure random tester for
this type of short-budget property-based testing, at least without substantial
additional effort.  The LOC heuristic also seems to be able to improve
code coverage for Java testing as well, when used to bias the Randoop
tool's generation method.  Outdated LOC estimates seem to have little (negative or positive) impact on effectiveness of
results, even across fairly large code changes, so long as the tested API itself is not altered.  Finally, for some of our SUTs, the utility of the LOC
heuristic extends to much larger testing budgets.

\subsection{Discussion}
\label{sec:disc}

Is the LOC heuristic likely to be universally effective for improving
testing in settings with expensive code coverage instrumentation?  No; it yields worse
results for some of our Python SUTs.  
For large budget testing, this would be a real problem. In
practice few, if any, test generation heuristics are close to
universally effective, and the chance that a given, usually useful,
heuristic may prove harmful,
which cannot be easily predicted or detected (since we may only
have time to run one technique) is frustrating with large test
budgets.  For instance, in our experiments, such established methods
as a coverage-driven GA and swarm testing both perform worse than
pure random testing in terms of branch coverage for 5 SUTs, and worse
in terms of fault detection for 3 and 2 SUTs, respectively.  This
obviously does not mean these are ``bad'' methods, merely that they
are \emph{heuristic}.  We note that the
corresponding ``worse than random'' numbers for the LOC heuristic are 4 for branch coverage
and 1 for fault detection, so a standard that would reject it would
also presumably dismiss other well-established test generation approaches.

An expert performing an automated code
audit for security testing has strong incentive to choose the most
powerful fuzzing technique, but may have little way of knowing up
front which method will perform best on a previously untested SUT, and
lack resources to try multiple methods with full effect.  In contrast, if a heuristic is
often effective for small test budgets, and if effectiveness tends to
be consistent for the same SUT over time, then it is easy to try several different techniques, measure coverage for
them all, and configure the testing to apply the best technique. Our experimental results show that in such a setting the
LOC heuristic would often be chosen, either by itself or in
combination with another method.

Moreover, we can \emph{sometimes} predict when the LOC heuristic will not be
effective.  When we examined the {\tt simplejson} harness, we
predicted that LOC would not work well.  There are only four action
classes that call any SUT code, and these four action classes only
call two different methods, {\tt dumps} and {\tt loads}.  The majority
of the interesting behavior during testing is the generation of Python
values to be encoded as JSON.  The
{\tt simplejson} harness includes a property that calls both {\tt
  loads} and {\tt dumps} (to check that basic encoding and decoding work
properly for all generated Python values).  Properties are checked after every action,
ensuring that the important functions to test are called.   Actions
call {\tt loads} and {\tt dumps} with various optional
parameters, but LOC does not help distinguish which of these are most
important to test, since they all rely on the same functions.  When
most testing of the SUT is accomplished by a property, not by actions, the LOC
heuristic is likely to be useless or even harmful.  

On the
other hand, {\tt bidict} is very similar to
{\tt sortedcontainers}, except that
{\tt sortedcontainers} has many more action classes (due to being larger and more complex).  We do not know why the LOC heuristic makes {\tt bidict}
testing less effective, but speculate that all of the high
probability actions calling the same function (via
wrappers) may be responsible.  The LOC
heuristic thus concentrates too much on the {\tt update} method.  The {\tt sortedcontainers} high
probability action classes are much more diverse, including list
slice and index modifications, the constructor for a {\tt
  SortedDict}, a copy method, set union, and dictionary keys
counting.  A similar problem may explain the poor performance on {\tt
  biopython}: more than 30\% of the probability distribution is split
between just two action classes, both of which call complex (and
potentially computationally expensive) algorithms with no impact on
object state, and both of which do not actually offer much in terms of coverage, since the harness has trouble producing non-trivial inputs that satisfy their preconditions (most inputs generated seem to be valid but uninteresting, the equivalent of empty lists).  Finally, {\tt arrow} has worse performance for LOC, as far as we can tell, almost entirely due to the fault LOC finds so much more frequently; unfortunately, due to the nature of the fault itself, it is hard to make TSTL not slower, even if we ignore the failing tests (restarting testing due to these faults is costly).

A better question might be, why is the LOC heuristic so effective, when it works?  One possibility is that during a
30 or 60 second test run, not all action classes are explored by pure
random testing.  The LOC heuristic ensures that the action classes
never chosen will usually be ones with a small LOC count---if you can't
cover everything, at least cover the ``big'' actions.  In some cases
this is critical; for example, in the C parser, making sure that every
test run at least attempts to parse a program is essential to
effective testing, and explains most of the difference between LOC and
pure random testing for 30-60 second budgets.  Alternative methods for
avoiding failing to cover important action classes (such as the bias
in our LOC sampling) are likely to impose a much larger overhead on
testing than the LOC heuristic.

However, this does not explain the results for every SUT, and
obviously does not explain the continued utility of the LOC heuristic
over longer runs as shown in Figures \ref{fig:graphC}-\ref{fig:graphsorted}.  For {\tt redis-py} all
action classes are easily covered after as little as five minutes of random
testing, on average.  The C parser's actions (with the exception of one action
that never calls any SUT code, and appears to only exist to reset
uninteresting completed programs that do not contain any conditionals)
are similarly usually covered in ten minutes or less.  Only {\tt
  sortedcontainers} poses a challenge for random testing, in terms of action-class coverage, due to a
very large number of action classes, and even for it, 25 minutes of
testing almost always suffices for complete action class coverage.
While the gap between pure random testing and the heuristic arguably
closes somewhere near the point when all classes have been tested, it does
not disappear, in any of these cases, and for {\tt redis-py} and {\tt
  sortedcontainers}, the gap never disappears, even after an hour of testing.

We
include the AVL and heap examples precisely because they feature
extremely small interfaces, with only a few test actions, and
saturate (or come close to saturating) coverage even during a 60 second run.  For
the heap example, pure random testing executes the least-taken action
class about 40 times on average, during 30 second runs, and for AVL
the only action class with a significant probability of not being
taken is the action of displaying an AVL tree (which is not in any way
helpful in
detecting the fault); the next least-frequently executed action
class is executed about 10 times during 30 seconds of random testing.  Why is LOC able to triple the fault
detection rates for these simple SUTs?  The best explanation we can
propose is that our assumptions discussed in the introduction to this paper are frequently true for
individual SUTs:  longer
functions modify system state more, and perform more complex
computation.  All things being equal, they contribute more to test
effectiveness, and calling such functions more frequently helps detect
faults, both by modifying system state in more complex ways, and
performing complex computations that expose erroneous state.
An alternative way of seeing the same effect is to observe that the
LOC heuristic decreases the frequency with which tests perform simple query
actions, such as AVL tree traversal, checking emptiness of
a container, or drawing a random byte in a cryptography library.  And, indeed, when we investigate the details of the faults found much more easily for {\tt sortedcontainers}, {\tt arrow}, {\tt pyfakefs}, and the toy examples, the bugs are located in unusually large functions that also have ways to execute very little code (conditions under which they do very little), just as we would expect.  Such code is probably quite common, in that ``do nothing for simple inputs (e.g., empty lists), do a lot for complex inputs (e.g., nested lists)'' is a common pattern in many algorithms.  LOC seems to help hit the first case.

More generally, one overall take-away from this paper should be that
test generation heuristics are not equally effective for all SUTs and
test harnesses; rather, performance varies widely by the structure of
the input and state space. This is not a novel observation, of course
\cite{RandOrGenetic,shamshiri2018random}. In a sense, this goes with the territory
of heuristics, vs. mathematically proven optimizations of testing
(alas, the latter are rarely possible) \cite{STTTHeur,ISSRE}.
Examining methods in isolation is also insufficient to obtain
maximally effective testing: while combining methods sometimes reduced
effectiveness, a combination of some method
with LOC was often the most effective approach.  Assuming effectiveness ranking is 
stable for each SUT over time (at least over days or weeks, which
seems highly likely), we believe that projects seeking effective automated test
generation should run simple experiments to determine good test
generation configurations once, and then re-use those settings,
perhaps parameterized by test budget (e.g., different settings may be
needed for testing during development, ten minute ``coffee break''
testing, ``lunch-hour'' testing, and overnight runs).  Because it is
performed most frequently, and is most useful for debugging (faults
are easiest to fix just after introduction), it is fortunate that
tuning very small test budgets is quite easy.  In our Python experiments, we
note that the most effective method seldom changed between results for
as few as 10 tests and for the full 100-test experiments. Effect
sizes that matter can be detected in less than an hour.

We therefore  propose a simple, one-time, method for choosing a standard
property-based testing approach for an SUT under development/test.
Run pure random testing, LOC alone, the GA, and swarm, 10 times each,
for 1 minute.  This requires 40 minutes, a reasonable cost for a
one-time decision that can improve small-budget testing for a long development period.  If LOC performs worse than pure random
testing, use whichever method is best (and perhaps combine GA and
swarm if both outperform pure random testing, though we have no
experimental data on this combination).  If LOC is better than pure random
testing, combine it with GA, or swarm, or perhaps both, if they also
improved on random testing.  Finally, and critically, if the approach chosen does not include use of the GA, \emph{run testing without coverage instrumentation} to take advantage of the higher throughput LOC and swarm allow.

Moreover, there is often no reason to find ``the best method.''  In
modern security fuzzing, there is a growing awareness that predicting
the best method is difficult, and ensemble methods are highly
effective \cite{chen2019enfuzz}; tools from firms performing security
audits are beginning to reflect this wisdom (\url{https://blog.trailofbits.com/2019/09/03/deepstate-now-supports-ensemble-fuzzing/}).  Running LOC, GA, GA+LOC, and swarm+LOC
for 15 seconds each might well be the best use of 60 seconds of test
budget for an SUT.  Certainly for larger test budgets, this is likely
to be the case, due to the diversity effect seen in security fuzzing,
and in our fault detection results.  This diversity effect, after all, is the inspiration for highly successful swarm verification \cite{swarmIEEE} and testing \cite{ISSTA12} methods, themselves.

Finally, despite the wide variance of heuristic performance, our
results for Python and Java were surprisingly similar, given that we
used the LOC heuristic in combination with quite different underlying
random testing methods, for different languages, different styles of
testing (property-based with a harness and more complex oracle
vs. automatic unit test generation for classes), and even different notions of LOC
(dynamic sampling vs. purely static, and with comments/blank lines
vs. code only).  We believe this provides a strong argument that LOC
does provide a good, if rough, measure of the ability of ``test
actions'' (broadly conceived) to explore SUT/CUT behavior.  In other
words, our results support our belief that, while there may be even better
answers to the ``{\tt f} or {\tt g}?'' question (though we suspect even
these would take function size into account as one among a number of factors), testing the function
with more LOC is a good, and practically useful, approach.

\section{Threats to Validity}

{\bf Internal Validity:} For our SUTs, we believe the causal relationships for primary RQs are unlikely to be spurious; we used 100 runs, and compared results using appropriate statistical tests that do not assume normality \cite{arcuri2014hitchhiker}.  We do not claim that the LOC heuristic is uniformly effective, only that, for the SUTs considered it often improves fault detection, branch, and statement coverage by a significant amount.  The Java experiments are highly preliminary, essentiallly exploratory, since they do not include fault detection results.

{\bf External Validity:} The primary threats are to external validity. The Python results are based on a limited set of programs with harnesses already existing in the TSTL repository when we began this investigation.  We did not modify the harnesses, and only used harnesses covering a realistic subset of library behavior (test harnesses that might be used in practice, and in three cases that have been used to report real faults).  Two of the subjects are essentially small toy examples.  The remaining SUTs are popular real-world Python libraries with a large number of GitHub stars or pip downloads, but only three of the projects ({\tt SymPy}, {\tt biopython}, and {\tt TensorFlow}) are extremely large.  However, in practice, property-based testing is usually focused on a smaller subset of a system, and the 500-5000 LOC size of most of the subjects (around 2KLOC for most) is likely a reasonable approximation of the size of SUT where small budget testing is likely to be highly effective (and reflects the size of many important Python libraries with subtle behaviors: Python is a highly compact language with similarities to Haskell in terms of density \cite{GopinathMutants}).  The Java projects were chosen \cite{ahmed_testedness,just2014defects4j,AutoTestFaults} to be representative open source Java projects, but may be subject to bias due to GitHub's unknown selection methods.  While the exact cross-method comparisons are unlikely to be preserved, we think it is highly unlikely that LOC is not at least frequently a useful bias to impose on any test method that chooses from a random set of actions representing method or function calls.

{\bf Construct Validity:}  The test generation methods use a common code base in TSTL, and the {\tt coverage.py} tool for collecting Python coverage data.  The threats to our results would arise from either (1) an incorrect implementation of one of our test generation methods, meaning that we evaluate a different testing approach than we claim to, or (2) an error in {\tt coverage.py} that somehow favors one method over another.   We inspected the implementations of the testing methods in TSTL carefully, and they have been tested on multiple SUTs, including simple ones where we were able to follow the data structures used by approaches and compare to expected results and tests.  The {\tt coverage.py} library is very widely used, and it is further unlikely that any subtle bugs in it favor one testing approach more than another. However, it is \emph{possible} that errors in these implementations did bias our results. We welcome independent re-implementations to check for the possibility of remaining consequentual errors.  The Java experiment relies only on the ability of Understand to count LOC and Randoop to bias probabilities using a method list: any errors in Randoop itself would only change the context of the comparison, not the impact of using LOC.


\section{Related Work}
\label{sec:related}

There is a long line of work investigating relationships between static code measures such as LOC and defects in code, though usually at the module or file level, and never in the context of test generation.  Radjenovi\'{c} et al. \cite{Radjenovic2013IST} provide a detailed literature review of metrics used for fault prediction.   Zhang \cite{Zhang2009ICSM} showed that 20\% of the largest modules studied contained 51-63\% of the defects. Ostrand et al. \cite{Ostrand2005TSE} showed that the largest 20\% of files contained between 59\% to 83\% of of the faults. Koru et al. \cite{Koru2009TSE} and Syer et al. \cite{Syer2015TSE} reported that defect proneness increases with module size, but at a slower rate. Other studies \cite{Fenton2000TSE,Olague2007TSE,Andersson2007TSE} also showed that LOC correlates with the number of faults. In general, all of this work aims to advise developers to keep code small, rather than to aid testing; it has never proven a highly useful method even for default prediction, compared to less generic techniques \cite{Menzies2002SEW, Fenton2000TSE, Graves2000TSE, Zimmermann2008ICSE}.
There are also a large number of metrics designed specifically for object oriented programs.  Some (referred to as CBO, WMC and RFC, in the relevant papers) have been proposed as useful predictors of pre-release faults \cite{Basili1996TSE, Briand2001ESE, Gyimothy2005TSE, Olague2007TSE, Pai2007TSE}, while other measures, such as LCOM, DIT, and NOC, did not perform well \cite{Zhou2006TSE, Gyimothy2005TSE, Olague2007TSE, Olague2007TSE, Pai2007TSE}.  Olague et al. \cite{Olague2007TSE} claimed that the QMOOD metrics~\cite{Bansiya2002TSE} were suitable for fault prediction, while the MOOD suite of metrics \cite{Abreu1994SQ, Brito1996ISMS} was not.  
Cohesion metrics (LCC and TCC) \cite{Bieman1995SEN} had modest effectiveness for predicting future faults \cite{Briand2001ESE, Marcus2008TSE}, and coupling metrics, proposed by Briand et al.~\cite{Briand1999ICSE} were good predictors of future faults \cite{Briand1999ICSE, Briand2001ESE, Briand2002, ElEmam2001TSE}.   Our work, rather than demonstrating a coarse, weak correlation between code entity size and defects detected, uses code size to drive test generation, improving code coverage and fault detection.  As discussed in the conclusion, it would be interesting to use some of the above measures, or other (semi-)static measures, than LOC to bias testing, or in combination with LOC, including ones not highly useful in isolation.  Further possible interesting measures to investigate include code changes/revision history \cite{Ekelund,Saha,Buchgeher,White2008,White2004,Engstrom2010,Wikstrand,Gligoric}, source file \cite{Saha,White2008,White2004,Hirzel,Carlson}, number of contributors \cite{Herzig}, class dependencies \cite{Skoglund}, component changes \cite{Zheng06,Zheng07}, estimated execution time \cite{Tahvili,Srivastava}, or filed-issue-related metrics \cite{Engstrom2010,Wikstrand,Herzig,Marijan13,Marijan15}.

LOC has sometimes been used as a independent variable or as an objective function in search-based-software engineering. Fatinegun et el.~\cite{Fatiregun2004SCAM} looked at heuristics to reduce the size of program, and Dolado et al.~\cite{Dolado2000TSE}  proposed a technique to estimate the final LOC size of a program.  Again, the purposes of these uses (or estimations) of LOC are completely different than our proposed heuristic to guide random test generation.

Previous approaches to tuning probabilities in random testing, such as Nighthawk \cite{AndrewsL07} and ABP \cite{ISSRE} learned probabilities based on coverage feedback, rather than assigning fixed probabilities based on a simple (and essentially static) metric of the tested code, the core novel concept presented in this paper.  Randoop \cite{Pacheco} and other feedback-based approaches \cite{FeedControl} arguably obtain part of their effectiveness from an indirect avoidance of short functions that do not modify state, but pay the cost of determining if a call produces no state change.  To our knowledge the \emph{size} of functions/methods has never been used as even a factor in the decision of which API calls to  make in  automated test generation.  Arguably, an approach such as that taken  by VUzzer \cite{rawat2017vuzzer}, a fuzzer where code regions with ``deep'' and ``interesting'' paths are prioritized, based on static analysis  of control features, bears some abstract, high-level, resemblance  to  our method, but the actual heuristics  used and settings are utterly different.   The methods with which we compare in this paper, a genetic algorithm-based approach, and the swarm approach, are based on alternative proposed methods for guiding this type of random testing, in particular evolutionary approaches such as EvoSuite \cite{FA11} and the swarm testing concept of configuration diversity \cite{ISSTA12}, which was originally inspired by the use of  diverse searches in model checking \cite{swarm,swarmVer,swarmIEEE}.  The swarm notion of diversity also informed our decision  to evaluate \emph{combinations}  of orthogonal heuristics, under  the  assumption that no single method for guiding testing is likely to be best in all, or even most,  cases.

\section{Conclusions and Future Work}

This paper argues that simply counting the
relative LOC of software components can provide valuable
information for use in automated test generation.  We show that biasing random testing
probabilities by the LOC counts of tested functions and methods can
improve the effectiveness of automated test generation for Python.  The LOC heuristic often produces large, statistically
significant, improvements in both code coverage and fault detection.
As future work, we propose to further investigate the LOC
heuristic, including for larger test budgets, given the promise shown
in a few longer runs.

More generally, the LOC heuristic opens up a new approach to biased
random test generation, based on the ``{\tt f} or {\tt g}?'' thought experiment.  For instance, one promising next step is to
modify the heuristic to also bias testing towards executing code that
has been the subject of a static analysis tool warning, is
less tested in existing tests, or is otherwise anomalous \cite{Ray:2016:NBC:2884781.2884848}; alternatively, we can use cyclomatic
complexity \cite{McCabe,JurgenCC} or another more ``sophisticated'' measure (e.g., number of
mutants) in place of simple LOC, or use various measures discussed in
Section \ref{sec:related} to refine the LOC estimate of desirability
of a test action.  For instance, for what is perhaps property-based unit testing's
most important goal---detecting errors newly introduced into code during
development---an integration with directed swarm testing
\cite{DirSwarm} to target recently changed code is both feasible and very promising.

\bibliographystyle{ACM-Reference-Format}
\bibliography{ase}


\begin{thebibliography}{119}


\ifx \showCODEN    \undefined \def \showCODEN     #1{\unskip}     \fi
\ifx \showDOI      \undefined \def \showDOI       #1{#1}\fi
\ifx \showISBNx    \undefined \def \showISBNx     #1{\unskip}     \fi
\ifx \showISBNxiii \undefined \def \showISBNxiii  #1{\unskip}     \fi
\ifx \showISSN     \undefined \def \showISSN      #1{\unskip}     \fi
\ifx \showLCCN     \undefined \def \showLCCN      #1{\unskip}     \fi
\ifx \shownote     \undefined \def \shownote      #1{#1}          \fi
\ifx \showarticletitle \undefined \def \showarticletitle #1{#1}   \fi
\ifx \showURL      \undefined \def \showURL       {\relax}        \fi
\providecommand\bibfield[2]{#2}
\providecommand\bibinfo[2]{#2}
\providecommand\natexlab[1]{#1}
\providecommand\showeprint[2][]{arXiv:#2}

\bibitem[\protect\citeauthoryear{Aburas and Groce}{Aburas and Groce}{2016}]%
        {GuidedGA}
\bibfield{author}{\bibinfo{person}{Ali Aburas} {and} \bibinfo{person}{Alex
  Groce}.} \bibinfo{year}{2016}\natexlab{}.
\newblock \showarticletitle{A Method Dependence Relations Guided Genetic
  Algorithm}. In \bibinfo{booktitle}{\emph{Search Based Software Engineering -
  8th International Symposium, {SSBSE} 2016, Raleigh, NC, USA, October 8-10,
  2016, Proceedings}}. \bibinfo{pages}{267--273}.
\newblock


\bibitem[\protect\citeauthoryear{Agrawal}{Agrawal}{1994}]%
        {Agrawal:1994:DSB:174675.175935}
\bibfield{author}{\bibinfo{person}{Hiralal Agrawal}.}
  \bibinfo{year}{1994}\natexlab{}.
\newblock \showarticletitle{Dominators, Super Blocks, and Program Coverage}. In
  \bibinfo{booktitle}{\emph{Proceedings of the 21st ACM SIGPLAN-SIGACT
  Symposium on Principles of Programming Languages}} (Portland, Oregon, USA)
  \emph{(\bibinfo{series}{POPL '94})}. \bibinfo{publisher}{ACM},
  \bibinfo{address}{New York, NY, USA}, \bibinfo{pages}{25--34}.
\newblock
\showISBNx{0-89791-636-0}
\urldef\tempurl%
\url{https://doi.org/10.1145/174675.175935}
\showDOI{\tempurl}


\bibitem[\protect\citeauthoryear{Ahmed, Gopinath, Brindescu, Groce, and
  Jensen}{Ahmed et~al\mbox{.}}{2016}]%
        {ahmed_testedness}
\bibfield{author}{\bibinfo{person}{Iftekhar Ahmed}, \bibinfo{person}{Rahul
  Gopinath}, \bibinfo{person}{Caius Brindescu}, \bibinfo{person}{Alex Groce},
  {and} \bibinfo{person}{Carlos Jensen}.} \bibinfo{year}{2016}\natexlab{}.
\newblock \showarticletitle{Can Testedness Be Effectively Measured?}. In
  \bibinfo{booktitle}{\emph{Proceedings of the 2016 24th ACM SIGSOFT
  International Symposium on Foundations of Software Engineering}} (Seattle,
  WA, USA) \emph{(\bibinfo{series}{FSE 2016})}. \bibinfo{publisher}{ACM},
  \bibinfo{address}{New York, NY, USA}, \bibinfo{pages}{547--558}.
\newblock
\showISBNx{978-1-4503-4218-6}
\urldef\tempurl%
\url{https://doi.org/10.1145/2950290.2950324}
\showDOI{\tempurl}


\bibitem[\protect\citeauthoryear{Alipour, Groce, Gopinath, and Christi}{Alipour
  et~al\mbox{.}}{2016}]%
        {DirSwarm}
\bibfield{author}{\bibinfo{person}{Mohammad~Amin Alipour},
  \bibinfo{person}{Alex Groce}, \bibinfo{person}{Rahul Gopinath}, {and}
  \bibinfo{person}{Arpit Christi}.} \bibinfo{year}{2016}\natexlab{}.
\newblock \showarticletitle{Generating Focused Random Tests Using Directed
  Swarm Testing}. In \bibinfo{booktitle}{\emph{Proceedings of the 25th
  International Symposium on Software Testing and Analysis}}
  (Saarbr\&\#252;cken, Germany) \emph{(\bibinfo{series}{ISSTA 2016})}.
  \bibinfo{publisher}{ACM}, \bibinfo{address}{New York, NY, USA},
  \bibinfo{pages}{70--81}.
\newblock
\showISBNx{978-1-4503-4390-9}
\urldef\tempurl%
\url{https://doi.org/10.1145/2931037.2931056}
\showDOI{\tempurl}


\bibitem[\protect\citeauthoryear{Andersson and Runeson}{Andersson and
  Runeson}{2007}]%
        {Andersson2007TSE}
\bibfield{author}{\bibinfo{person}{C. Andersson} {and} \bibinfo{person}{P.
  Runeson}.} \bibinfo{year}{2007}\natexlab{}.
\newblock \showarticletitle{A Replicated Quantitative Analysis of Fault
  Distributions in Complex Software Systems}.
\newblock \bibinfo{journal}{\emph{IEEE Transactions on Software Engineering}}
  \bibinfo{volume}{33}, \bibinfo{number}{5} (\bibinfo{date}{May}
  \bibinfo{year}{2007}), \bibinfo{pages}{273--286}.
\newblock
\showISSN{0098-5589}
\urldef\tempurl%
\url{https://doi.org/10.1109/TSE.2007.1005}
\showDOI{\tempurl}


\bibitem[\protect\citeauthoryear{Andrews, Li, and Menzies}{Andrews
  et~al\mbox{.}}{2007}]%
        {AndrewsL07}
\bibfield{author}{\bibinfo{person}{James Andrews}, \bibinfo{person}{Felix Li},
  {and} \bibinfo{person}{Tim Menzies}.} \bibinfo{year}{2007}\natexlab{}.
\newblock \showarticletitle{Nighthawk: A Two-Level Genetic-Random Unit Test
  Data Generator}. In \bibinfo{booktitle}{\emph{Automated Software
  Engineering}}. \bibinfo{pages}{144--153}.
\newblock


\bibitem[\protect\citeauthoryear{Andrews, Zhang, and Groce}{Andrews
  et~al\mbox{.}}{2010}]%
        {AndrewsTR}
\bibfield{author}{\bibinfo{person}{Jamie Andrews}, \bibinfo{person}{Yihao~Ross
  Zhang}, {and} \bibinfo{person}{Alex Groce}.} \bibinfo{year}{2010}\natexlab{}.
\newblock \bibinfo{booktitle}{\emph{Comparing Automated Unit Testing
  Strategies}}.
\newblock \bibinfo{type}{{T}echnical {R}eport} 736.
  \bibinfo{institution}{Department of Computer Science, University of Western
  Ontario}.
\newblock


\bibitem[\protect\citeauthoryear{Andrews, Briand, and Labiche}{Andrews
  et~al\mbox{.}}{2005}]%
        {mutant}
\bibfield{author}{\bibinfo{person}{James~H. Andrews}, \bibinfo{person}{L.~C.
  Briand}, {and} \bibinfo{person}{Y. Labiche}.}
  \bibinfo{year}{2005}\natexlab{}.
\newblock \showarticletitle{Is Mutation an Appropriate Tool for Testing
  Experiments?}. In \bibinfo{booktitle}{\emph{International Conference on
  Software Engineering}}. \bibinfo{pages}{402--411}.
\newblock


\bibitem[\protect\citeauthoryear{Arcuri and Briand}{Arcuri and Briand}{2014}]%
        {arcuri2014hitchhiker}
\bibfield{author}{\bibinfo{person}{Andrea Arcuri} {and} \bibinfo{person}{Lionel
  Briand}.} \bibinfo{year}{2014}\natexlab{}.
\newblock \showarticletitle{A hitchhiker's guide to statistical tests for
  assessing randomized algorithms in software engineering}.
\newblock \bibinfo{journal}{\emph{Software Testing, Verification and
  Reliability}} \bibinfo{volume}{24}, \bibinfo{number}{3}
  (\bibinfo{year}{2014}), \bibinfo{pages}{219--250}.
\newblock


\bibitem[\protect\citeauthoryear{Arcuri, Iqbal, and Briand}{Arcuri
  et~al\mbox{.}}{2010}]%
        {RandFormal}
\bibfield{author}{\bibinfo{person}{Andrea Arcuri}, \bibinfo{person}{Muhammad
  Zohaib~Z. Iqbal}, {and} \bibinfo{person}{Lionel~C. Briand}.}
  \bibinfo{year}{2010}\natexlab{}.
\newblock \showarticletitle{Formal Analysis of the effectiveness and
  predictability of random testing}. In \bibinfo{booktitle}{\emph{International
  Symposium on Software Testing and Analysis}}. \bibinfo{pages}{219--230}.
\newblock


\bibitem[\protect\citeauthoryear{Bansiya and Davis}{Bansiya and Davis}{2002}]%
        {Bansiya2002TSE}
\bibfield{author}{\bibinfo{person}{J. Bansiya} {and} \bibinfo{person}{C.~G.
  Davis}.} \bibinfo{year}{2002}\natexlab{}.
\newblock \showarticletitle{A hierarchical model for object-oriented design
  quality assessment}.
\newblock \bibinfo{journal}{\emph{IEEE Transactions on Software Engineering}}
  \bibinfo{volume}{28}, \bibinfo{number}{1} (\bibinfo{date}{Jan}
  \bibinfo{year}{2002}), \bibinfo{pages}{4--17}.
\newblock
\showISSN{0098-5589}
\urldef\tempurl%
\url{https://doi.org/10.1109/32.979986}
\showDOI{\tempurl}


\bibitem[\protect\citeauthoryear{Basili, Briand, and Melo}{Basili
  et~al\mbox{.}}{1996}]%
        {Basili1996TSE}
\bibfield{author}{\bibinfo{person}{V.~R. Basili}, \bibinfo{person}{L.~C.
  Briand}, {and} \bibinfo{person}{W.~L. Melo}.}
  \bibinfo{year}{1996}\natexlab{}.
\newblock \showarticletitle{A validation of object-oriented design metrics as
  quality indicators}.
\newblock \bibinfo{journal}{\emph{IEEE Transactions on Software Engineering}}
  \bibinfo{volume}{22}, \bibinfo{number}{10} (\bibinfo{date}{Oct}
  \bibinfo{year}{1996}), \bibinfo{pages}{751--761}.
\newblock
\showISSN{0098-5589}
\urldef\tempurl%
\url{https://doi.org/10.1109/32.544352}
\showDOI{\tempurl}


\bibitem[\protect\citeauthoryear{Batchelder}{Batchelder}{2015}]%
        {Coveragepy}
\bibfield{author}{\bibinfo{person}{Ned Batchelder}.}
  \bibinfo{year}{2015}\natexlab{}.
\newblock \bibinfo{title}{Coverage.py}.
\newblock
  \bibinfo{howpublished}{\url{https://coverage.readthedocs.org/en/coverage-4.0.1/}}.
\newblock


\bibitem[\protect\citeauthoryear{Bieman and Kang}{Bieman and Kang}{1995}]%
        {Bieman1995SEN}
\bibfield{author}{\bibinfo{person}{James~M. Bieman} {and}
  \bibinfo{person}{Byung-Kyoo Kang}.} \bibinfo{year}{1995}\natexlab{}.
\newblock \showarticletitle{Cohesion and Reuse in an Object-oriented System}.
\newblock \bibinfo{journal}{\emph{SIGSOFT Softw. Eng. Notes}}
  \bibinfo{volume}{20}, \bibinfo{number}{SI} (\bibinfo{date}{Aug.}
  \bibinfo{year}{1995}), \bibinfo{pages}{259--262}.
\newblock
\showISSN{0163-5948}
\urldef\tempurl%
\url{https://doi.org/10.1145/223427.211856}
\showDOI{\tempurl}


\bibitem[\protect\citeauthoryear{B\"{o}hme and Paul}{B\"{o}hme and
  Paul}{2014}]%
        {AutoEfficiency}
\bibfield{author}{\bibinfo{person}{Marcel B\"{o}hme} {and}
  \bibinfo{person}{Soumya Paul}.} \bibinfo{year}{2014}\natexlab{}.
\newblock \showarticletitle{On the Efficiency of Automated Testing}. In
  \bibinfo{booktitle}{\emph{Proceedings of the 22nd ACM SIGSOFT International
  Symposium on Foundations of Software Engineering}} (Hong Kong, China)
  \emph{(\bibinfo{series}{FSE 2014})}. \bibinfo{publisher}{ACM},
  \bibinfo{address}{New York, NY, USA}, \bibinfo{pages}{632--642}.
\newblock
\showISBNx{978-1-4503-3056-5}
\urldef\tempurl%
\url{https://doi.org/10.1145/2635868.2635923}
\showDOI{\tempurl}


\bibitem[\protect\citeauthoryear{Briand and W\"{u}st}{Briand and
  W\"{u}st}{2002}]%
        {Briand2002}
\bibfield{author}{\bibinfo{person}{Lionel~C. Briand} {and}
  \bibinfo{person}{J\"{u}rgen W\"{u}st}.} \bibinfo{year}{2002}\natexlab{}.
\newblock \showarticletitle{Empirical Studies of Quality Models in
  Object-Oriented Systems}.
\newblock \bibinfo{series}{Advances in Computers}, Vol.~\bibinfo{volume}{56}.
  \bibinfo{publisher}{Elsevier}, \bibinfo{pages}{97 -- 166}.
\newblock
\showISSN{0065-2458}
\urldef\tempurl%
\url{https://doi.org/10.1016/S0065-2458(02)80005-5}
\showDOI{\tempurl}


\bibitem[\protect\citeauthoryear{Briand, W\"{u}st, Ikonomovski, and
  Lounis}{Briand et~al\mbox{.}}{1999}]%
        {Briand1999ICSE}
\bibfield{author}{\bibinfo{person}{Lionel~C. Briand},
  \bibinfo{person}{J\"{u}rgen W\"{u}st}, \bibinfo{person}{Stefan~V.
  Ikonomovski}, {and} \bibinfo{person}{Hakim Lounis}.}
  \bibinfo{year}{1999}\natexlab{}.
\newblock \showarticletitle{Investigating Quality Factors in Object-oriented
  Designs: An Industrial Case Study}. In \bibinfo{booktitle}{\emph{Proceedings
  of the 21st International Conference on Software Engineering}} (Los Angeles,
  California, USA) \emph{(\bibinfo{series}{ICSE '99})}.
  \bibinfo{publisher}{ACM}, \bibinfo{address}{New York, NY, USA},
  \bibinfo{pages}{345--354}.
\newblock
\showISBNx{1-58113-074-0}
\urldef\tempurl%
\url{https://doi.org/10.1145/302405.302654}
\showDOI{\tempurl}


\bibitem[\protect\citeauthoryear{Briand, W{\"u}st, and Lounis}{Briand
  et~al\mbox{.}}{2001}]%
        {Briand2001ESE}
\bibfield{author}{\bibinfo{person}{Lionel~C. Briand},
  \bibinfo{person}{J{\"u}rgen W{\"u}st}, {and} \bibinfo{person}{Hakim Lounis}.}
  \bibinfo{year}{2001}\natexlab{}.
\newblock \showarticletitle{Replicated Case Studies for Investigating Quality
  Factors in Object-Oriented Designs}.
\newblock \bibinfo{journal}{\emph{Empirical Software Engineering}}
  \bibinfo{volume}{6}, \bibinfo{number}{1} (\bibinfo{year}{2001}),
  \bibinfo{pages}{11--58}.
\newblock
\showISSN{1573-7616}
\urldef\tempurl%
\url{https://doi.org/10.1023/A:1009815306478}
\showDOI{\tempurl}


\bibitem[\protect\citeauthoryear{{Buchgeher}, {Ernstbrunner}, {Ramler}, and
  {Lusser}}{{Buchgeher} et~al\mbox{.}}{2013}]%
        {Buchgeher}
\bibfield{author}{\bibinfo{person}{G. {Buchgeher}}, \bibinfo{person}{C.
  {Ernstbrunner}}, \bibinfo{person}{R. {Ramler}}, {and} \bibinfo{person}{M.
  {Lusser}}.} \bibinfo{year}{2013}\natexlab{}.
\newblock \showarticletitle{Towards Tool-Support for Test Case Selection in
  Manual Regression Testing}. In \bibinfo{booktitle}{\emph{2013 IEEE Sixth
  International Conference on Software Testing, Verification and Validation
  Workshops}}. \bibinfo{pages}{74--79}.
\newblock
\showISSN{null}
\urldef\tempurl%
\url{https://doi.org/10.1109/ICSTW.2013.16}
\showDOI{\tempurl}


\bibitem[\protect\citeauthoryear{Cadar, Dunbar, and Engler}{Cadar
  et~al\mbox{.}}{2008}]%
        {KLEE}
\bibfield{author}{\bibinfo{person}{Cristian Cadar}, \bibinfo{person}{Daniel
  Dunbar}, {and} \bibinfo{person}{Dawson Engler}.}
  \bibinfo{year}{2008}\natexlab{}.
\newblock \showarticletitle{{KLEE}: Unassisted and Automatic Generation of
  High-Coverage Tests for Complex Systems Programs}. In
  \bibinfo{booktitle}{\emph{Operating System Design and Implementation}}.
  \bibinfo{pages}{209--224}.
\newblock


\bibitem[\protect\citeauthoryear{{Carlson}, {Do}, and {Denton}}{{Carlson}
  et~al\mbox{.}}{2011}]%
        {Carlson}
\bibfield{author}{\bibinfo{person}{R. {Carlson}}, \bibinfo{person}{H. {Do}},
  {and} \bibinfo{person}{A. {Denton}}.} \bibinfo{year}{2011}\natexlab{}.
\newblock \showarticletitle{A clustering approach to improving test case
  prioritization: An industrial case study}. In \bibinfo{booktitle}{\emph{2011
  27th IEEE International Conference on Software Maintenance (ICSM)}}.
  \bibinfo{pages}{382--391}.
\newblock
\showISSN{1063-6773}
\urldef\tempurl%
\url{https://doi.org/10.1109/ICSM.2011.6080805}
\showDOI{\tempurl}


\bibitem[\protect\citeauthoryear{Chen, Jiang, Ma, Liang, Wang, Zhou, Jiao, and
  Su}{Chen et~al\mbox{.}}{2019}]%
        {chen2019enfuzz}
\bibfield{author}{\bibinfo{person}{Yuanliang Chen}, \bibinfo{person}{Yu Jiang},
  \bibinfo{person}{Fuchen Ma}, \bibinfo{person}{Jie Liang},
  \bibinfo{person}{Mingzhe Wang}, \bibinfo{person}{Chijin Zhou},
  \bibinfo{person}{Xun Jiao}, {and} \bibinfo{person}{Zhuo Su}.}
  \bibinfo{year}{2019}\natexlab{}.
\newblock \showarticletitle{EnFuzz: Ensemble fuzzing with seed synchronization
  among diverse fuzzers}. In \bibinfo{booktitle}{\emph{28th $\{$USENIX$\}$
  Security Symposium ($\{$USENIX$\}$ Security 19)}}.
  \bibinfo{pages}{1967--1983}.
\newblock


\bibitem[\protect\citeauthoryear{Chilakamarri and Elbaum}{Chilakamarri and
  Elbaum}{2004}]%
        {chilakamarri2004reducing}
\bibfield{author}{\bibinfo{person}{Kalyan-Ram Chilakamarri} {and}
  \bibinfo{person}{Sebastian Elbaum}.} \bibinfo{year}{2004}\natexlab{}.
\newblock \showarticletitle{Reducing coverage collection overhead with
  disposable instrumentation}. In \bibinfo{booktitle}{\emph{Software
  Reliability Engineering, 2004. ISSRE 2004. 15th International Symposium on}}.
  IEEE, \bibinfo{pages}{233--244}.
\newblock


\bibitem[\protect\citeauthoryear{{CI}}{{CI}}{[n.d.]}]%
        {TravisDoc}
\bibfield{author}{\bibinfo{person}{{Travis} {CI}}.}
  \bibinfo{year}{[n.d.]}\natexlab{}.
\newblock \bibinfo{title}{Customizing the Build: Build Timeouts}.
\newblock
  \bibinfo{howpublished}{\url{https://docs.travis-ci.com/user/customizing-the-build/\#Build-Timeouts}}.
\newblock


\bibitem[\protect\citeauthoryear{Claessen and Hughes}{Claessen and
  Hughes}{2000}]%
        {ClaessenH00}
\bibfield{author}{\bibinfo{person}{Koen Claessen} {and} \bibinfo{person}{John
  Hughes}.} \bibinfo{year}{2000}\natexlab{}.
\newblock \showarticletitle{{QuickCheck}: a lightweight tool for random testing
  of Haskell programs}. In \bibinfo{booktitle}{\emph{ICFP}}.
  \bibinfo{pages}{268--279}.
\newblock


\bibitem[\protect\citeauthoryear{Dolado}{Dolado}{2000}]%
        {Dolado2000TSE}
\bibfield{author}{\bibinfo{person}{J.~J. Dolado}.}
  \bibinfo{year}{2000}\natexlab{}.
\newblock \showarticletitle{A validation of the component-based method for
  software size estimation}.
\newblock \bibinfo{journal}{\emph{IEEE Transactions on Software Engineering}}
  \bibinfo{volume}{26}, \bibinfo{number}{10} (\bibinfo{date}{Oct}
  \bibinfo{year}{2000}), \bibinfo{pages}{1006--1021}.
\newblock
\showISSN{0098-5589}
\urldef\tempurl%
\url{https://doi.org/10.1109/32.879821}
\showDOI{\tempurl}


\bibitem[\protect\citeauthoryear{Dwyer, Person, and Elbaum}{Dwyer
  et~al\mbox{.}}{2006}]%
        {PathSensitive}
\bibfield{author}{\bibinfo{person}{Matthew~B. Dwyer}, \bibinfo{person}{Suzette
  Person}, {and} \bibinfo{person}{Sebastian Elbaum}.}
  \bibinfo{year}{2006}\natexlab{}.
\newblock \showarticletitle{Controlling factors in evaluating path-sensitive
  error detection techniques}. In \bibinfo{booktitle}{\emph{Foundations of
  Software Engineering}}. \bibinfo{pages}{92--104}.
\newblock


\bibitem[\protect\citeauthoryear{e~Abreu and Carapu\c{c}a}{e~Abreu and
  Carapu\c{c}a}{1994}]%
        {Abreu1994SQ}
\bibfield{author}{\bibinfo{person}{Fernando~Brito e Abreu} {and}
  \bibinfo{person}{Rog\'{e}rio Carapu\c{c}a}.} \bibinfo{year}{1994}\natexlab{}.
\newblock \showarticletitle{Object-Oriented Software Engineering: Measuring and
  Controlling the Development Process}. In \bibinfo{booktitle}{\emph{Proc.
  Int'l Conf. Software Quality ({QSIC})}}.
\newblock


\bibitem[\protect\citeauthoryear{e~Abreu and Melo}{e~Abreu and Melo}{1996}]%
        {Brito1996ISMS}
\bibfield{author}{\bibinfo{person}{F.~Brito e Abreu} {and} \bibinfo{person}{W.
  Melo}.} \bibinfo{year}{1996}\natexlab{}.
\newblock \showarticletitle{Evaluating the impact of object-oriented design on
  software quality}. In \bibinfo{booktitle}{\emph{Proceedings of the 3rd
  International Software Metrics Symposium}}. \bibinfo{pages}{90--99}.
\newblock
\urldef\tempurl%
\url{https://doi.org/10.1109/METRIC.1996.492446}
\showDOI{\tempurl}


\bibitem[\protect\citeauthoryear{{Ekelund} and {Engström}}{{Ekelund} and
  {Engström}}{2015}]%
        {Ekelund}
\bibfield{author}{\bibinfo{person}{E.~D. {Ekelund}} {and} \bibinfo{person}{E.
  {Engström}}.} \bibinfo{year}{2015}\natexlab{}.
\newblock \showarticletitle{Efficient regression testing based on test history:
  An industrial evaluation}. In \bibinfo{booktitle}{\emph{2015 IEEE
  International Conference on Software Maintenance and Evolution (ICSME)}}.
  \bibinfo{pages}{449--457}.
\newblock
\showISSN{null}
\urldef\tempurl%
\url{https://doi.org/10.1109/ICSM.2015.7332496}
\showDOI{\tempurl}


\bibitem[\protect\citeauthoryear{El~Emam, Benlarbi, Goel, and Rai}{El~Emam
  et~al\mbox{.}}{2001}]%
        {ElEmam2001TSE}
\bibfield{author}{\bibinfo{person}{Kalhed El~Emam}, \bibinfo{person}{Sa\"{\i}da
  Benlarbi}, \bibinfo{person}{Nishith Goel}, {and} \bibinfo{person}{Shesh~N.
  Rai}.} \bibinfo{year}{2001}\natexlab{}.
\newblock \showarticletitle{The Confounding Effect of Class Size on the
  Validity of Object-Oriented Metrics}.
\newblock \bibinfo{journal}{\emph{IEEE Trans. Softw. Eng.}}
  \bibinfo{volume}{27}, \bibinfo{number}{7} (\bibinfo{date}{July}
  \bibinfo{year}{2001}), \bibinfo{pages}{630--650}.
\newblock
\showISSN{0098-5589}
\urldef\tempurl%
\url{https://doi.org/10.1109/32.935855}
\showDOI{\tempurl}


\bibitem[\protect\citeauthoryear{{Engström}, {Runeson}, and
  {Wikstrand}}{{Engström} et~al\mbox{.}}{2010}]%
        {Engstrom2010}
\bibfield{author}{\bibinfo{person}{E. {Engström}}, \bibinfo{person}{P.
  {Runeson}}, {and} \bibinfo{person}{G. {Wikstrand}}.}
  \bibinfo{year}{2010}\natexlab{}.
\newblock \showarticletitle{An Empirical Evaluation of Regression Testing Based
  on Fix-Cache Recommendations}. In \bibinfo{booktitle}{\emph{2010 Third
  International Conference on Software Testing, Verification and Validation}}.
  \bibinfo{pages}{75--78}.
\newblock
\showISSN{2159-4848}
\urldef\tempurl%
\url{https://doi.org/10.1109/ICST.2010.40}
\showDOI{\tempurl}


\bibitem[\protect\citeauthoryear{Fatiregun, Harman, and Hierons}{Fatiregun
  et~al\mbox{.}}{2004}]%
        {Fatiregun2004SCAM}
\bibfield{author}{\bibinfo{person}{D. Fatiregun}, \bibinfo{person}{M. Harman},
  {and} \bibinfo{person}{R.~M. Hierons}.} \bibinfo{year}{2004}\natexlab{}.
\newblock \showarticletitle{Evolving transformation sequences using genetic
  algorithms}. In \bibinfo{booktitle}{\emph{Source Code Analysis and
  Manipulation, Fourth IEEE International Workshop on}}.
  \bibinfo{pages}{65--74}.
\newblock
\urldef\tempurl%
\url{https://doi.org/10.1109/SCAM.2004.11}
\showDOI{\tempurl}


\bibitem[\protect\citeauthoryear{Fenton and Ohlsson}{Fenton and
  Ohlsson}{2000}]%
        {Fenton2000TSE}
\bibfield{author}{\bibinfo{person}{N.~E. Fenton} {and} \bibinfo{person}{N.
  Ohlsson}.} \bibinfo{year}{2000}\natexlab{}.
\newblock \showarticletitle{Quantitative analysis of faults and failures in a
  complex software system}.
\newblock \bibinfo{journal}{\emph{IEEE Transactions on Software Engineering}}
  \bibinfo{volume}{26}, \bibinfo{number}{8} (\bibinfo{date}{Aug}
  \bibinfo{year}{2000}), \bibinfo{pages}{797--814}.
\newblock
\showISSN{0098-5589}
\urldef\tempurl%
\url{https://doi.org/10.1109/32.879815}
\showDOI{\tempurl}


\bibitem[\protect\citeauthoryear{Fowler}{Fowler}{2010}]%
        {Fow10}
\bibfield{author}{\bibinfo{person}{M. Fowler}.}
  \bibinfo{year}{2010}\natexlab{}.
\newblock \bibinfo{booktitle}{\emph{{Domain-Specific Languages}}}.
\newblock \bibinfo{publisher}{Addison-Wesley Professional}.
\newblock


\bibitem[\protect\citeauthoryear{Fraser and Arcuri}{Fraser and Arcuri}{2011}]%
        {FA11}
\bibfield{author}{\bibinfo{person}{Gordon Fraser} {and} \bibinfo{person}{Andrea
  Arcuri}.} \bibinfo{year}{2011}\natexlab{}.
\newblock \showarticletitle{{EvoSuite}: automatic test suite generation for
  object-oriented software}. In \bibinfo{booktitle}{\emph{Proceedings of the
  19th ACM SIGSOFT Symposium and the 13th European Conference on Foundations of
  Software Engineering}} \emph{(\bibinfo{series}{ESEC/FSE '11})}.
  \bibinfo{publisher}{ACM}, \bibinfo{pages}{416--419}.
\newblock


\bibitem[\protect\citeauthoryear{Gay}{Gay}{2018}]%
        {GayCoverage}
\bibfield{author}{\bibinfo{person}{Gregory Gay}.}
  \bibinfo{year}{2018}\natexlab{}.
\newblock \showarticletitle{To Call, or Not to Call: Contrasting Direct and
  Indirect Branch Coverage in Test Generation}. In
  \bibinfo{booktitle}{\emph{Proceedings of the 11th International Workshop on
  Search-Based Software Testing}} (Gothenburg, Sweden)
  \emph{(\bibinfo{series}{SBST '18})}. \bibinfo{publisher}{ACM},
  \bibinfo{address}{New York, NY, USA}, \bibinfo{pages}{43--50}.
\newblock
\showISBNx{978-1-4503-5741-8}
\urldef\tempurl%
\url{https://doi.org/10.1145/3194718.3194719}
\showDOI{\tempurl}


\bibitem[\protect\citeauthoryear{Gligoric, Groce, Zhang, Sharma, Alipour, and
  Marinov}{Gligoric et~al\mbox{.}}{2013}]%
        {ISSTA13}
\bibfield{author}{\bibinfo{person}{Milos Gligoric}, \bibinfo{person}{Alex
  Groce}, \bibinfo{person}{Chaoqiang Zhang}, \bibinfo{person}{Rohan Sharma},
  \bibinfo{person}{Amin Alipour}, {and} \bibinfo{person}{Darko Marinov}.}
  \bibinfo{year}{2013}\natexlab{}.
\newblock \showarticletitle{Comparing Non-Adequate Test Suites using Coverage
  Criteria}. In \bibinfo{booktitle}{\emph{International Symposium on Software
  Testing and Analysis}}. \bibinfo{pages}{302--313}.
\newblock


\bibitem[\protect\citeauthoryear{Gligoric, Negara, Legunsen, and
  Marinov}{Gligoric et~al\mbox{.}}{2014}]%
        {Gligoric}
\bibfield{author}{\bibinfo{person}{Milos Gligoric}, \bibinfo{person}{Stas
  Negara}, \bibinfo{person}{Owolabi Legunsen}, {and} \bibinfo{person}{Darko
  Marinov}.} \bibinfo{year}{2014}\natexlab{}.
\newblock \showarticletitle{An Empirical Evaluation and Comparison of Manual
  and Automated Test Selection}. In \bibinfo{booktitle}{\emph{Proceedings of
  the 29th ACM/IEEE International Conference on Automated Software
  Engineering}} (Vasteras, Sweden) \emph{(\bibinfo{series}{ASE ’14})}.
  \bibinfo{publisher}{Association for Computing Machinery},
  \bibinfo{address}{New York, NY, USA}, \bibinfo{pages}{361–372}.
\newblock
\showISBNx{9781450330138}
\urldef\tempurl%
\url{https://doi.org/10.1145/2642937.2643019}
\showDOI{\tempurl}


\bibitem[\protect\citeauthoryear{Godefroid, Klarlund, and Sen}{Godefroid
  et~al\mbox{.}}{2005}]%
        {GodefroidKS05}
\bibfield{author}{\bibinfo{person}{Patrice Godefroid}, \bibinfo{person}{Nils
  Klarlund}, {and} \bibinfo{person}{Koushik Sen}.}
  \bibinfo{year}{2005}\natexlab{}.
\newblock \showarticletitle{{DART:} directed automated random testing}. In
  \bibinfo{booktitle}{\emph{Programming Language Design and Implementation}}.
  \bibinfo{pages}{213--223}.
\newblock


\bibitem[\protect\citeauthoryear{Goodman}{Goodman}{2016}]%
        {TrailBitsSeeded}
\bibfield{author}{\bibinfo{person}{Peter Goodman}.}
  \bibinfo{year}{2016}\natexlab{}.
\newblock \bibinfo{title}{A fuzzer and a symbolic executor walk into a cloud}.
\newblock
  \bibinfo{howpublished}{\url{https://blog.trailofbits.com/2016/08/02/engineering-solutions-to-hard-program-analysis-problems/}}.
\newblock


\bibitem[\protect\citeauthoryear{Gopinath, Jensen, and Groce}{Gopinath
  et~al\mbox{.}}{2014a}]%
        {SuiteEval}
\bibfield{author}{\bibinfo{person}{Rahul Gopinath}, \bibinfo{person}{Carlos
  Jensen}, {and} \bibinfo{person}{Alex Groce}.}
  \bibinfo{year}{2014}\natexlab{a}.
\newblock \showarticletitle{Code Coverage for Suite Evaluation by Developers}.
  In \bibinfo{booktitle}{\emph{Proceedings of the 36th International Conference
  on Software Engineering}} (Hyderabad, India) \emph{(\bibinfo{series}{ICSE
  2014})}. \bibinfo{publisher}{ACM}, \bibinfo{address}{New York, NY, USA},
  \bibinfo{pages}{72--82}.
\newblock
\showISBNx{978-1-4503-2756-5}
\urldef\tempurl%
\url{https://doi.org/10.1145/2568225.2568278}
\showDOI{\tempurl}


\bibitem[\protect\citeauthoryear{Gopinath, Jensen, and Groce}{Gopinath
  et~al\mbox{.}}{2014b}]%
        {GopinathMutants}
\bibfield{author}{\bibinfo{person}{Rahul Gopinath}, \bibinfo{person}{Carlos
  Jensen}, {and} \bibinfo{person}{Alex Groce}.}
  \bibinfo{year}{2014}\natexlab{b}.
\newblock \showarticletitle{Mutations: How Close Are They to Real Faults?}. In
  \bibinfo{booktitle}{\emph{International Symposium on Software Reliability
  Engineering}}. \bibinfo{pages}{189--200}.
\newblock


\bibitem[\protect\citeauthoryear{Graves, Karr, Marron, and Siy}{Graves
  et~al\mbox{.}}{2000}]%
        {Graves2000TSE}
\bibfield{author}{\bibinfo{person}{Todd~L. Graves}, \bibinfo{person}{Alan~F.
  Karr}, \bibinfo{person}{J.~S. Marron}, {and} \bibinfo{person}{Harvey Siy}.}
  \bibinfo{year}{2000}\natexlab{}.
\newblock \showarticletitle{Predicting Fault Incidence Using Software Change
  History}.
\newblock \bibinfo{journal}{\emph{IEEE Transactions on Software Engineering}}
  \bibinfo{volume}{26}, \bibinfo{number}{7} (\bibinfo{date}{July}
  \bibinfo{year}{2000}), \bibinfo{pages}{653--661}.
\newblock
\showISSN{0098-5589}
\urldef\tempurl%
\url{https://doi.org/10.1109/32.859533}
\showDOI{\tempurl}


\bibitem[\protect\citeauthoryear{Groce and Erwig}{Groce and Erwig}{2012}]%
        {WODACommon}
\bibfield{author}{\bibinfo{person}{Alex Groce} {and} \bibinfo{person}{Martin
  Erwig}.} \bibinfo{year}{2012}\natexlab{}.
\newblock \showarticletitle{Finding Common Ground: Choose, Assert, and Assume}.
  In \bibinfo{booktitle}{\emph{International Workshop on Dynamic Analysis}}.
  \bibinfo{pages}{12--17}.
\newblock


\bibitem[\protect\citeauthoryear{Groce, Fern, Erwig, Pinto, Bauer, and
  Alipour}{Groce et~al\mbox{.}}{2012a}]%
        {ISOLA12}
\bibfield{author}{\bibinfo{person}{Alex Groce}, \bibinfo{person}{Alan Fern},
  \bibinfo{person}{Martin Erwig}, \bibinfo{person}{Jervis Pinto},
  \bibinfo{person}{Tim Bauer}, {and} \bibinfo{person}{Amin Alipour}.}
  \bibinfo{year}{2012}\natexlab{a}.
\newblock \showarticletitle{Learning-Based Test Programming for Programmers}.
  In \bibinfo{booktitle}{\emph{International Symposium on Leveraging
  Applications of Formal Methods, Verification and Validation}}.
  \bibinfo{pages}{752--786}.
\newblock


\bibitem[\protect\citeauthoryear{Groce, Fern, Pinto, Bauer, Alipour, Erwig, and
  Lopez}{Groce et~al\mbox{.}}{2012b}]%
        {ISSRE}
\bibfield{author}{\bibinfo{person}{Alex Groce}, \bibinfo{person}{Alan Fern},
  \bibinfo{person}{Jervis Pinto}, \bibinfo{person}{Tim Bauer},
  \bibinfo{person}{Amin Alipour}, \bibinfo{person}{Martin Erwig}, {and}
  \bibinfo{person}{Camden Lopez}.} \bibinfo{year}{2012}\natexlab{b}.
\newblock \showarticletitle{Lightweight Automated Testing with Adaptation-Based
  Programming}. In \bibinfo{booktitle}{\emph{IEEE International Symposium on
  Software Reliability Engineering}}. \bibinfo{pages}{161--170}.
\newblock


\bibitem[\protect\citeauthoryear{Groce, Holzmann, and Joshi}{Groce
  et~al\mbox{.}}{2007}]%
        {ICSEDiff}
\bibfield{author}{\bibinfo{person}{Alex Groce}, \bibinfo{person}{Gerard
  Holzmann}, {and} \bibinfo{person}{Rajeev Joshi}.}
  \bibinfo{year}{2007}\natexlab{}.
\newblock \showarticletitle{Randomized Differential Testing as a Prelude to
  Formal Verification}. In \bibinfo{booktitle}{\emph{International Conference
  on Software Engineering}}. \bibinfo{pages}{621--631}.
\newblock


\bibitem[\protect\citeauthoryear{Groce and Pinto}{Groce and Pinto}{2015}]%
        {NFM15}
\bibfield{author}{\bibinfo{person}{Alex Groce} {and} \bibinfo{person}{Jervis
  Pinto}.} \bibinfo{year}{2015}\natexlab{}.
\newblock \showarticletitle{A Little Language for Testing}. In
  \bibinfo{booktitle}{\emph{NASA Formal Methods Symposium}}.
  \bibinfo{pages}{204--218}.
\newblock


\bibitem[\protect\citeauthoryear{Groce, Pinto, Azimi, Mittal, Holmes, and
  Kellar}{Groce et~al\mbox{.}}{2015}]%
        {tstl}
\bibfield{author}{\bibinfo{person}{Alex Groce}, \bibinfo{person}{Jervis Pinto},
  \bibinfo{person}{Pooria Azimi}, \bibinfo{person}{Pranjal Mittal},
  \bibinfo{person}{Josie Holmes}, {and} \bibinfo{person}{Kevin Kellar}.}
  \bibinfo{year}{2015}\natexlab{}.
\newblock \bibinfo{title}{{TSTL}: the template scripting testing language}.
\newblock \bibinfo{howpublished}{\url{https://github.com/agroce/tstl}}.
\newblock


\bibitem[\protect\citeauthoryear{Groce and Visser}{Groce and Visser}{2004}]%
        {STTTHeur}
\bibfield{author}{\bibinfo{person}{Alex Groce} {and} \bibinfo{person}{Willem
  Visser}.} \bibinfo{year}{2004}\natexlab{}.
\newblock \showarticletitle{Heuristics for model checking {Java} programs}.
\newblock \bibinfo{journal}{\emph{Software Tools for Technology Transfer}}
  \bibinfo{volume}{6(4)} (\bibinfo{year}{2004}), \bibinfo{pages}{260--276}.
\newblock


\bibitem[\protect\citeauthoryear{Groce, Zhang, Eide, Chen, and Regehr}{Groce
  et~al\mbox{.}}{2012c}]%
        {ISSTA12}
\bibfield{author}{\bibinfo{person}{Alex Groce}, \bibinfo{person}{Chaoqiang
  Zhang}, \bibinfo{person}{Eric Eide}, \bibinfo{person}{Yang Chen}, {and}
  \bibinfo{person}{John Regehr}.} \bibinfo{year}{2012}\natexlab{c}.
\newblock \showarticletitle{Swarm Testing}. In
  \bibinfo{booktitle}{\emph{International Symposium on Software Testing and
  Analysis}}. \bibinfo{pages}{78--88}.
\newblock


\bibitem[\protect\citeauthoryear{Gyimothy, Ferenc, and Siket}{Gyimothy
  et~al\mbox{.}}{2005}]%
        {Gyimothy2005TSE}
\bibfield{author}{\bibinfo{person}{T. Gyimothy}, \bibinfo{person}{R. Ferenc},
  {and} \bibinfo{person}{I. Siket}.} \bibinfo{year}{2005}\natexlab{}.
\newblock \showarticletitle{Empirical validation of object-oriented metrics on
  open source software for fault prediction}.
\newblock \bibinfo{journal}{\emph{IEEE Transactions on Software Engineering}}
  \bibinfo{volume}{31}, \bibinfo{number}{10} (\bibinfo{date}{Oct}
  \bibinfo{year}{2005}), \bibinfo{pages}{897--910}.
\newblock
\showISSN{0098-5589}
\urldef\tempurl%
\url{https://doi.org/10.1109/TSE.2005.112}
\showDOI{\tempurl}


\bibitem[\protect\citeauthoryear{Hamlet}{Hamlet}{1994}]%
        {Hamlet94}
\bibfield{author}{\bibinfo{person}{Richard Hamlet}.}
  \bibinfo{year}{1994}\natexlab{}.
\newblock \showarticletitle{Random testing}.
\newblock In \bibinfo{booktitle}{\emph{Encyclopedia of Software Engineering}}.
  \bibinfo{publisher}{Wiley}, \bibinfo{pages}{970--978}.
\newblock


\bibitem[\protect\citeauthoryear{Harman and O'Hearn}{Harman and
  O'Hearn}{2018}]%
        {HarmanSCAM}
\bibfield{author}{\bibinfo{person}{Mark Harman} {and} \bibinfo{person}{Peter
  O'Hearn}.} \bibinfo{year}{2018}\natexlab{}.
\newblock \showarticletitle{From Start-Ups To Scale-Ups: Open Problems,
  Challenges And Myths In Static And Dynamic Program Analysis For Testing And
  Verification}. In \bibinfo{booktitle}{\emph{IEEE International Working
  Conference on Source Code Analysis and Manipulation}}.
\newblock


\bibitem[\protect\citeauthoryear{Herzig, Greiler, Czerwonka, and Murphy}{Herzig
  et~al\mbox{.}}{2015}]%
        {Herzig}
\bibfield{author}{\bibinfo{person}{Kim Herzig}, \bibinfo{person}{Michaela
  Greiler}, \bibinfo{person}{Jacek Czerwonka}, {and} \bibinfo{person}{Brendan
  Murphy}.} \bibinfo{year}{2015}\natexlab{}.
\newblock \showarticletitle{The Art of Testing Less without Sacrificing
  Quality}. In \bibinfo{booktitle}{\emph{Proceedings of the 37th International
  Conference on Software Engineering - Volume 1}} (Florence, Italy)
  \emph{(\bibinfo{series}{ICSE ’15})}. \bibinfo{publisher}{IEEE Press},
  \bibinfo{pages}{483–493}.
\newblock
\showISBNx{9781479919345}


\bibitem[\protect\citeauthoryear{Hirzel and Klaeren}{Hirzel and
  Klaeren}{2016}]%
        {Hirzel}
\bibfield{author}{\bibinfo{person}{Matthias Hirzel} {and}
  \bibinfo{person}{Herbert Klaeren}.} \bibinfo{year}{2016}\natexlab{}.
\newblock \showarticletitle{Graph-Walk-based Selective Regression Testing of
  Web Applications Created with {Google} Web Toolkit}. In
  \bibinfo{booktitle}{\emph{Gemeinsamer Tagungsband der Workshops der Tagung
  Software Engineering 2016 {(SE} 2016), Wien, 23.-26. Februar 2016}}.
  \bibinfo{pages}{55--69}.
\newblock
\urldef\tempurl%
\url{http://ceur-ws.org/Vol-1559/paper05.pdf}
\showURL{%
\tempurl}


\bibitem[\protect\citeauthoryear{Holmes, Groce, Pinto, Mittal, Azimi, Kellar,
  and O'Brien}{Holmes et~al\mbox{.}}{2017}]%
        {tstlsttt}
\bibfield{author}{\bibinfo{person}{Josie Holmes}, \bibinfo{person}{Alex Groce},
  \bibinfo{person}{Jervis Pinto}, \bibinfo{person}{Pranjal Mittal},
  \bibinfo{person}{Pooria Azimi}, \bibinfo{person}{Kevin Kellar}, {and}
  \bibinfo{person}{James O'Brien}.} \bibinfo{year}{2017}\natexlab{}.
\newblock \showarticletitle{{TSTL:} the Template Scripting Testing Language}.
\newblock \bibinfo{journal}{\emph{International Journal on Software Tools for
  Technology Transfer}} (\bibinfo{year}{2017}).
\newblock
\newblock
\shownote{Accepted for publication.}


\bibitem[\protect\citeauthoryear{Holzmann, Joshi, and Groce}{Holzmann
  et~al\mbox{.}}{2008a}]%
        {swarmVer}
\bibfield{author}{\bibinfo{person}{Gerard Holzmann}, \bibinfo{person}{Rajeev
  Joshi}, {and} \bibinfo{person}{Alex Groce}.}
  \bibinfo{year}{2008}\natexlab{a}.
\newblock \showarticletitle{Swarm Verification}. In
  \bibinfo{booktitle}{\emph{Automated Software Engineering}}.
  \bibinfo{pages}{1--6}.
\newblock


\bibitem[\protect\citeauthoryear{Holzmann, Joshi, and Groce}{Holzmann
  et~al\mbox{.}}{2008b}]%
        {swarm}
\bibfield{author}{\bibinfo{person}{Gerard Holzmann}, \bibinfo{person}{Rajeev
  Joshi}, {and} \bibinfo{person}{Alex Groce}.}
  \bibinfo{year}{2008}\natexlab{b}.
\newblock \showarticletitle{Tackling Large Verification Problems with the Swarm
  Tool}. In \bibinfo{booktitle}{\emph{SPIN Workshop on Model Checking of
  Software}}. \bibinfo{pages}{134--143}.
\newblock


\bibitem[\protect\citeauthoryear{Holzmann, Joshi, and Groce}{Holzmann
  et~al\mbox{.}}{2011}]%
        {swarmIEEE}
\bibfield{author}{\bibinfo{person}{Gerard Holzmann}, \bibinfo{person}{Rajeev
  Joshi}, {and} \bibinfo{person}{Alex Groce}.} \bibinfo{year}{2011}\natexlab{}.
\newblock \showarticletitle{Swarm Verification Techniques}.
\newblock \bibinfo{journal}{\emph{IEEE Transactions on Software Engineering}}
  \bibinfo{volume}{37}, \bibinfo{number}{6} (\bibinfo{year}{2011}),
  \bibinfo{pages}{845--857}.
\newblock


\bibitem[\protect\citeauthoryear{Inozemtseva}{Inozemtseva}{[n.d.]}]%
        {notcorweb}
\bibfield{author}{\bibinfo{person}{Laura Inozemtseva}.}
  \bibinfo{year}{[n.d.]}\natexlab{}.
\newblock \bibinfo{title}{Supplemental results for "Coverage is not
  Correlated..."}.
\newblock
  \bibinfo{howpublished}{\url{http://inozemtseva.com/research/2014/icse/coverage}}.
\newblock
\newblock
\shownote{Viewed May 2017; site appears to no longer be available, though
  referred to in text of the paper.}


\bibitem[\protect\citeauthoryear{Inozemtseva and Holmes}{Inozemtseva and
  Holmes}{2014}]%
        {NotCorrelated}
\bibfield{author}{\bibinfo{person}{Laura Inozemtseva} {and}
  \bibinfo{person}{Reid Holmes}.} \bibinfo{year}{2014}\natexlab{}.
\newblock \showarticletitle{Coverage is Not Strongly Correlated with Test Suite
  Effectiveness}. In \bibinfo{booktitle}{\emph{Proceedings of the 36th
  International Conference on Software Engineering}} (Hyderabad, India)
  \emph{(\bibinfo{series}{ICSE 2014})}. \bibinfo{publisher}{ACM},
  \bibinfo{address}{New York, NY, USA}, \bibinfo{pages}{435--445}.
\newblock
\showISBNx{978-1-4503-2756-5}
\urldef\tempurl%
\url{https://doi.org/10.1145/2568225.2568271}
\showDOI{\tempurl}


\bibitem[\protect\citeauthoryear{Just, Jalali, and Ernst}{Just
  et~al\mbox{.}}{2014}]%
        {just2014defects4j}
\bibfield{author}{\bibinfo{person}{Ren{\'e} Just}, \bibinfo{person}{Darioush
  Jalali}, {and} \bibinfo{person}{Michael~D Ernst}.}
  \bibinfo{year}{2014}\natexlab{}.
\newblock \showarticletitle{{Defects4J}: A database of existing faults to
  enable controlled testing studies for Java programs}. In
  \bibinfo{booktitle}{\emph{Proceedings of the 2014 International Symposium on
  Software Testing and Analysis}}. ACM, \bibinfo{pages}{437--440}.
\newblock


\bibitem[\protect\citeauthoryear{Kaneoka}{Kaneoka}{2017}]%
        {Kazuki}
\bibfield{author}{\bibinfo{person}{Kazuki Kaneoka}.}
  \bibinfo{year}{2017}\natexlab{}.
\newblock \bibinfo{booktitle}{\emph{Feedback-Based Random Test Generator for
  {TSTL}}}.
\newblock \bibinfo{type}{{T}echnical {R}eport} MS thesis.
  \bibinfo{institution}{Oregon State University}.
\newblock


\bibitem[\protect\citeauthoryear{Klees, Ruef, Cooper, Wei, and Hicks}{Klees
  et~al\mbox{.}}{2018}]%
        {Hicks18}
\bibfield{author}{\bibinfo{person}{George Klees}, \bibinfo{person}{Andrew
  Ruef}, \bibinfo{person}{Benji Cooper}, \bibinfo{person}{Shiyi Wei}, {and}
  \bibinfo{person}{Michael Hicks}.} \bibinfo{year}{2018}\natexlab{}.
\newblock \showarticletitle{Evaluating Fuzz Testing}. In
  \bibinfo{booktitle}{\emph{Proceedings of the 2018 ACM SIGSAC Conference on
  Computer and Communications Security}} (Toronto, Canada)
  \emph{(\bibinfo{series}{CCS '18})}. \bibinfo{publisher}{ACM},
  \bibinfo{address}{New York, NY, USA}, \bibinfo{pages}{2123--2138}.
\newblock
\showISBNx{978-1-4503-5693-0}
\urldef\tempurl%
\url{https://doi.org/10.1145/3243734.3243804}
\showDOI{\tempurl}


\bibitem[\protect\citeauthoryear{Koru, Zhang, Emam, and Liu}{Koru
  et~al\mbox{.}}{2009}]%
        {Koru2009TSE}
\bibfield{author}{\bibinfo{person}{A.~G. Koru}, \bibinfo{person}{D. Zhang},
  \bibinfo{person}{K.~El Emam}, {and} \bibinfo{person}{H. Liu}.}
  \bibinfo{year}{2009}\natexlab{}.
\newblock \showarticletitle{An Investigation into the Functional Form of the
  Size-Defect Relationship for Software Modules}.
\newblock \bibinfo{journal}{\emph{IEEE Transactions on Software Engineering}}
  \bibinfo{volume}{35}, \bibinfo{number}{2} (\bibinfo{date}{March}
  \bibinfo{year}{2009}), \bibinfo{pages}{293--304}.
\newblock
\showISSN{0098-5589}
\urldef\tempurl%
\url{https://doi.org/10.1109/TSE.2008.90}
\showDOI{\tempurl}


\bibitem[\protect\citeauthoryear{Landman, Serebrenik, Bouwers, and
  Vinju}{Landman et~al\mbox{.}}{2016}]%
        {JurgenCC}
\bibfield{author}{\bibinfo{person}{Davy Landman}, \bibinfo{person}{Alexander
  Serebrenik}, \bibinfo{person}{Eric Bouwers}, {and} \bibinfo{person}{Jurgen~J.
  Vinju}.} \bibinfo{year}{2016}\natexlab{}.
\newblock \showarticletitle{Empirical analysis of the relationship between {CC}
  and {SLOC} in a large corpus of Java methods and {C} functions}.
\newblock \bibinfo{journal}{\emph{Journal of Software: Evolution and Process}}
  \bibinfo{volume}{28}, \bibinfo{number}{7} (\bibinfo{year}{2016}),
  \bibinfo{pages}{589--618}.
\newblock
\urldef\tempurl%
\url{https://doi.org/10.1002/smr.1760}
\showDOI{\tempurl}


\bibitem[\protect\citeauthoryear{MacIver}{MacIver}{2013}]%
        {Hypothesis}
\bibfield{author}{\bibinfo{person}{David~R. MacIver}.}
  \bibinfo{year}{2013}\natexlab{}.
\newblock \bibinfo{title}{Hypothesis: Test faster, fix more}.
\newblock \bibinfo{howpublished}{\url{http://hypothesis.works/}}.
\newblock


\bibitem[\protect\citeauthoryear{MacIver}{MacIver}{2016}]%
        {hypheaps}
\bibfield{author}{\bibinfo{person}{David~R. MacIver}.}
  \bibinfo{year}{2016}\natexlab{}.
\newblock \bibinfo{title}{Rule Based Stateful Testing}.
\newblock
  \bibinfo{howpublished}{\url{http://hypothesis.works/articles/rule-based-stateful-testing/}}.
\newblock


\bibitem[\protect\citeauthoryear{MacIver}{MacIver}{2017}]%
        {covdiff1}
\bibfield{author}{\bibinfo{person}{David~R. MacIver}.}
  \bibinfo{year}{2017}\natexlab{}.
\newblock \bibinfo{title}{{Python Coverage} could be fast}.
\newblock
  \bibinfo{howpublished}{\url{https://www.drmaciver.com/2017/09/python-coverage-could-be-fast/}}.
\newblock


\bibitem[\protect\citeauthoryear{MacIver}{MacIver}{ober}]%
        {covdiff2}
\bibfield{author}{\bibinfo{person}{David~R. MacIver}.}
  \bibinfo{year}{October}\natexlab{}.
\newblock \bibinfo{title}{Coverage adds a lot of overhead when the base test is
  fast}.
\newblock
  \bibinfo{howpublished}{\url{https://github.com/HypothesisWorks/hypothesis/issues/914}}.
\newblock


\bibitem[\protect\citeauthoryear{MacIver and {PyPI}}{MacIver and
  {PyPI}}{[n.d.]}]%
        {HypAdopt}
\bibfield{author}{\bibinfo{person}{David~R. MacIver} {and}
  \bibinfo{person}{{PyPI}}.} \bibinfo{year}{[n.d.]}\natexlab{}.
\newblock \bibinfo{title}{Usage stats for hypothesis on {PyPI}}.
\newblock
  \bibinfo{howpublished}{\url{https://libraries.io/pypi/hypothesis/usage}}.
\newblock


\bibitem[\protect\citeauthoryear{Marcus, Poshyvanyk, and Ferenc}{Marcus
  et~al\mbox{.}}{2008}]%
        {Marcus2008TSE}
\bibfield{author}{\bibinfo{person}{Andrian Marcus}, \bibinfo{person}{Denys
  Poshyvanyk}, {and} \bibinfo{person}{Rudolf Ferenc}.}
  \bibinfo{year}{2008}\natexlab{}.
\newblock \showarticletitle{Using the Conceptual Cohesion of Classes for Fault
  Prediction in Object-Oriented Systems}.
\newblock \bibinfo{journal}{\emph{IEEE Trans. Softw. Eng.}}
  \bibinfo{volume}{34}, \bibinfo{number}{2} (\bibinfo{date}{March}
  \bibinfo{year}{2008}), \bibinfo{pages}{287--300}.
\newblock
\showISSN{0098-5589}
\urldef\tempurl%
\url{https://doi.org/10.1109/TSE.2007.70768}
\showDOI{\tempurl}


\bibitem[\protect\citeauthoryear{{Marijan}}{{Marijan}}{2015}]%
        {Marijan15}
\bibfield{author}{\bibinfo{person}{D. {Marijan}}.}
  \bibinfo{year}{2015}\natexlab{}.
\newblock \showarticletitle{Multi-perspective Regression Test Prioritization
  for Time-Constrained Environments}. In \bibinfo{booktitle}{\emph{2015 IEEE
  International Conference on Software Quality, Reliability and Security}}.
  \bibinfo{pages}{157--162}.
\newblock
\showISSN{null}
\urldef\tempurl%
\url{https://doi.org/10.1109/QRS.2015.31}
\showDOI{\tempurl}


\bibitem[\protect\citeauthoryear{{Marijan}, {Gotlieb}, and {Sen}}{{Marijan}
  et~al\mbox{.}}{2013}]%
        {Marijan13}
\bibfield{author}{\bibinfo{person}{D. {Marijan}}, \bibinfo{person}{A.
  {Gotlieb}}, {and} \bibinfo{person}{S. {Sen}}.}
  \bibinfo{year}{2013}\natexlab{}.
\newblock \showarticletitle{Test Case Prioritization for Continuous Regression
  Testing: An Industrial Case Study}. In \bibinfo{booktitle}{\emph{2013 IEEE
  International Conference on Software Maintenance}}.
  \bibinfo{pages}{540--543}.
\newblock
\showISSN{1063-6773}
\urldef\tempurl%
\url{https://doi.org/10.1109/ICSM.2013.91}
\showDOI{\tempurl}


\bibitem[\protect\citeauthoryear{Marinescu and Cadar}{Marinescu and
  Cadar}{2012}]%
        {Marinescu:2012:MTS:2337223.2337308}
\bibfield{author}{\bibinfo{person}{Paul~Dan Marinescu} {and}
  \bibinfo{person}{Cristian Cadar}.} \bibinfo{year}{2012}\natexlab{}.
\newblock \showarticletitle{make test-zesti: a symbolic execution solution for
  improving regression testing}. In \bibinfo{booktitle}{\emph{International
  Conference on Software Engineering}}. \bibinfo{pages}{716--726}.
\newblock


\bibitem[\protect\citeauthoryear{McCabe}{McCabe}{1976}]%
        {McCabe}
\bibfield{author}{\bibinfo{person}{T.~J. McCabe}.}
  \bibinfo{year}{1976}\natexlab{}.
\newblock \showarticletitle{A Complexity Measure}.
\newblock \bibinfo{journal}{\emph{IEEE Trans. Softw. Eng.}}
  \bibinfo{volume}{2}, \bibinfo{number}{4} (\bibinfo{date}{July}
  \bibinfo{year}{1976}), \bibinfo{pages}{308--320}.
\newblock
\showISSN{0098-5589}
\urldef\tempurl%
\url{https://doi.org/10.1109/TSE.1976.233837}
\showDOI{\tempurl}


\bibitem[\protect\citeauthoryear{McKeeman}{McKeeman}{1998}]%
        {Differential}
\bibfield{author}{\bibinfo{person}{William McKeeman}.}
  \bibinfo{year}{1998}\natexlab{}.
\newblock \showarticletitle{Differential testing for software}.
\newblock \bibinfo{journal}{\emph{Digital Technical Journal of Digital
  Equipment Corporation}}  \bibinfo{volume}{10(1)} (\bibinfo{year}{1998}),
  \bibinfo{pages}{100--107}.
\newblock


\bibitem[\protect\citeauthoryear{McMinn}{McMinn}{2004}]%
        {McMinn04search-basedsoftware}
\bibfield{author}{\bibinfo{person}{Phil McMinn}.}
  \bibinfo{year}{2004}\natexlab{}.
\newblock \showarticletitle{Search-based Software Test Data Generation: A
  Survey}.
\newblock \bibinfo{journal}{\emph{Software Testing, Verification and
  Reliability}}  \bibinfo{volume}{14} (\bibinfo{year}{2004}),
  \bibinfo{pages}{105--156}.
\newblock


\bibitem[\protect\citeauthoryear{Menzies, Stefano, Chapman, and McGill}{Menzies
  et~al\mbox{.}}{2002}]%
        {Menzies2002SEW}
\bibfield{author}{\bibinfo{person}{T. Menzies}, \bibinfo{person}{J.~S.~Di
  Stefano}, \bibinfo{person}{M. Chapman}, {and} \bibinfo{person}{K. McGill}.}
  \bibinfo{year}{2002}\natexlab{}.
\newblock \showarticletitle{Metrics that matter}. In
  \bibinfo{booktitle}{\emph{27th Annual NASA Goddard/IEEE Software Engineering
  Workshop, 2002. Proceedings.}} \bibinfo{pages}{51--57}.
\newblock
\urldef\tempurl%
\url{https://doi.org/10.1109/SEW.2002.1199449}
\showDOI{\tempurl}


\bibitem[\protect\citeauthoryear{Nilsson, Auckland, Sumner, and
  Sahayam}{Nilsson et~al\mbox{.}}{2016}]%
        {ScalaCheckDoc}
\bibfield{author}{\bibinfo{person}{Rickard Nilsson}, \bibinfo{person}{Shane
  Auckland}, \bibinfo{person}{Mark Sumner}, {and} \bibinfo{person}{Sanjiv
  Sahayam}.} \bibinfo{year}{2016}\natexlab{}.
\newblock \bibinfo{title}{ScalaCheck User Guide}.
\newblock
  \bibinfo{howpublished}{\url{https://github.com/rickynils/scalacheck/blob/master/doc/UserGuide.md}}.
\newblock


\bibitem[\protect\citeauthoryear{Offutt and Abdurazik}{Offutt and
  Abdurazik}{2000}]%
        {Mut2000}
\bibfield{author}{\bibinfo{person}{J. Offutt} {and} \bibinfo{person}{A.
  Abdurazik}.} \bibinfo{year}{2000}\natexlab{}.
\newblock In \bibinfo{booktitle}{\emph{Mutation 2000: Mutation Testing in the
  Twentieth and the Twenty First Centuries}}.
\newblock


\bibitem[\protect\citeauthoryear{Ohmann, Brown, Neelakandan, Linderoth, and
  Liblit}{Ohmann et~al\mbox{.}}{2016}]%
        {ohmann2016optimizing}
\bibfield{author}{\bibinfo{person}{Peter Ohmann},
  \bibinfo{person}{David~Bingham Brown}, \bibinfo{person}{Naveen Neelakandan},
  \bibinfo{person}{Jeff Linderoth}, {and} \bibinfo{person}{Ben Liblit}.}
  \bibinfo{year}{2016}\natexlab{}.
\newblock \showarticletitle{Optimizing customized program coverage}. In
  \bibinfo{booktitle}{\emph{Automated Software Engineering (ASE), 2016 31st
  IEEE/ACM International Conference on}}. IEEE, \bibinfo{pages}{27--38}.
\newblock


\bibitem[\protect\citeauthoryear{Olague, Etzkorn, Gholston, and
  Quattlebaum}{Olague et~al\mbox{.}}{2007}]%
        {Olague2007TSE}
\bibfield{author}{\bibinfo{person}{Hector~M. Olague}, \bibinfo{person}{Letha~H.
  Etzkorn}, \bibinfo{person}{Sampson Gholston}, {and} \bibinfo{person}{Stephen
  Quattlebaum}.} \bibinfo{year}{2007}\natexlab{}.
\newblock \showarticletitle{Empirical Validation of Three Software Metrics
  Suites to Predict Fault-Proneness of Object-Oriented Classes Developed Using
  Highly Iterative or Agile Software Development Processes}.
\newblock \bibinfo{journal}{\emph{IEEE Transactions on Software Engineering}}
  \bibinfo{volume}{33}, \bibinfo{number}{6} (\bibinfo{date}{June}
  \bibinfo{year}{2007}), \bibinfo{pages}{402--419}.
\newblock
\showISSN{0098-5589}
\urldef\tempurl%
\url{https://doi.org/10.1109/TSE.2007.1015}
\showDOI{\tempurl}


\bibitem[\protect\citeauthoryear{Ostrand, Weyuker, and Bell}{Ostrand
  et~al\mbox{.}}{2005}]%
        {Ostrand2005TSE}
\bibfield{author}{\bibinfo{person}{T.~J. Ostrand}, \bibinfo{person}{E.~J.
  Weyuker}, {and} \bibinfo{person}{R.~M. Bell}.}
  \bibinfo{year}{2005}\natexlab{}.
\newblock \showarticletitle{Predicting the location and number of faults in
  large software systems}.
\newblock \bibinfo{journal}{\emph{IEEE Transactions on Software Engineering}}
  \bibinfo{volume}{31}, \bibinfo{number}{4} (\bibinfo{date}{April}
  \bibinfo{year}{2005}), \bibinfo{pages}{340--355}.
\newblock
\showISSN{0098-5589}
\urldef\tempurl%
\url{https://doi.org/10.1109/TSE.2005.49}
\showDOI{\tempurl}


\bibitem[\protect\citeauthoryear{Pacheco, Lahiri, Ernst, and Ball}{Pacheco
  et~al\mbox{.}}{2007}]%
        {Pacheco}
\bibfield{author}{\bibinfo{person}{Carlos Pacheco},
  \bibinfo{person}{Shuvendu~K. Lahiri}, \bibinfo{person}{Michael~D. Ernst},
  {and} \bibinfo{person}{Thomas Ball}.} \bibinfo{year}{2007}\natexlab{}.
\newblock \showarticletitle{Feedback-directed Random Test Generation}. In
  \bibinfo{booktitle}{\emph{International Conference on Software Engineering}}.
  \bibinfo{pages}{75--84}.
\newblock


\bibitem[\protect\citeauthoryear{Pai and Dugan}{Pai and Dugan}{2007}]%
        {Pai2007TSE}
\bibfield{author}{\bibinfo{person}{G.~J. Pai} {and} \bibinfo{person}{J.~Bechta
  Dugan}.} \bibinfo{year}{2007}\natexlab{}.
\newblock \showarticletitle{Empirical Analysis of Software Fault Content and
  Fault Proneness Using Bayesian Methods}.
\newblock \bibinfo{journal}{\emph{IEEE Transactions on Software Engineering}}
  \bibinfo{volume}{33}, \bibinfo{number}{10} (\bibinfo{date}{Oct}
  \bibinfo{year}{2007}), \bibinfo{pages}{675--686}.
\newblock
\showISSN{0098-5589}
\urldef\tempurl%
\url{https://doi.org/10.1109/TSE.2007.70722}
\showDOI{\tempurl}


\bibitem[\protect\citeauthoryear{Papadakis and Sagonas}{Papadakis and
  Sagonas}{2011}]%
        {PROPER}
\bibfield{author}{\bibinfo{person}{Manolis Papadakis} {and}
  \bibinfo{person}{Konstantinos Sagonas}.} \bibinfo{year}{2011}\natexlab{}.
\newblock \showarticletitle{A {PropEr} Integration of Types and Function
  Specifications with Property-Based Testing}. In
  \bibinfo{booktitle}{\emph{Proceedings of the 2011 ACM SIGPLAN Erlang
  Workshop}}. \bibinfo{publisher}{ACM Press}, \bibinfo{address}{New York, NY},
  \bibinfo{pages}{39--50}.
\newblock


\bibitem[\protect\citeauthoryear{Person, Yang, Rungta, and Khurshid}{Person
  et~al\mbox{.}}{2011}]%
        {Person:2011:DIS:1993498.1993558}
\bibfield{author}{\bibinfo{person}{Suzette Person}, \bibinfo{person}{Guowei
  Yang}, \bibinfo{person}{Neha Rungta}, {and} \bibinfo{person}{Sarfraz
  Khurshid}.} \bibinfo{year}{2011}\natexlab{}.
\newblock \showarticletitle{Directed Incremental Symbolic Execution}. In
  \bibinfo{booktitle}{\emph{Proceedings of the 32Nd ACM SIGPLAN Conference on
  Programming Language Design and Implementation}} \emph{(\bibinfo{series}{PLDI
  '11})}. \bibinfo{pages}{504--515}.
\newblock


\bibitem[\protect\citeauthoryear{Radjenovi\'{c}, Heri\v{c}ko, Torkar, and
  \v{Z}ivkovi\v{c}}{Radjenovi\'{c} et~al\mbox{.}}{2013}]%
        {Radjenovic2013IST}
\bibfield{author}{\bibinfo{person}{Danijel Radjenovi\'{c}},
  \bibinfo{person}{Marjan Heri\v{c}ko}, \bibinfo{person}{Richard Torkar}, {and}
  \bibinfo{person}{Ale\v{s} \v{Z}ivkovi\v{c}}.}
  \bibinfo{year}{2013}\natexlab{}.
\newblock \showarticletitle{Software Fault Prediction Metrics}.
\newblock \bibinfo{journal}{\emph{Information and Software Technologies}}
  \bibinfo{volume}{55}, \bibinfo{number}{8} (\bibinfo{date}{Aug.}
  \bibinfo{year}{2013}), \bibinfo{pages}{1397--1418}.
\newblock
\showISSN{0950-5849}
\urldef\tempurl%
\url{https://doi.org/10.1016/j.infsof.2013.02.009}
\showDOI{\tempurl}


\bibitem[\protect\citeauthoryear{Rawat, Jain, Kumar, Cojocar, Giuffrida, and
  Bos}{Rawat et~al\mbox{.}}{2017}]%
        {rawat2017vuzzer}
\bibfield{author}{\bibinfo{person}{Sanjay Rawat}, \bibinfo{person}{Vivek Jain},
  \bibinfo{person}{Ashish Kumar}, \bibinfo{person}{Lucian Cojocar},
  \bibinfo{person}{Cristiano Giuffrida}, {and} \bibinfo{person}{Herbert Bos}.}
  \bibinfo{year}{2017}\natexlab{}.
\newblock \showarticletitle{{VUzzer}: Application-aware Evolutionary Fuzzing.}.
  In \bibinfo{booktitle}{\emph{Network and Distributed Security Symposium
  (NDSS)}}.
\newblock


\bibitem[\protect\citeauthoryear{Ray, Hellendoorn, Godhane, Tu, Bacchelli, and
  Devanbu}{Ray et~al\mbox{.}}{2016}]%
        {Ray:2016:NBC:2884781.2884848}
\bibfield{author}{\bibinfo{person}{Baishakhi Ray}, \bibinfo{person}{Vincent
  Hellendoorn}, \bibinfo{person}{Saheel Godhane}, \bibinfo{person}{Zhaopeng
  Tu}, \bibinfo{person}{Alberto Bacchelli}, {and} \bibinfo{person}{Premkumar
  Devanbu}.} \bibinfo{year}{2016}\natexlab{}.
\newblock \showarticletitle{On the "Naturalness" of Buggy Code}. In
  \bibinfo{booktitle}{\emph{Proceedings of the 38th International Conference on
  Software Engineering}} (Austin, Texas) \emph{(\bibinfo{series}{ICSE '16})}.
  \bibinfo{publisher}{ACM}, \bibinfo{address}{New York, NY, USA},
  \bibinfo{pages}{428--439}.
\newblock
\showISBNx{978-1-4503-3900-1}
\urldef\tempurl%
\url{https://doi.org/10.1145/2884781.2884848}
\showDOI{\tempurl}


\bibitem[\protect\citeauthoryear{Saha, Zhang, Khurshid, and Perry}{Saha
  et~al\mbox{.}}{2015}]%
        {Saha}
\bibfield{author}{\bibinfo{person}{Ripon~K Saha}, \bibinfo{person}{Lingming
  Zhang}, \bibinfo{person}{Sarfraz Khurshid}, {and} \bibinfo{person}{Dewayne~E
  Perry}.} \bibinfo{year}{2015}\natexlab{}.
\newblock \showarticletitle{An information retrieval approach for regression
  test prioritization based on program changes}. In
  \bibinfo{booktitle}{\emph{2015 IEEE/ACM 37th IEEE International Conference on
  Software Engineering}}, Vol.~\bibinfo{volume}{1}. IEEE,
  \bibinfo{pages}{268--279}.
\newblock


\bibitem[\protect\citeauthoryear{{Scientific Toolworks, Inc.}}{{Scientific
  Toolworks, Inc.}}{2017}]%
        {_understand_2017}
\bibfield{author}{\bibinfo{person}{{Scientific Toolworks, Inc.}}}
  \bibinfo{year}{2017}\natexlab{}.
\newblock \bibinfo{title}{{{Understand}}\texttrademark{} {{Static Code Analysis
  Tool}}}.
\newblock \bibinfo{howpublished}{\url{https://scitools.com/}}.
\newblock


\bibitem[\protect\citeauthoryear{Seonghoon}{Seonghoon}{2015}]%
        {rustcov}
\bibfield{author}{\bibinfo{person}{Kang Seonghoon}.}
  \bibinfo{year}{2015}\natexlab{}.
\newblock \bibinfo{title}{Tutorial: How to collect test coverages for Rust
  project}.
\newblock
  \bibinfo{howpublished}{\url{https://users.rust-lang.org/t/tutorial-how-to-collect-test-coverages-for-rust-project/650}}.
\newblock


\bibitem[\protect\citeauthoryear{Shamshiri, Just, Rojas, Fraser, McMinn, and
  Arcuri}{Shamshiri et~al\mbox{.}}{2015a}]%
        {AutoTestFaults}
\bibfield{author}{\bibinfo{person}{Sina Shamshiri}, \bibinfo{person}{Rene
  Just}, \bibinfo{person}{Jose~Miguel Rojas}, \bibinfo{person}{Gordon Fraser},
  \bibinfo{person}{Phil McMinn}, {and} \bibinfo{person}{Andrea Arcuri}.}
  \bibinfo{year}{2015}\natexlab{a}.
\newblock \showarticletitle{Do automatically generated unit tests find real
  faults? an empirical study of effectiveness and challenges (t)}. In
  \bibinfo{booktitle}{\emph{Automated Software Engineering (ASE), 2015 30th
  IEEE/ACM International Conference on}}. IEEE, \bibinfo{pages}{201--211}.
\newblock


\bibitem[\protect\citeauthoryear{Shamshiri, Rojas, Fraser, and
  McMinn}{Shamshiri et~al\mbox{.}}{2015b}]%
        {RandOrGenetic}
\bibfield{author}{\bibinfo{person}{Sina Shamshiri},
  \bibinfo{person}{Jos\'{e}~Miguel Rojas}, \bibinfo{person}{Gordon Fraser},
  {and} \bibinfo{person}{Phil McMinn}.} \bibinfo{year}{2015}\natexlab{b}.
\newblock \showarticletitle{Random or Genetic Algorithm Search for
  Object-Oriented Test Suite Generation?}. In
  \bibinfo{booktitle}{\emph{Proceedings of the 2015 Annual Conference on
  Genetic and Evolutionary Computation}} (Madrid, Spain)
  \emph{(\bibinfo{series}{GECCO ’15})}. \bibinfo{publisher}{Association for
  Computing Machinery}, \bibinfo{address}{New York, NY, USA},
  \bibinfo{pages}{1367–1374}.
\newblock
\showISBNx{9781450334723}
\urldef\tempurl%
\url{https://doi.org/10.1145/2739480.2754696}
\showDOI{\tempurl}


\bibitem[\protect\citeauthoryear{Shamshiri, Rojas, Gazzola, Fraser, McMinn,
  Mariani, and Arcuri}{Shamshiri et~al\mbox{.}}{2018}]%
        {shamshiri2018random}
\bibfield{author}{\bibinfo{person}{Sina Shamshiri},
  \bibinfo{person}{Jos{\'e}~Miguel Rojas}, \bibinfo{person}{Luca Gazzola},
  \bibinfo{person}{Gordon Fraser}, \bibinfo{person}{Phil McMinn},
  \bibinfo{person}{Leonardo Mariani}, {and} \bibinfo{person}{Andrea Arcuri}.}
  \bibinfo{year}{2018}\natexlab{}.
\newblock \showarticletitle{Random or evolutionary search for object-oriented
  test suite generation?}
\newblock \bibinfo{journal}{\emph{Software Testing, Verification and
  Reliability}} \bibinfo{volume}{28}, \bibinfo{number}{4}
  (\bibinfo{year}{2018}), \bibinfo{pages}{e1660}.
\newblock


\bibitem[\protect\citeauthoryear{{Skoglund} and {Runeson}}{{Skoglund} and
  {Runeson}}{2005}]%
        {Skoglund}
\bibfield{author}{\bibinfo{person}{M. {Skoglund}} {and} \bibinfo{person}{P.
  {Runeson}}.} \bibinfo{year}{2005}\natexlab{}.
\newblock \showarticletitle{A case study of the class firewall regression test
  selection technique on a large scale distributed software system}. In
  \bibinfo{booktitle}{\emph{2005 International Symposium on Empirical Software
  Engineering, 2005.}} \bibinfo{pages}{10 pp.--}.
\newblock
\showISSN{null}
\urldef\tempurl%
\url{https://doi.org/10.1109/ISESE.2005.1541816}
\showDOI{\tempurl}


\bibitem[\protect\citeauthoryear{Srivastava and Thiagarajan}{Srivastava and
  Thiagarajan}{2002}]%
        {Srivastava}
\bibfield{author}{\bibinfo{person}{Amitabh Srivastava} {and}
  \bibinfo{person}{Jay Thiagarajan}.} \bibinfo{year}{2002}\natexlab{}.
\newblock \showarticletitle{Effectively Prioritizing Tests in Development
  Environment}.
\newblock \bibinfo{journal}{\emph{SIGSOFT Softw. Eng. Notes}}
  \bibinfo{volume}{27}, \bibinfo{number}{4} (\bibinfo{date}{July}
  \bibinfo{year}{2002}), \bibinfo{pages}{97–106}.
\newblock
\showISSN{0163-5948}
\urldef\tempurl%
\url{https://doi.org/10.1145/566171.566187}
\showDOI{\tempurl}


\bibitem[\protect\citeauthoryear{Staats, Whalen, and Heimdahl}{Staats
  et~al\mbox{.}}{2011}]%
        {progTestOracle}
\bibfield{author}{\bibinfo{person}{Matt Staats}, \bibinfo{person}{{Michael W.}
  Whalen}, {and} \bibinfo{person}{{Mats P.E.} Heimdahl}.}
  \bibinfo{year}{2011}\natexlab{}.
\newblock \showarticletitle{Programs, tests, and oracles: The foundations of
  testing revisited}. In \bibinfo{booktitle}{\emph{ICSE 2011 - 33rd
  International Conference on Software Engineering, Proceedings of the
  Conference}} \emph{(\bibinfo{series}{Proceedings - International Conference
  on Software Engineering})}. \bibinfo{pages}{391--400}.
\newblock
\showISBNx{9781450304450}
\urldef\tempurl%
\url{https://doi.org/10.1145/1985793.1985847}
\showDOI{\tempurl}
\newblock
\shownote{33rd International Conference on Software Engineering, ICSE 2011 ;
  Conference date: 21-05-2011 Through 28-05-2011.}


\bibitem[\protect\citeauthoryear{Syer, Nagappan, Adams, and Hassan}{Syer
  et~al\mbox{.}}{2015}]%
        {Syer2015TSE}
\bibfield{author}{\bibinfo{person}{M.~D. Syer}, \bibinfo{person}{M. Nagappan},
  \bibinfo{person}{B. Adams}, {and} \bibinfo{person}{A.~E. Hassan}.}
  \bibinfo{year}{2015}\natexlab{}.
\newblock \showarticletitle{Replicating and Re-Evaluating the Theory of
  Relative Defect-Proneness}.
\newblock \bibinfo{journal}{\emph{IEEE Transactions on Software Engineering}}
  \bibinfo{volume}{41}, \bibinfo{number}{2} (\bibinfo{date}{Feb}
  \bibinfo{year}{2015}), \bibinfo{pages}{176--197}.
\newblock
\showISSN{0098-5589}
\urldef\tempurl%
\url{https://doi.org/10.1109/TSE.2014.2361131}
\showDOI{\tempurl}


\bibitem[\protect\citeauthoryear{Tahvili, Afzal, Saadatmand, Bohlin, Sundmark,
  and Larsson}{Tahvili et~al\mbox{.}}{2016}]%
        {Tahvili}
\bibfield{author}{\bibinfo{person}{Sahar Tahvili}, \bibinfo{person}{Wasif
  Afzal}, \bibinfo{person}{Mehrdad Saadatmand}, \bibinfo{person}{Markus
  Bohlin}, \bibinfo{person}{Daniel Sundmark}, {and} \bibinfo{person}{Stig
  Larsson}.} \bibinfo{year}{2016}\natexlab{}.
\newblock \showarticletitle{Towards earlier fault detection by value-driven
  prioritization of test cases using fuzzy TOPSIS}.
\newblock In \bibinfo{booktitle}{\emph{Information Technology: New
  Generations}}. \bibinfo{publisher}{Springer}, \bibinfo{pages}{745--759}.
\newblock


\bibitem[\protect\citeauthoryear{Tikir and Hollingsworth}{Tikir and
  Hollingsworth}{2002}]%
        {Tikir:2002:EIC:566172.566186}
\bibfield{author}{\bibinfo{person}{Mustafa~M. Tikir} {and}
  \bibinfo{person}{Jeffrey~K. Hollingsworth}.} \bibinfo{year}{2002}\natexlab{}.
\newblock \showarticletitle{Efficient Instrumentation for Code Coverage
  Testing}. In \bibinfo{booktitle}{\emph{Proceedings of the 2002 ACM SIGSOFT
  International Symposium on Software Testing and Analysis}} (Roma, Italy)
  \emph{(\bibinfo{series}{ISSTA '02})}. \bibinfo{publisher}{ACM},
  \bibinfo{address}{New York, NY, USA}, \bibinfo{pages}{86--96}.
\newblock
\showISBNx{1-58113-562-9}
\urldef\tempurl%
\url{https://doi.org/10.1145/566172.566186}
\showDOI{\tempurl}


\bibitem[\protect\citeauthoryear{Tomassi, Dmeiri, Wang, Bhowmick, Liu, Devanbu,
  Vasilescu, and Rubio{-}Gonz{\'{a}}lez}{Tomassi et~al\mbox{.}}{2019}]%
        {DBLP:conf/icse/DmeiriTWBLDVR19}
\bibfield{author}{\bibinfo{person}{David~A. Tomassi}, \bibinfo{person}{Naji
  Dmeiri}, \bibinfo{person}{Yichen Wang}, \bibinfo{person}{Antara Bhowmick},
  \bibinfo{person}{Yen{-}Chuan Liu}, \bibinfo{person}{Premkumar~T. Devanbu},
  \bibinfo{person}{Bogdan Vasilescu}, {and} \bibinfo{person}{Cindy
  Rubio{-}Gonz{\'{a}}lez}.} \bibinfo{year}{2019}\natexlab{}.
\newblock \showarticletitle{BugSwarm: mining and continuously growing a dataset
  of reproducible failures and fixes}. In \bibinfo{booktitle}{\emph{{ICSE}}}.
  \bibinfo{publisher}{{IEEE} / {ACM}}, \bibinfo{pages}{339--349}.
\newblock


\bibitem[\protect\citeauthoryear{user1689822}{user1689822}{2012}]%
        {avltree}
\bibfield{author}{\bibinfo{person}{user1689822}.}
  \bibinfo{year}{2012}\natexlab{}.
\newblock \bibinfo{title}{python {AVL} tree insertion}.
\newblock
  \bibinfo{howpublished}{\url{http://stackoverflow.com/questions/12537986/python-avl-tree-insertion}}.
\newblock


\bibitem[\protect\citeauthoryear{White, Jaber, Robinson, and Rajlich}{White
  et~al\mbox{.}}{2008}]%
        {White2008}
\bibfield{author}{\bibinfo{person}{Lee White}, \bibinfo{person}{Khaled Jaber},
  \bibinfo{person}{Brian Robinson}, {and} \bibinfo{person}{V\'{a}clav
  Rajlich}.} \bibinfo{year}{2008}\natexlab{}.
\newblock \showarticletitle{Extended Firewall for Regression Testing: An
  Experience Report}.
\newblock \bibinfo{journal}{\emph{J. Softw. Maint. Evol.}}
  \bibinfo{volume}{20}, \bibinfo{number}{6} (\bibinfo{date}{Nov.}
  \bibinfo{year}{2008}), \bibinfo{pages}{419–433}.
\newblock
\showISSN{1532-060X}


\bibitem[\protect\citeauthoryear{{White} and {Robinson}}{{White} and
  {Robinson}}{2004}]%
        {White2004}
\bibfield{author}{\bibinfo{person}{L. {White}} {and} \bibinfo{person}{B.
  {Robinson}}.} \bibinfo{year}{2004}\natexlab{}.
\newblock \showarticletitle{Industrial real-time regression testing and
  analysis using firewalls}. In \bibinfo{booktitle}{\emph{20th IEEE
  International Conference on Software Maintenance, 2004. Proceedings.}}
  \bibinfo{pages}{18--27}.
\newblock
\showISSN{1063-6773}
\urldef\tempurl%
\url{https://doi.org/10.1109/ICSM.2004.1357786}
\showDOI{\tempurl}


\bibitem[\protect\citeauthoryear{{Wikstrand}, {Feldt}, {Gorantla}, {Zhe}, and
  {White}}{{Wikstrand} et~al\mbox{.}}{2009}]%
        {Wikstrand}
\bibfield{author}{\bibinfo{person}{G. {Wikstrand}}, \bibinfo{person}{R.
  {Feldt}}, \bibinfo{person}{J.~K. {Gorantla}}, \bibinfo{person}{W. {Zhe}},
  {and} \bibinfo{person}{C. {White}}.} \bibinfo{year}{2009}\natexlab{}.
\newblock \showarticletitle{Dynamic Regression Test Selection Based on a File
  Cache An Industrial Evaluation}. In \bibinfo{booktitle}{\emph{2009
  International Conference on Software Testing Verification and Validation}}.
  \bibinfo{pages}{299--302}.
\newblock
\showISSN{2159-4848}
\urldef\tempurl%
\url{https://doi.org/10.1109/ICST.2009.42}
\showDOI{\tempurl}


\bibitem[\protect\citeauthoryear{Yang, Li, and Weiss}{Yang
  et~al\mbox{.}}{2007}]%
        {yang2007survey}
\bibfield{author}{\bibinfo{person}{Qian Yang}, \bibinfo{person}{J~Jenny Li},
  {and} \bibinfo{person}{David~M Weiss}.} \bibinfo{year}{2007}\natexlab{}.
\newblock \showarticletitle{A survey of coverage-based testing tools}.
\newblock \bibinfo{journal}{\emph{Comput. J.}} \bibinfo{volume}{52},
  \bibinfo{number}{5} (\bibinfo{year}{2007}), \bibinfo{pages}{589--597}.
\newblock


\bibitem[\protect\citeauthoryear{Yatoh, Sakamoto, Ishikawa, and Honiden}{Yatoh
  et~al\mbox{.}}{2015}]%
        {FeedControl}
\bibfield{author}{\bibinfo{person}{Kohsuke Yatoh}, \bibinfo{person}{Kazunori
  Sakamoto}, \bibinfo{person}{Fuyuki Ishikawa}, {and} \bibinfo{person}{Shinichi
  Honiden}.} \bibinfo{year}{2015}\natexlab{}.
\newblock \showarticletitle{Feedback-controlled Random Test Generation}. In
  \bibinfo{booktitle}{\emph{Proceedings of the 2015 International Symposium on
  Software Testing and Analysis}} (Baltimore, MD, USA)
  \emph{(\bibinfo{series}{ISSTA 2015})}. \bibinfo{publisher}{ACM},
  \bibinfo{address}{New York, NY, USA}, \bibinfo{pages}{316--326}.
\newblock
\showISBNx{978-1-4503-3620-8}
\urldef\tempurl%
\url{https://doi.org/10.1145/2771783.2771805}
\showDOI{\tempurl}


\bibitem[\protect\citeauthoryear{Zalewski}{Zalewski}{2014}]%
        {aflfuzz}
\bibfield{author}{\bibinfo{person}{Michal Zalewski}.}
  \bibinfo{year}{2014}\natexlab{}.
\newblock \bibinfo{title}{american fuzzy lop (2.35b)}.
\newblock \bibinfo{howpublished}{\url{http://lcamtuf.coredump.cx/afl/}}.
\newblock


\bibitem[\protect\citeauthoryear{Zhang, Groce, and Alipour}{Zhang
  et~al\mbox{.}}{2014}]%
        {issta14}
\bibfield{author}{\bibinfo{person}{Chaoqiang Zhang}, \bibinfo{person}{Alex
  Groce}, {and} \bibinfo{person}{Mohammad~Amin Alipour}.}
  \bibinfo{year}{2014}\natexlab{}.
\newblock \showarticletitle{Using Test Case Reduction and Prioritization to
  Improve Symbolic Execution}. In \bibinfo{booktitle}{\emph{International
  Symposium on Software Testing and Analysis}}. \bibinfo{pages}{160--170}.
\newblock


\bibitem[\protect\citeauthoryear{Zhang}{Zhang}{2009}]%
        {Zhang2009ICSM}
\bibfield{author}{\bibinfo{person}{H. Zhang}.} \bibinfo{year}{2009}\natexlab{}.
\newblock \showarticletitle{An investigation of the relationships between lines
  of code and defects}. In \bibinfo{booktitle}{\emph{2009 IEEE International
  Conference on Software Maintenance}}. \bibinfo{pages}{274--283}.
\newblock
\showISSN{1063-6773}
\urldef\tempurl%
\url{https://doi.org/10.1109/ICSM.2009.5306304}
\showDOI{\tempurl}


\bibitem[\protect\citeauthoryear{Zheng, Robinson, Williams, and Smiley}{Zheng
  et~al\mbox{.}}{2006}]%
        {Zheng06}
\bibfield{author}{\bibinfo{person}{Jiang Zheng}, \bibinfo{person}{Brian
  Robinson}, \bibinfo{person}{Laurie Williams}, {and} \bibinfo{person}{Karen
  Smiley}.} \bibinfo{year}{2006}\natexlab{}.
\newblock \showarticletitle{Applying Regression Test Selection for COTS-Based
  Applications}. In \bibinfo{booktitle}{\emph{Proceedings of the 28th
  International Conference on Software Engineering}} (Shanghai, China)
  \emph{(\bibinfo{series}{ICSE ’06})}. \bibinfo{publisher}{Association for
  Computing Machinery}, \bibinfo{address}{New York, NY, USA},
  \bibinfo{pages}{512–522}.
\newblock
\showISBNx{1595933751}
\urldef\tempurl%
\url{https://doi.org/10.1145/1134285.1134357}
\showDOI{\tempurl}


\bibitem[\protect\citeauthoryear{Zheng, Williams, and Robinson}{Zheng
  et~al\mbox{.}}{2007}]%
        {Zheng07}
\bibfield{author}{\bibinfo{person}{Jiang Zheng}, \bibinfo{person}{Laurie
  Williams}, {and} \bibinfo{person}{Brian Robinson}.}
  \bibinfo{year}{2007}\natexlab{}.
\newblock \showarticletitle{Pallino: Automation to Support Regression Test
  Selection for Cots-Based Applications}. In
  \bibinfo{booktitle}{\emph{Proceedings of the Twenty-Second IEEE/ACM
  International Conference on Automated Software Engineering}} (Atlanta,
  Georgia, USA) \emph{(\bibinfo{series}{ASE ’07})}.
  \bibinfo{publisher}{Association for Computing Machinery},
  \bibinfo{address}{New York, NY, USA}, \bibinfo{pages}{224–233}.
\newblock
\showISBNx{9781595938824}
\urldef\tempurl%
\url{https://doi.org/10.1145/1321631.1321665}
\showDOI{\tempurl}


\bibitem[\protect\citeauthoryear{Zhou and Leung}{Zhou and Leung}{2006}]%
        {Zhou2006TSE}
\bibfield{author}{\bibinfo{person}{Yuming Zhou} {and} \bibinfo{person}{Hareton
  Leung}.} \bibinfo{year}{2006}\natexlab{}.
\newblock \showarticletitle{Empirical Analysis of Object-Oriented Design
  Metrics for Predicting High and Low Severity Faults}.
\newblock \bibinfo{journal}{\emph{IEEE Transactions on Software Engineering}}
  \bibinfo{volume}{32}, \bibinfo{number}{10} (\bibinfo{date}{Oct}
  \bibinfo{year}{2006}), \bibinfo{pages}{771--789}.
\newblock
\showISSN{0098-5589}
\urldef\tempurl%
\url{https://doi.org/10.1109/TSE.2006.102}
\showDOI{\tempurl}


\bibitem[\protect\citeauthoryear{Zimmermann and Nagappan}{Zimmermann and
  Nagappan}{2008}]%
        {Zimmermann2008ICSE}
\bibfield{author}{\bibinfo{person}{T. Zimmermann} {and} \bibinfo{person}{N.
  Nagappan}.} \bibinfo{year}{2008}\natexlab{}.
\newblock \showarticletitle{Predicting defects using network analysis on
  dependency graphs}. In \bibinfo{booktitle}{\emph{2008 ACM/IEEE 30th
  International Conference on Software Engineering}}.
  \bibinfo{pages}{531--540}.
\newblock
\showISSN{0270-5257}
\urldef\tempurl%
\url{https://doi.org/10.1145/1368088.1368161}
\showDOI{\tempurl}


\end{thebibliography}

\end{document}